\documentclass[%
 reprint,
 superscriptaddress,
nofootinbib,
 amsmath,amssymb,
 aps,
floatfix,
]{revtex4-2}
\usepackage{tabularx, makecell, booktabs} 
\usepackage{graphicx}
\usepackage{dcolumn}
\usepackage{changepage}
\usepackage{bm}
\newcolumntype{L}{>{$}l<{$}} 
\usepackage{hyperref}
\usepackage{tablefootnote} 

\usepackage{siunitx} 

\begin{document}

\preprint{APS/123-QED}

\title{Improved measurements of the TeV--PeV extragalactic neutrino spectrum from joint analyses of IceCube tracks and cascades}

\affiliation{III. Physikalisches Institut, RWTH Aachen University, D-52056 Aachen, Germany}
\affiliation{Department of Physics, University of Adelaide, Adelaide, 5005, Australia}
\affiliation{Dept. of Physics and Astronomy, University of Alaska Anchorage, 3211 Providence Dr., Anchorage, AK 99508, USA}
\affiliation{School of Physics and Center for Relativistic Astrophysics, Georgia Institute of Technology, Atlanta, GA 30332, USA}
\affiliation{Dept. of Physics, Southern University, Baton Rouge, LA 70813, USA}
\affiliation{Dept. of Physics, University of California, Berkeley, CA 94720, USA}
\affiliation{Lawrence Berkeley National Laboratory, Berkeley, CA 94720, USA}
\affiliation{Institut f{\"u}r Physik, Humboldt-Universit{\"a}t zu Berlin, D-12489 Berlin, Germany}
\affiliation{Fakult{\"a}t f{\"u}r Physik {\&} Astronomie, Ruhr-Universit{\"a}t Bochum, D-44780 Bochum, Germany}
\affiliation{Universit{\'e} Libre de Bruxelles, Science Faculty CP230, B-1050 Brussels, Belgium}
\affiliation{Vrije Universiteit Brussel (VUB), Dienst ELEM, B-1050 Brussels, Belgium}
\affiliation{Dept. of Physics, Simon Fraser University, Burnaby, BC V5A 1S6, Canada}
\affiliation{Department of Physics and Laboratory for Particle Physics and Cosmology, Harvard University, Cambridge, MA 02138, USA}
\affiliation{Dept. of Physics, Massachusetts Institute of Technology, Cambridge, MA 02139, USA}
\affiliation{Dept. of Physics and The International Center for Hadron Astrophysics, Chiba University, Chiba 263-8522, Japan}
\affiliation{Department of Physics, Loyola University Chicago, Chicago, IL 60660, USA}
\affiliation{Dept. of Physics and Astronomy, University of Canterbury, Private Bag 4800, Christchurch, New Zealand}
\affiliation{Dept. of Physics, University of Maryland, College Park, MD 20742, USA}
\affiliation{Dept. of Astronomy, Ohio State University, Columbus, OH 43210, USA}
\affiliation{Dept. of Physics and Center for Cosmology and Astro-Particle Physics, Ohio State University, Columbus, OH 43210, USA}
\affiliation{Niels Bohr Institute, University of Copenhagen, DK-2100 Copenhagen, Denmark}
\affiliation{Dept. of Physics, TU Dortmund University, D-44221 Dortmund, Germany}
\affiliation{Dept. of Physics and Astronomy, Michigan State University, East Lansing, MI 48824, USA}
\affiliation{Dept. of Physics, University of Alberta, Edmonton, Alberta, T6G 2E1, Canada}
\affiliation{Erlangen Centre for Astroparticle Physics, Friedrich-Alexander-Universit{\"a}t Erlangen-N{\"u}rnberg, D-91058 Erlangen, Germany}
\affiliation{Physik-department, Technische Universit{\"a}t M{\"u}nchen, D-85748 Garching, Germany}
\affiliation{D{\'e}partement de physique nucl{\'e}aire et corpusculaire, Universit{\'e} de Gen{\`e}ve, CH-1211 Gen{\`e}ve, Switzerland}
\affiliation{Dept. of Physics and Astronomy, University of Gent, B-9000 Gent, Belgium}
\affiliation{Dept. of Physics and Astronomy, University of California, Irvine, CA 92697, USA}
\affiliation{Karlsruhe Institute of Technology, Institute for Astroparticle Physics, D-76021 Karlsruhe, Germany}
\affiliation{Karlsruhe Institute of Technology, Institute of Experimental Particle Physics, D-76021 Karlsruhe, Germany}
\affiliation{Dept. of Physics, Engineering Physics, and Astronomy, Queen's University, Kingston, ON K7L 3N6, Canada}
\affiliation{Department of Physics {\&} Astronomy, University of Nevada, Las Vegas, NV 89154, USA}
\affiliation{Nevada Center for Astrophysics, University of Nevada, Las Vegas, NV 89154, USA}
\affiliation{Dept. of Physics and Astronomy, University of Kansas, Lawrence, KS 66045, USA}
\affiliation{Centre for Cosmology, Particle Physics and Phenomenology - CP3, Universit{\'e} catholique de Louvain, Louvain-la-Neuve, Belgium}
\affiliation{Department of Physics, Mercer University, Macon, GA 31207-0001, USA}
\affiliation{Dept. of Astronomy, University of Wisconsin{\textemdash}Madison, Madison, WI 53706, USA}
\affiliation{Dept. of Physics and Wisconsin IceCube Particle Astrophysics Center, University of Wisconsin{\textemdash}Madison, Madison, WI 53706, USA}
\affiliation{Institute of Physics, University of Mainz, Staudinger Weg 7, D-55099 Mainz, Germany}
\affiliation{Department of Physics, Marquette University, Milwaukee, WI 53201, USA}
\affiliation{Institut f{\"u}r Kernphysik, Universit{\"a}t M{\"u}nster, D-48149 M{\"u}nster, Germany}
\affiliation{Bartol Research Institute and Dept. of Physics and Astronomy, University of Delaware, Newark, DE 19716, USA}
\affiliation{Dept. of Physics, Yale University, New Haven, CT 06520, USA}
\affiliation{Columbia Astrophysics and Nevis Laboratories, Columbia University, New York, NY 10027, USA}
\affiliation{Dept. of Physics, University of Oxford, Parks Road, Oxford OX1 3PU, United Kingdom}
\affiliation{Dipartimento di Fisica e Astronomia Galileo Galilei, Universit{\`a} Degli Studi di Padova, I-35122 Padova PD, Italy}
\affiliation{Dept. of Physics, Drexel University, 3141 Chestnut Street, Philadelphia, PA 19104, USA}
\affiliation{Physics Department, South Dakota School of Mines and Technology, Rapid City, SD 57701, USA}
\affiliation{Dept. of Physics, University of Wisconsin, River Falls, WI 54022, USA}
\affiliation{Dept. of Physics and Astronomy, University of Rochester, Rochester, NY 14627, USA}
\affiliation{Department of Physics and Astronomy, University of Utah, Salt Lake City, UT 84112, USA}
\affiliation{Dept. of Physics, Chung-Ang University, Seoul 06974, Republic of Korea}
\affiliation{Oskar Klein Centre and Dept. of Physics, Stockholm University, SE-10691 Stockholm, Sweden}
\affiliation{Dept. of Physics and Astronomy, Stony Brook University, Stony Brook, NY 11794-3800, USA}
\affiliation{Dept. of Physics, Sungkyunkwan University, Suwon 16419, Republic of Korea}
\affiliation{Institute of Physics, Academia Sinica, Taipei, 11529, Taiwan}
\affiliation{Dept. of Physics and Astronomy, University of Alabama, Tuscaloosa, AL 35487, USA}
\affiliation{Dept. of Astronomy and Astrophysics, Pennsylvania State University, University Park, PA 16802, USA}
\affiliation{Dept. of Physics, Pennsylvania State University, University Park, PA 16802, USA}
\affiliation{Dept. of Physics and Astronomy, Uppsala University, Box 516, SE-75120 Uppsala, Sweden}
\affiliation{Dept. of Physics, University of Wuppertal, D-42119 Wuppertal, Germany}
\affiliation{Deutsches Elektronen-Synchrotron DESY, Platanenallee 6, D-15738 Zeuthen, Germany}

\author{R. Abbasi}
\affiliation{Department of Physics, Loyola University Chicago, Chicago, IL 60660, USA}
\author{M. Ackermann}
\affiliation{Deutsches Elektronen-Synchrotron DESY, Platanenallee 6, D-15738 Zeuthen, Germany}
\author{J. Adams}
\affiliation{Dept. of Physics and Astronomy, University of Canterbury, Private Bag 4800, Christchurch, New Zealand}
\author{S. K. Agarwalla}
\thanks{also at Institute of Physics, Sachivalaya Marg, Sainik School Post, Bhubaneswar 751005, India}
\affiliation{Dept. of Physics and Wisconsin IceCube Particle Astrophysics Center, University of Wisconsin{\textemdash}Madison, Madison, WI 53706, USA}
\author{J. A. Aguilar}
\affiliation{Universit{\'e} Libre de Bruxelles, Science Faculty CP230, B-1050 Brussels, Belgium}
\author{M. Ahlers}
\affiliation{Niels Bohr Institute, University of Copenhagen, DK-2100 Copenhagen, Denmark}
\author{J.M. Alameddine}
\affiliation{Dept. of Physics, TU Dortmund University, D-44221 Dortmund, Germany}
\author{S. Ali}
\affiliation{Dept. of Physics and Astronomy, University of Kansas, Lawrence, KS 66045, USA}
\author{N. M. Amin}
\affiliation{Bartol Research Institute and Dept. of Physics and Astronomy, University of Delaware, Newark, DE 19716, USA}
\author{K. Andeen}
\affiliation{Department of Physics, Marquette University, Milwaukee, WI 53201, USA}
% ...

\date{\today}
\collaboration{IceCube Collaboration}
\noaffiliation
%
%
\begin{abstract}

The IceCube South Pole Neutrino Observatory has discovered the presence of a diffuse astrophysical neutrino flux at energies of TeV and beyond using neutrino induced muon tracks and cascade events from neutrino interactions. We present two analyses sensitive to neutrino events in the energy range \SI{1}{TeV} to \SI{10}{PeV}, using more than 10 years of IceCube data.  
Both analyses consistently reject a neutrino spectrum following a single power-law with significance  $>4\,\sigma$ in favor of a broken power law. We describe the methods implemented in the two analyses, the spectral constraints obtained, and the validation of the robustness of the results. Additionally, we report the detection of a muon neutrino in the MESE sample with an energy of $11.4^{+2.46}_{-2.53} $\,\si{PeV}, the highest energy neutrino observed by IceCube to date. The results presented here show insights into the spectral shape of astrophysical neutrinos, which has important implications for inferring their production processes in a multi-messenger picture.
\end{abstract}
\maketitle

\section{Introduction}
The IceCube Neutrino Observatory is a neutrino detector at the South Pole~\cite{Aartsen:2016_Inst}, which uses the Antarctic glacier as a target and a medium to detect Cherenkov light from neutrino interactions with nucleons in ice. An array of 5160 digital optical modules (DOMs) is deployed along 86 strings, at depths of 1450~m to 2450~m below the surface of the ice. The DOMs capture Cherenkov radiation emitted by charged secondary particles created when neutrinos undergo deep inelastic scattering (DIS) with nucleons in the ice. This radiation is recorded as a 'pulse', which is a measurement of the charge and the arrival time of the detected light.
One of the primary science goals of IceCube is to study the spectrum of high energy astrophysical neutrinos, first detected in 2013~\cite{IceCube:2013_HE,PRL_PeV,IceCube:2014_3HE}. These neutrinos are produced in astrophysical sources via the interaction of high-energy cosmic rays (CRs) with matter or photons in the environment of these sources or during propagation of CRs through space. 
The total observed neutrino flux reflects the superposition of all contributing sources and thus constitutes a calorimetric measurement of the total production of high-energy neutrinos in the Universe \cite{MultiComponentAstroFlux}.
Features in the energy spectrum are particularly interesting as they can indicate changes in the contributing source populations or regions of production~\cite{murase_hidden_2016}.
Measurements of the total neutrino energy spectrum can therefore shed light on the populations of their sources, as well as provide a window into the mechanisms of CR acceleration. 
There are also models that predict the contribution of new physics, such as neutrinos produced from the annihilation of dark matter~\cite{DarkMatterAnnihilation}, that could leave a characteristic imprint on the observed neutrino spectrum. A precise measurement of the spectrum may allow the identification and distinction of the various contributions. Although Fermi acceleration models predict a single power-law spectrum, more detailed models of neutrino production in different sources often predict spectra with distinct features, such as a spectral break.  Models where muons lose or gain energy before decaying will also exhibit spectral features. Spectral features could also help elucidate the relative contribution to the cosmic neutrino flux from $\gamma$-opaque and $\gamma$-transparent sources~\cite{murase_hidden_2016,fang_tev_2022}.

\begin{figure*}[hbt]
    \begin{minipage}[b]{0.24\linewidth}
     \centering
        \includegraphics[width=1\linewidth]{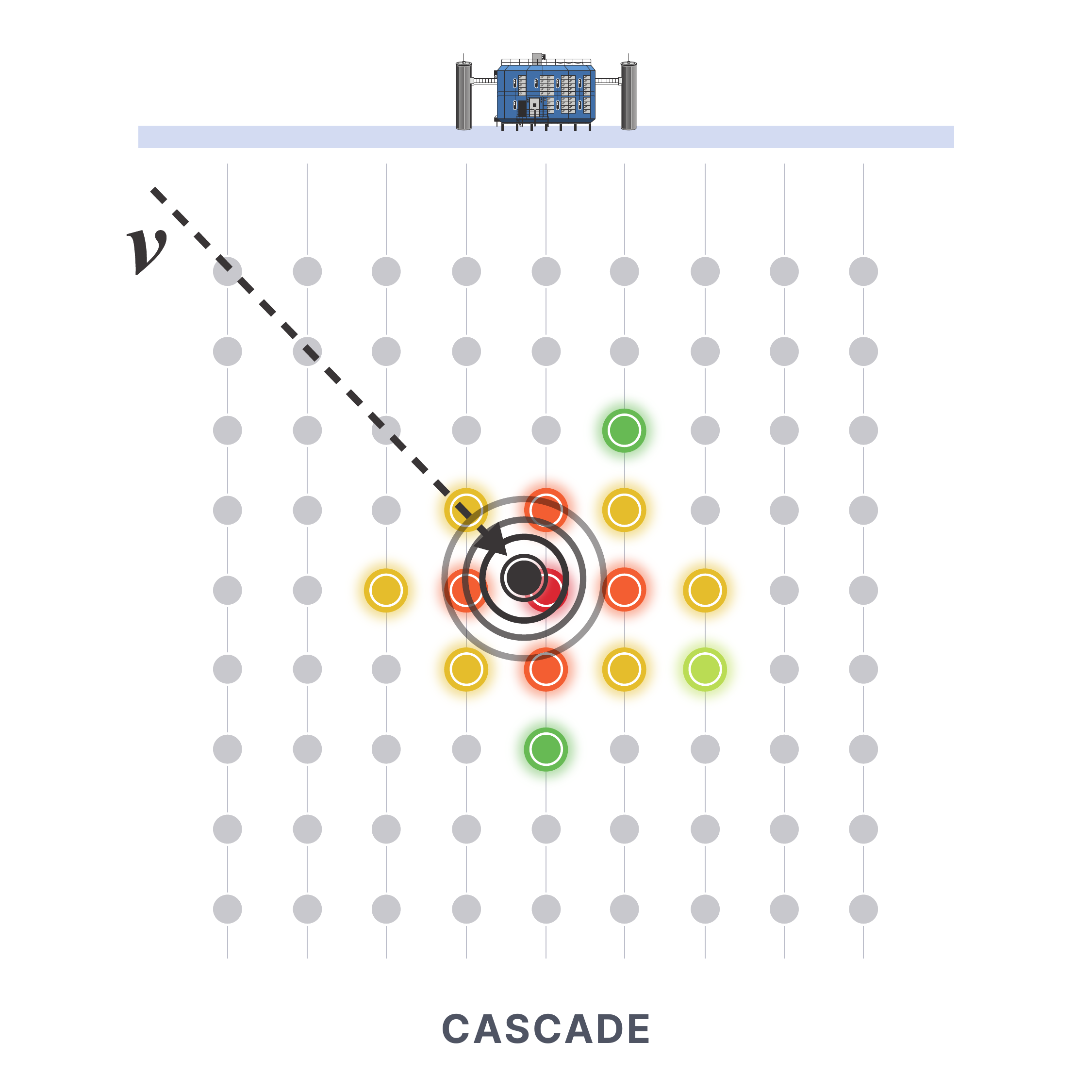}  
    \end{minipage}
    \begin{minipage}[b]{0.24\linewidth}
     \centering
        \includegraphics[width=1\linewidth]{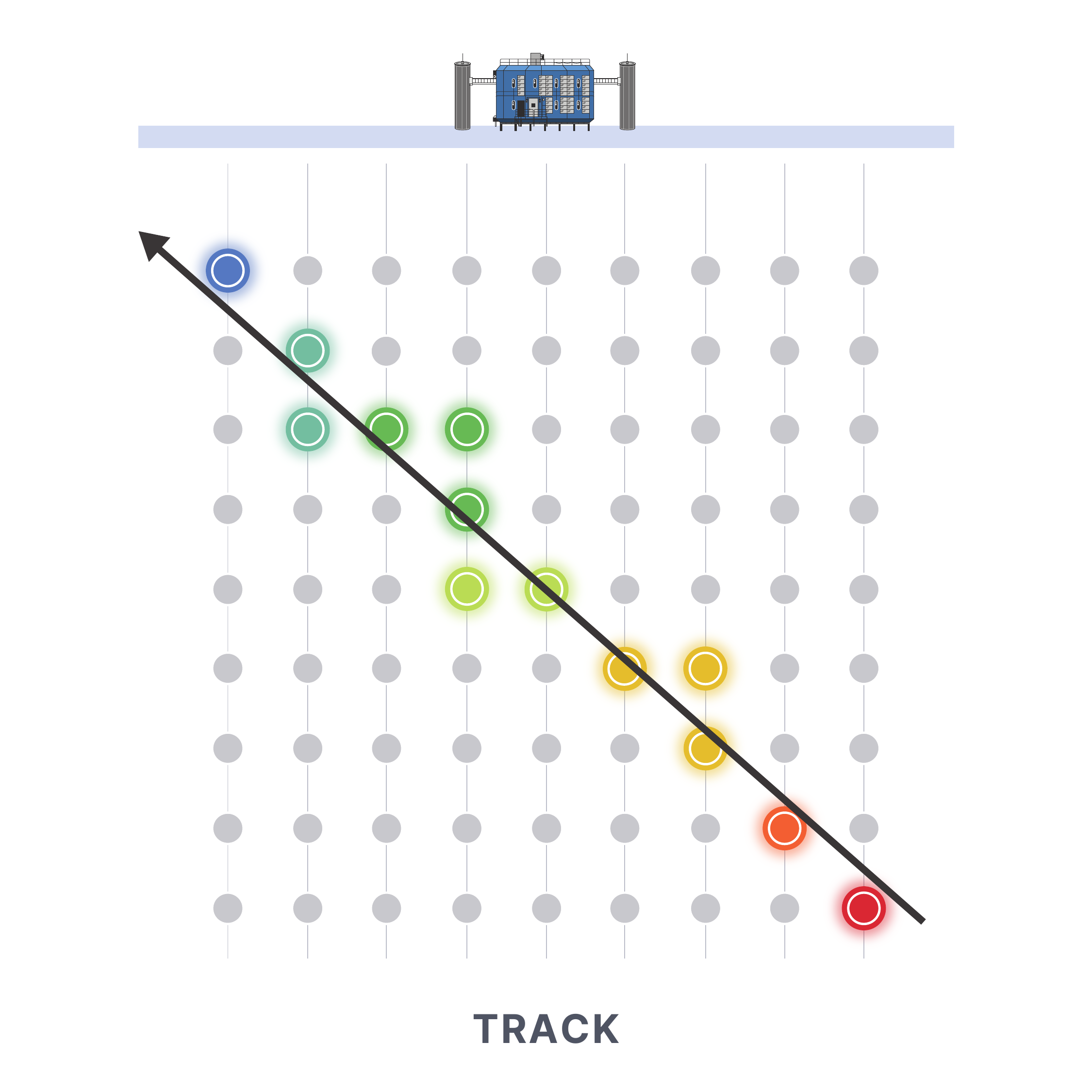}  
    \end{minipage}
    \begin{minipage}[b]{0.24\linewidth}
     \centering
        \includegraphics[width=1\linewidth]{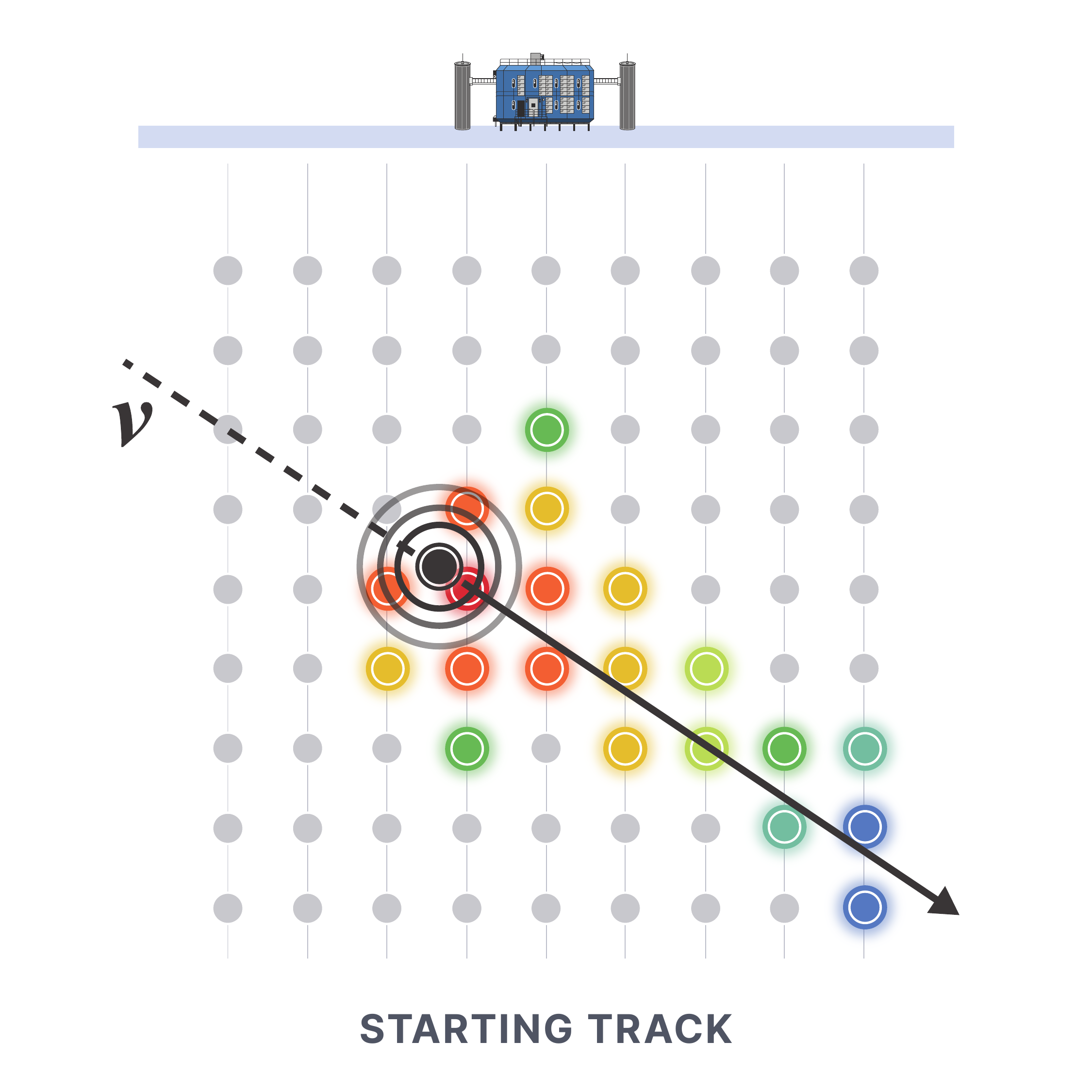}  
    \end{minipage}
    \begin{minipage}[b]{0.24\linewidth}
     \centering
        \includegraphics[width=1\linewidth]{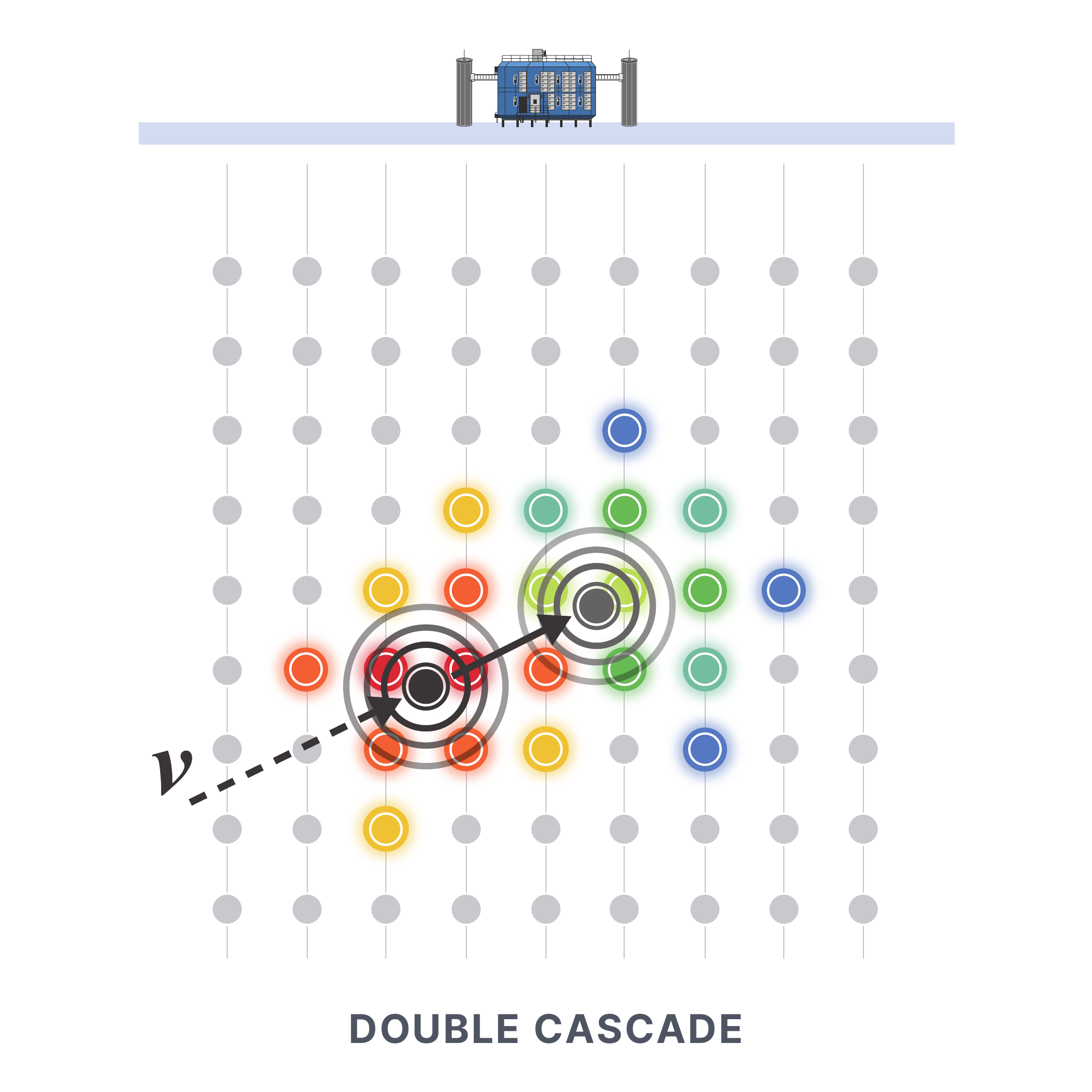}  
    \end{minipage}
    \caption{A schematic of the various event morphologies typically observed by IceCube. The dark grey circle represents the neutrino interaction vertex, with the second double cascade vertex representing tau lepton decay. The other circles represent DOMs which see light. The early pulses are in red, while later pulses are first in yellow, then green and finally blue.}
    \label{fig:event_types}
\end{figure*}

The observed neutrino events in IceCube are typically classified into two morphologies, `tracks' and `cascades'. 
Tracks are observed when muon neutrinos or antineutrinos undergo a charged current (CC) DIS interaction, resulting in an elongated pattern of Cherenkov light photons following the trajectory of the outgoing muon. Daughter muons may also be created by the decay of a $\tau$ lepton, formed by the CC interaction of a $\nu_\tau$, with a small branching fraction of $\sim 17\%$. 
As TeV muons propagate for several kilometers in ice, track events are observable even when the interaction vertex is far from the detector, leading to an effective detection volume that extends far beyond the geometrical detector boundary. 
Cascade events are generated when an electron neutrino or antineutrino undergoes a CC interaction or when a neutrino or antineutrino of any flavor undergoes a neutral current (NC) interaction.  
The resulting electromagnetic, hadronic or mixed shower has a longitudinal extension of typically less than $\sim \SI{10}{m}$, a region which is small compared to the \SI{125}{m} horizontal spacing between detector strings. It therefore results in a more spherical emission of Cherenkov light around the cascade's position.
Cascades can also result from tau neutrino charged-current (CC) interactions, provided the tau lepton lacks sufficient energy to travel enough distance to produce a second cascade from its decay that is distinguishable from the initial cascade. In cases where the two cascades can be separated the events can be classified as `double cascade' events. Such a classification requires a specialized reconstruction and is not applied for the events presented in this paper. 
While cascades produced from CC interactions deposit the majority of the neutrino energy inside the detector, those from NC interactions have a fraction of missing energy carried away by the outgoing neutrino. 
The total number of Cherenkov photons emitted in these events is proportional to the energy deposited within the detector \cite{IceCube:2013dkx,IceCube:2024csv}. 
Track events have superior angular resolution when compared to cascades, due to the long lever arm of the track. 
However, their energy resolution is limited because in most cases only a fraction of the total energy is deposited within the detector volume. 
Cascades on the other hand have better energy resolution as most or all of the deposited energy of the NC or CC event, respectively, is contained within the detector boundaries, and thus visible to the DOMs. 
Through-going tracks (tracks that originate outside the detector volume) are detected at a much higher rate than cascades. This is due to the larger flux of atmospheric muon neutrinos with this morphology, as well as cosmic-ray induced atmospheric muons entering the detector from above. Another important factor is the larger effective volume for through-going tracks, as their interaction vertex is outside the detector.
 `Starting events' form an important subset of both morphologies, defined as events where the neutrino interaction vertex is contained within the detector volume. This gives access to the initial interaction vertex of the neutrino, and is beneficial for distinguishing neutrino events from atmospheric muon events. As the initial hadronic cascade from the neutrino interaction lies within the detector volume, the energy resolution is improved with respect to through going tracks. 
An illustration of the different event types is shown in Fig.~\ref{fig:event_types}.

IceCube has measured the spectrum of astrophysical neutrinos in multiple detection channels, depending on the morphology of the events in the sample. 
These include the first detection of high-energy astrophysical neutrinos in 2013~\cite{IceCube:2013_HE}, using a sample of high-energy starting events (HESE). The HESE measurement found evidence of a flux of astrophysical neutrinos above \SI{60}{TeV} statistically compatible with a single power law (SPL, $\Phi\propto E^{-\gamma}$), a measurement which was repeated with an extended HESE dataset with additional years of data~\cite{HESE7.5}. 
Further measurements of the astrophysical neutrino flux were performed using a sample of through-going muon neutrino tracks (Northern tracks)~\cite{NorthernTracks} observed in 9.5 years of IceCube data, along with a measurement using cascade events dominated by electron and tau neutrino flavors with 6 years of IceCube data~\cite{SBUCascades}, and, most recently, ESTES,
a sample of starting tracks with a lower energy threshold than the HESE searches~\cite{ESTES}.
These measurements are compatible with an SPL spectrum. 

Here we present an updated measurement of the astrophysical flux with two complementary analyses: 
\begin{itemize}
    \item  The \textbf{MESE} analysis, which updates the study reported in~\cite{MESE_2yr}, focuses on starting events. The study of starting events at TeV energies has numerous advantages, including sensitivity to all neutrino flavors and events from the entire sky. The dataset consistently selects cascades and tracks through the same background vetoing techniques. Extending the veto-based approach of HESE to lower energies 
allows us to study the cosmic neutrino spectrum across a wide energy range, of almost four orders of magnitude. This was the motivation behind the Medium Energy Starting Events (MESE) sample~\cite{MESE_2yr}, which has been updated~\cite{MESE_ICRC} with improved veto techniques and additional years of IceCube data, as detailed in Sec.~\ref{sec:MESE_Selection}.
    \item The \textbf{Combined Fit (CF)} analysis 
    combines a sample of contained cascades \cite{SBUCascades}, with 5 additional
years of data~\cite{naab2023measurementastrophysicaldiffuseneutrino}, and a sample of through-going tracks \cite{NorthernTracks}
    that were
    used previously to measure the spectrum separately. The advantages of performing a joint fit with cascades and tracks were demonstrated in a previous publication~\cite{icecube_collaboration_combined_2015}, where the high statistics of the tracks sample and the low muon content of the cascades sample help strengthen the analysis. This idea is also followed in the new CF analysis, which is described further in Sec.~\ref{sec:combinedfit}.
\end{itemize}

Both analyses were developed separately, to measure the neutrino spectrum. We report the tested astrophysical spectral shapes, also discussed in our companion paper~\cite{IceCube:2025tgp}.
We also report the highest energy neutrino detected by IceCube so far, in this paper. IceCube has conducted studies exclusively on neutrino events in the PeV scale~\cite{IceCube:2025ary}, with results consistent with background simulations, and constrained the spectrum at extremely high energies with the highest energy sample of events~\cite{EHE_PRL}. The highest energy event reported here is also contained in the sample used in~\cite{HESE12} and was issued as a public alert via the General Coordinates Network~\cite{2019GCN.24028....1I}.

In Sec.~\ref{sec:datasamples} we describe the data samples used for the two analyses, including the simulations used for the studies. Sec.~\ref{sec:methods} presents the analysis methodology and the obtained results are presented in Sec.~\ref{sec:results}. We further present the cross-checks performed for the two analyses in Sec.~\ref{sec:crosschecks} concluding with a discussion in Sec.~\ref{sec:discussion}.

\section{Data Samples}
\label{sec:datasamples}
In this section, we discuss the different data samples used in the analyses, as well as the simulation used. The new MESE selection, in particular, will be explained in detail in this section. For the 
CF analysis, the respective data selections for tracks and cascades remain unchanged with respect to previous publications, and we refer to the detailed descriptions in \cite{NorthernTracks,IceCube:2016umi} for the track selection and \cite{SBUCascades} for the cascade selection. The number of cascade and track events in each analysis is shown in Tab.~\ref{tab:event_numbers}.
\begin{table}[h!]
\vspace{-5mm}
\caption {Number of events in each data sample (E $>$ 1 TeV)}
\begin{tabular}{c|c|c|c}
 & MESE & CF & Overlap \\ \hline
Cascades & 4949 & 10569 & 2514 \\ \hline
Tracks & 4908 & 231486 & 1799
\end{tabular}
\label{tab:event_numbers}
\vspace{-6mm}
\end{table}
\subsection{Signal and Background Simulations}
\label{sec:detandsim}
For the analyses presented here, the main background comes from leptons produced in cosmic-ray air showers, consisting of atmospheric muons and atmospheric neutrino fluxes. The atmospheric neutrino background arises from both the decay of pions and kaons, referred to as the conventional flux, and the decay of charmed mesons, known as the prompt flux. Both the signal, which is the astrophysical neutrino flux,  and the background are modeled with the help of Monte Carlo (MC) simulations.
Both analyses consistently use the same set of MC simulations to describe the physics processes of the observed events.
The \textsc{NuGen} software~\cite{NuGen}  is used to simulate the 
DIS interactions of neutrinos in the detector volume. For the interactions, the Cooper-Sarkar-Mertsch-Sarkar (CSMS) model~\cite{CSMS} for the neutrino-nucleon cross sections is used with the Earth's density described by the Preliminary Reference Earth Model (PREM). 
Using the \textsc{MCEq} software package~\cite{fedynitch_calculation_2015,MCEq} atmospheric neutrino flux weights are calculated according to model predictions from the Gaisser H4a primary CR flux model~\cite{gaisser_spectrum_2012}, and assuming hadronic interactions as described by the Sibyll 2.3c model~\cite{riehn_hadronic_2018}. Cosmic-ray muons, which form another major portion of the background 
are simulated with the \textsc{MuonGun} package~\cite{jvs_thesis} which generates samples of single muons with properties derived from simulations of cosmic-ray air showers generated with \textsc{CORSIKA}~\cite{corsika}, using Sibyll 2.1~\cite{SIBYLL2.1} as the hadronic interaction model. 
The use of \textsc{MuonGun} significantly reduces the computing time for the MC generation compared to the direct simulation of CR air showers with \textsc{CORSIKA} and thus enables us to generate the background estimation with sufficient statistical power. 
The secondary particles produced in the neutrino interactions and the atmospheric muons are propagated through the detector, to obtain the light pulses recorded in the DOMs. 
We use \textsc{PROPOSAL}~\cite{Proposal} for the propagation of secondary muons and taus, and ``Cascade Monte Carlo" (\textsc{CMC}) \cite{IceTray_CMC} for electromagnetic cascades, with an appropriate scaling factor applied for hadronic cascades. This energy-dependent scaling factor, derived from phenomenological parameterizations, accounts for the fact that only a fraction of the hadron energy appears as visible electromagnetic cascades~\citep{Internal_note_cmc, jvs_thesis}.
The ray-tracing software package \textsc{CLSim}~\cite{clsim} is used for generating and propagating the photons in the ice that eventually get detected in the DOMs. 
We apply flux weighting schemes to the MC-generated neutrinos (generated with a spectral index of 1 or 1.5) and the muons to efficiently obtain adequate statistics for each component and energy range contributing to the observed events: single muon, conventional atmospheric neutrino, prompt atmospheric neutrino, and astrophysical neutrino fluxes. Note that we do not explicitly account for muon bundles (several muons boosted along the trajectory of the primary) since comparisons of full \textsc{CORSIKA} and \textsc{MuonGun} simulations have shown that they get efficiently rejected at the final level of the event selection, as studied and verified during the development of the contained cascades sample used in the CF analysis~\cite{SBUCascades} and the MESE event selection described below.

\subsection{MESE Selection Procedure}\label{sec:MESE_Selection}
The main goal of the event selection is to reduce the background from muons produced in CR air showers in the data to a level where one can conduct measurements with the neutrinos that interact within the detector volume. The event selection is developed based on the MC samples of neutrinos and muons, as described in Sec.~\ref{sec:detandsim}.
A pre-selection cut is applied to reject events with a low number of hits (photons captured by the DOMs), which cannot be reliably reconstructed. 
These pre-selection cuts are: the requirement that the event triggers one of the IceCube event-filter streams~\cite{Aartsen:2016_Inst}, a minimum deposited-charge threshold of 100 photoelectrons (pe), and a minimum number of 3 strings to have hits on them. These events are further used in the event selection.

\subsubsection{Outer-layer veto} 
The first stage of the event selection rejects events that deposit charge at the outer edges of the detector. This region is called a `veto' layer.
The veto layer is defined as: all strings at the edge of the detector, a 90\,m layer at the top of the detector, and a single layer of DOMs at the bottom of the detector. An additional 120\,m thick layer of horizontal veto region is also included. This region is below a band of high dust  concentration in the glacial ice~\cite{SouthPoleIceMeasurement2013}. The purpose of this additional veto layer is to catch the muons that sneak through the dust undetected and can appear as starting events in the clear ice below this layer of dust.  
Events with a total deposited charge $>\,6000$ pe are tagged as HESE events within this sample. 
These HESE events are required to deposit a maximum charge of 3\,pe or less during the initial phase of the event in the outer layer of the detector, consistent with the requirement in previous selections~\cite{HESE7.5}.
Lower-charge events ($<\,6000$\,pe) have a strict requirement of zero charge in this veto layer within a time window of \SI{3}{\micro\second} with respect to the vertex time. 
The vertex time is approximated as the time at which the cumulative charge in the event reaches a certain threshold. The threshold charge that defines the vertex time is 250\,pe for HESE-tagged events, and is adjusted for lower-charge events as 1/24th of the total charge.
This outer layer veto condition vetoes a majority of the bright and/or bundle muons and suppresses the background by 4 orders of magnitude (see Fig.~\ref{fig:MESE_Rates}).
 
\subsubsection{Down-going track veto} 
\begin{figure}[t]
\includegraphics[width=1\linewidth]{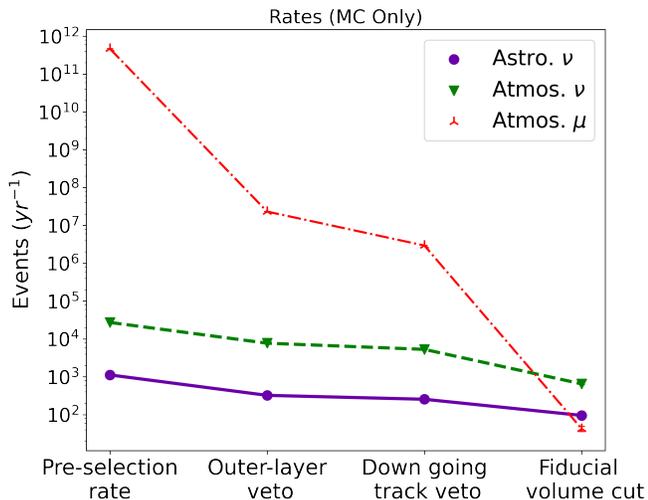}
\caption{\textbf{MESE Expected Rates:} The expected rates of astrophysical neutrinos and background atmospheric neutrinos and muons, derived from MC, after each stage of the event selection. Here we assume a single power law astrophysical flux with normalization $\phi_0\,=\,2.06\times\rm{10^{-18}GeV^{-1}cm^{-2}s^{-1}sr^{-1}}$, and spectral index $\gamma$ = 2.46 (from the 2-year MESE analysis~\cite{MESE_2yr}) and the Gaisser H4a with Sibyll 2.3c atmospheric flux model. The cuts (see text) suppress the muon flux by ten orders of magnitude at the final level, while retaining a larger proportion of signal neutrinos. 
}
\label{fig:MESE_Rates}
\end{figure}

The second step removes muon tracks which sneak undetected through the outer layer veto. 
For this, we calculate the propagation time of light from the reconstructed neutrino interaction vertex inside the detector to each DOM. This  restricts the allowed time range of detected photons to be consistent with the expectation from a neutrino interaction. We select pulses that are determined to \textit{not} be causally connected to the event vertex. The vertex used for the down going track veto is obtained from a cascade reconstruction performed at this stage of the selection.
It is then determined if the selected non-causal pulses can be associated with an incoming muon track by performing track reconstructions along 104 down-going trajectories (centered on a HEALPIX grid~\cite{healpix} with nside~=~4) passing through the vertex. 
We look within a \SI{100}{m} distance from the reconstructed track direction to identify the \textit{veto pulses}. Those that fall within a time window of \SI{-15}{ns} to \SI{1000}{ns} with respect to the propagation time from the interaction vertex are retained as veto pulses.
The track-fit associated with the largest summed charge from these veto pulses is chosen as the best-fit veto track.
For events with charge $> \SI{1000}{pe}$ ($< \SI{1000}{pe}$) we require the veto charge to be less than \SI{2}{pe} (\SI{0.5}{pe}) for the event to be retained in the event sample. This procedure reduces the atmospheric muon rate by an order of magnitude, while retaining a majority of the neutrino events as shown in Fig.~\ref{fig:MESE_Rates}. 

At this stage we also classify events as cascades and tracks, using a deep neural network classifier~\citep{DNN_TheoGlauch,dissertationTheo} and perform separate energy and angular reconstructions for them. A maximum-likelihood fit is performed to obtain the reconstructed energy and direction of the cascade events~\cite{IceCube:2024csv}. For track events, we first perform a direction reconstruction using a maximum likelihood method that assumes an infinite muon track and approximates all the deposited light to be emitted by the Cherenkov emission from the muon~\cite{IceCube:2013dkx}. 
We subsequently perform an  additional likelihood fit for energy reconstruction which reconstructs stochastic energy losses along the track direction~\citep{IceCube:2013dkx,wallace2016millipede}.
We split the track into consecutive small segments along its direction and fit the light emission from each segment as a point-like emission. The summed energy for all segments
provides a good energy estimate for starting tracks, since the initial hadronic cascade also gets well reconstructed with this method.

The following list of cuts are also applied at this level to eliminate pile-up events from coincident air showers. These are cases where multiple events occur within the detector at the same time, which are difficult to reconstruct. We first compute the average charge-weighted distance from the DOMs to the reconstructed event direction\footnote{The charge-weighted distance is defined as
$    D_{\mathrm{avg}} = \frac{\Sigma_i d_{\perp,i}\times q_i}{\Sigma_i q_i}$,
where $q$ represents the charge on each DOM and $d_{\perp}$ is the closest distance from the DOM to the reconstructed direction.} (for both the cascade and the track hypotheses).
If the hits in the detector are from two coincident events, $D_{\mathrm{avg}}$ would be quite large compared to causally connected hits from a single event. 
We therefore remove events with $D_{\mathrm{avg}} >$ \SI{150}{m} with respect to the cascade reconstruction and $D_{\mathrm{avg}} >$ \SI{110}{m} for the track reconstruction. 
The next coincidence cut checks the opening angles between different track fits, using different algorithms and seed guesses, applied to the event. 
If the opening angle is greater than 30$^\circ$, this is likely a coincident event and is also removed. 
Lastly, we apply a cut based on the reduced log-likelihood of the reconstruction, a proxy for the goodness-of-fit of the reconstruction. As these reconstructions attempt to fit a single event within the detector, the goodness of the fit is worse for coincident events. 
\begin{figure*}[t!]
    \begin{minipage}[b]{0.49\linewidth}
     \centering
        \includegraphics[width=0.95\linewidth]{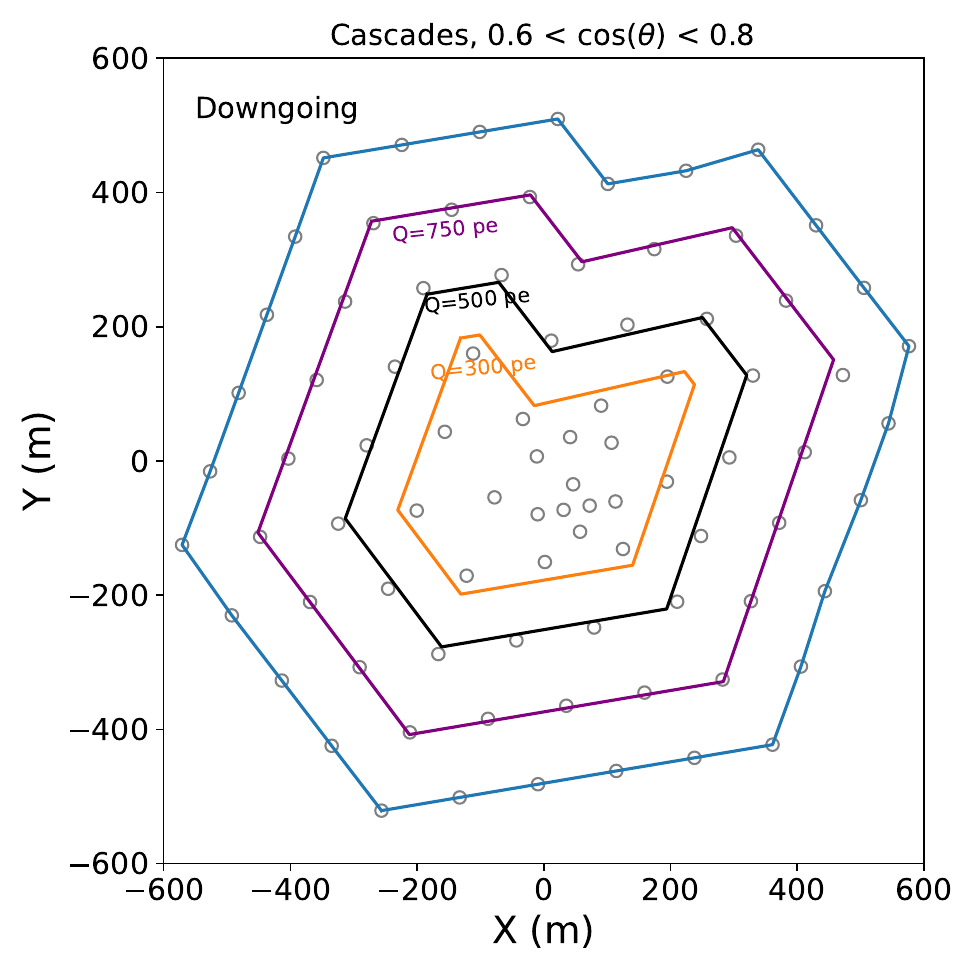}  
    \end{minipage}
    \begin{minipage}[b]{0.49\linewidth}
     \centering
        \includegraphics[width=0.95\linewidth]{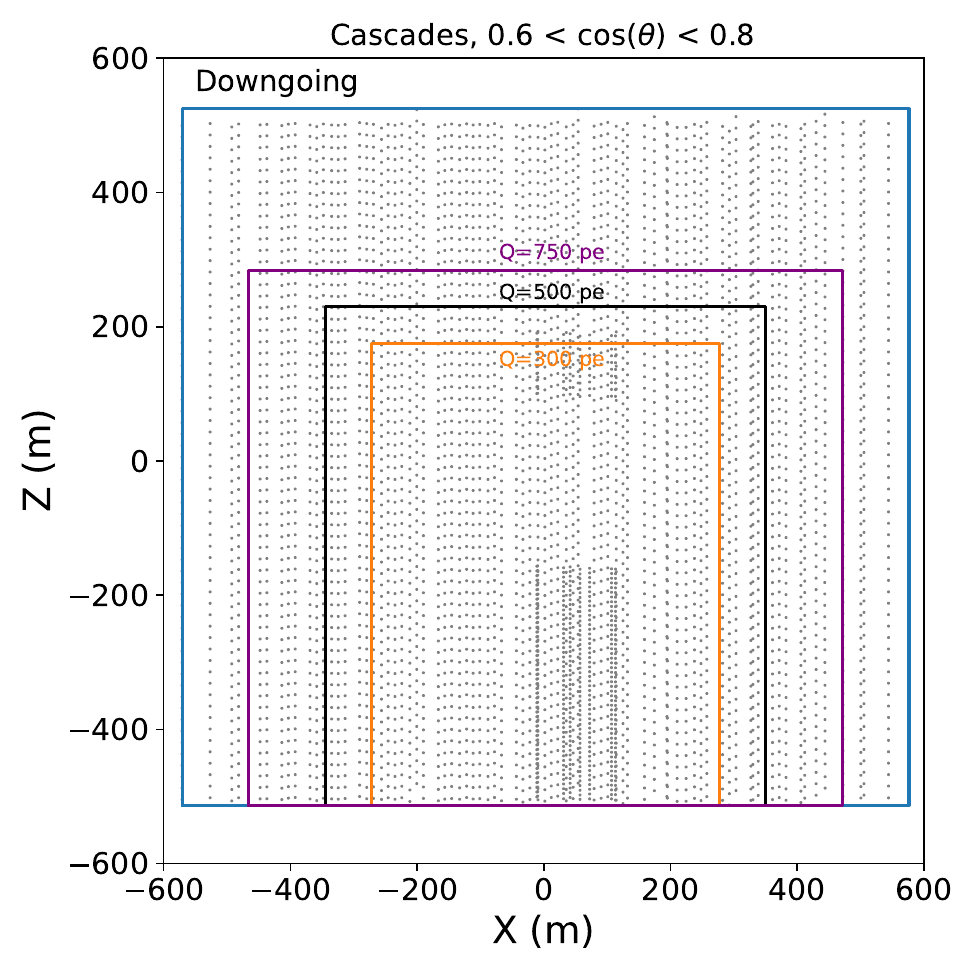}  
    \end{minipage}
    \begin{minipage}[b]{0.49\linewidth}
     \centering
        \includegraphics[width=0.95\linewidth]{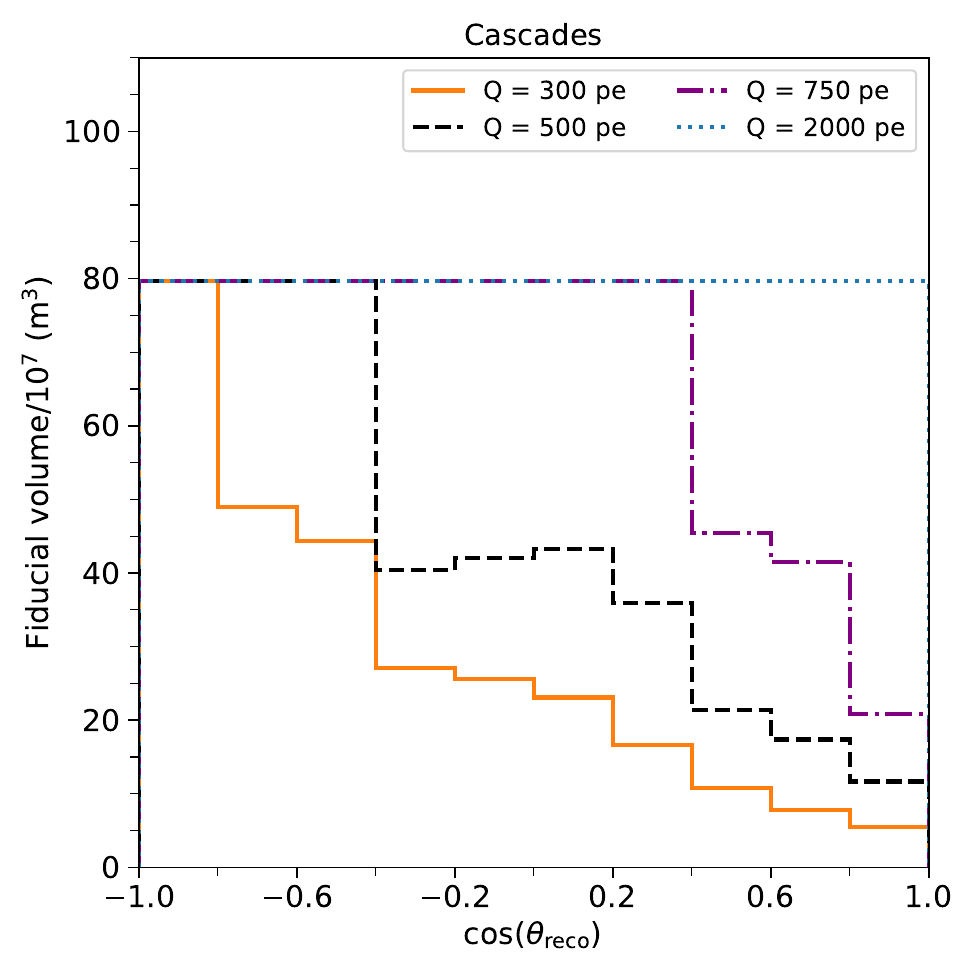}  
    \end{minipage}
    \begin{minipage}[b]{0.49\linewidth}
     \centering
        \includegraphics[width=0.95\linewidth]{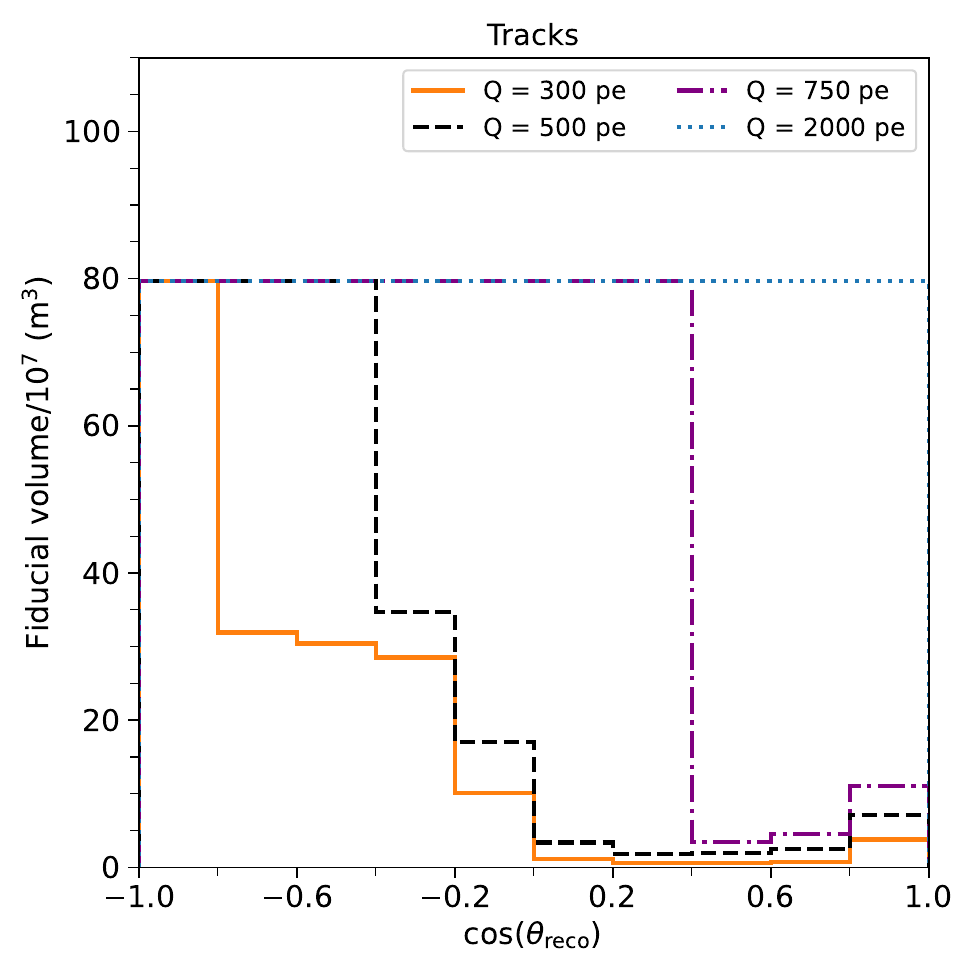}  
    \end{minipage}
\caption{\textbf{MESE Fiducial Volume Cuts:} The selection cut in MESE rejecting a majority of the dim muons, at the final stage of the event selection. The cut depends on the charge, the zenith
angle, and the morphology of the event. We use continuous functions of charge in each zenith bin for this. The cuts shown here are representations at specific charges. A down-going event has more stringent cuts than an up-going
event. Down-going tracks in particular are subject to strict cuts. The two panels at the top show the views from the top and the side of the detector. The allowed margins are shown for cascades events that fall within a given cos($\theta_{\mathrm{reco}}$) bin and have various total deposited charges. X, Y, and Z represent the distance in a cartesian coordinate system, where the origin is at the center of the detector. The bottom two panels show the allowed fiducial volume for cascades (left) and tracks (right) with a given total deposited charge and arriving from different directions. A maximum fiducial volume of $80\times 10^7$~m$^3$ is allowed for all events.} 

\label{fig:FiducialVol}
\end{figure*}

\subsubsection{Fiducial volume cut} The final stage of the event selection suppresses the remaining muon background within the sample. 
The cut achieves a reduction of the rate of low-energy muons producing little light in the outer detector (``dim'' muons) by requiring that the vertex of the lower energy events lies deeper inside the detector, see Fig.~\ref{fig:FiducialVol}. A dim muon is more likely to sneak through a single veto layer due to its reduced light yield and the stochastic nature of its energy loss, which may result in insufficient light deposition for detection.

A veto margin is defined by the amount of distance from the top edge or side edge of the detector that is treated as a veto layer.
The probability of a muon mimicking a bright neutrino interaction while being undetected strongly decreases with the increasing observed event energy. Therefore,
higher-energy events required a smaller veto margin than lower-energy events.

The applied fiducial volume cut depends on total charge ($Q_{\mathrm{tot}}$), reconstructed zenith ($\theta_{\mathrm{reco}}$), and morphology as explained in the following. 
The veto margins ($M$) to the side and top are parameterized as  
a function of charge in each cos\,$\theta_{\mathrm{reco}}$ bin. This function is shown in Eq.~\ref{eq:FiducialCut}
\begin{equation}
M\,=\,a + b\,[c-(\mathrm{log}_{10}(Q_{\mathrm{tot}}/\mathrm{pe}))]^{1/d},
\label{eq:FiducialCut}
\end{equation}
where $a$, $b$, $c$, and $d$ are fixed parameters in each zenith bin. These parameters are obtained by fitting the polynomial in Eq.~\ref{eq:FiducialCut} to MC events. The desired veto margin cut for a given charge is obtained by maximizing the ratio of the atmospheric neutrino to atmospheric muon rates, from simulations, which keeps the cut agnostic to the astrophysical flux model. 
These margins are zenith dependent as muons mainly enter the sample from the down-going sky, and we expect different muon rates from each direction due to the difference in the overburden that the muons have to travel through. 
Therefore, we apply stricter 
cuts to the near vertically down-going bins when compared to the horizontal directions.
The amount of charge a muon deposits in the top/side of the detector depends on the direction from which it arrives. 
For example, the deposited charge on the side margin for a horizontal muon is correlated with its energy. Similarly, for the vertical muons it is the charge deposited in the top margin that shows a correlation with the energy.
A small contamination of the event sample due to misreconstructed atmospheric muons is expected. However, their rate is very low compared to the rate of down-going events. 
Therefore, a softer cut on events reconstructed in the up-going direction retains a majority of the events from these directions. The zenith-angle dependence of the fiducial volume cuts rejects a majority of the muon background while retaining a large portion of signal events. This boosts our event rates and improves our sensitivity, especially at energies below \SI{10}{TeV}.

Furthermore, the fiducial volume cut is implemented to be morphology dependent. We expect more contamination of muons in the starting tracks channel than in the starting cascades channel, especially in the down-going direction. 
These cuts are applied in ten cos($\theta_{\mathrm{reco}}$) bins, corresponding to the ten analysis bins used for MESE (see Sec.~\ref{sec:methods} for details on the analysis bins), for both cascades and tracks independently. 
Additionally, we require a minimum thickness of \SI{100}{m} in the top margin, to reduce the possibility of down-going muons accidentally failing the veto condition. We also require that the vertex of the event lies within the detector volume itself, to ensure that the events are truly starting events. These fiducial volume cuts are applied only on the events \textit{not} tagged as HESE in the event selection. Figure~\ref{fig:FiducialVol} shows the side and top views of the fiducial volume cuts at discrete charge values for cascades in a given zenith bin (cos($\theta_{\mathrm{reco}}$) $\in (0.6,0.8)$). The overall dependence on cos($\theta_{\mathrm{reco}}$) for some example values of the deposited charge in the detector is also shown in the figure, for cascades and tracks.
The fiducial volume cut is responsible for a significant reduction in muon rates during the event selection process, suppressing muons by $\sim 5$ orders of magnitude, as shown in Fig.~\ref{fig:MESE_Rates}. The cut thresholds for each of the selection variables were obtained by comparing signal and background rates using simulated events. To validate the consistency of these  variables, we compared a subsample of data (10\% of two years) to simulations using nominal values for the fit parameters, to verify overall agreement. 
We checked this agreement at each stage of the cuts for the MESE selection, for the charge distributions along with the energy and direction reconstructions once those are performed. This ensured that no features in our selection were introduced while we performed the cuts. We additionally checked for agreement between simulations and the sub-sample data by removing a single cut on the margins (side/top) at a time, and verifying that the simulation can explain the data observed. This subsample of data was later discarded from the final analysis to avoid bias.
All events that pass these veto stages are retained in the final sample. Events during data-taking periods with least 5035 active DOMs are used for the MESE analysis described in Sec.~\ref{sec:methods}.

\subsection{Combined Fit (CF)}
\label{sec:combinedfit}
The CF analysis was developed as a framework to jointly analyze event samples within IceCube in a consistent manner. Here, specifically, the analysis combines the through-going tracks data sample (Northern tracks)~\cite{IceCube:2016umi,NorthernTracks} and the cascades data sample (contained cascades)~\cite{SBUCascades}. The tracks data sample utilizes the Earth as a shield to remove atmospheric muons and retain only tracks from neutrino events, and therefore focuses on events from the Northern Hemisphere. The cascades data sample picks out cascades contained within the detectors with the help of a boosted decision tree and therefore reduces the atmospheric-muon background naturally. Both data samples are well established and are used in multiple IceCube analyses~\cite{IceCube:2022der,IceCube:2018ndw,IceCube:2018tkk,IceCube:2019lzm}, and particularly some of the previous measurements of the diffuse astrophysical neutrino spectrum were based on these samples. 
The samples differ strongly in the experimental signatures, making them ideal for use in combination.
The selections result in a small overlap of 70 common events, which is \SI{0.5}{\percent} of the cascade signal sample and negligible in the through-going track sample.
By combining the two samples consistently in this analysis we can fully exploit the complementarity of good energy resolution for cascades with high statistics of through-going muons which strongly constrains detector and background-rate uncertainties. 
The analysis presented here combines these two samples.
In this analysis, the data from the preliminary configuration of IceCube with only 59 strings has been excluded from the track sample and the cascade sample has been extended to include 10.5 years of IceCube data. All data have been uniformly reprocessed with improved detector calibrations~\cite{2020JInst..15P6032A}. For achieving fully disjoint samples, events which pass the cascade selection are excluded from the through-going track sample where their contribution is marginal.
This analysis also contains a significant methodological improvement over a previous combined fit performed using tracks and cascades data \cite{icecube_collaboration_combined_2015}
by utilizing a fully consistent treatment of systematic uncertainties and jointly produced MC data samples.

\section{Analysis Methodology}
\label{sec:methods}

We use a binned forward folding method to fit both the MESE sample and the samples used for CF, which is performed using a common software tool (see Sec.~\ref{sec:nnmfit}). Data counts in a 2D binned representation of reconstructed energy and cos($\theta_{\mathrm{reco}}$) are compared to expectation from both background and signal calculated from simulations under several hypotheses with free parameters (see Sec.~\ref{sec:AstrophysicalFitModels} for the tested hypotheses with a description of the free parameters). This corresponds to a binned likelihood of the form 
\begin{equation}
\mathcal{L}\,=\,\prod_{i=1}^{n_{\mathrm{bins}}} \mathcal{L}_{i,\mathcal{H(\eta)}},
\label{eq:likelihood}
\end{equation}
where $n_{\mathrm{bins}}$ runs over all 2D-analysis bins. $\mathcal{L}_{i}$ is a modified Poisson likelihood under a signal hypothesis, $\mathcal{H}$ and $\eta $ are the free and nuisance parameters. The modified Poisson likelihood for the i-th bin accounts for the limited MC statistics available for calculating the bin expectations, and incorporates the additional uncertainty related to this as described in~\cite{Leff}. 
The components that enter the likelihood include the expected counts of astrophysical neutrinos, according to the given flux model, which is the signal term and is expressed via the normalization, spectral index, and additional parameters where applicable. We also include the background components: conventional neutrino flux, the prompt neutrino flux, and the muon background. They are included as separate normalization parameters for each component.
All considered systematic uncertainties are included as nuisance parameters in the likelihood (see Sec.~\ref{sec:nuisanceparams} and Sec.~\ref{sec:snowstorm}). The sensitivity of the analysis to the different flux models was validated by performing injection-recovery tests, where a given flux model is injected and we attempt to recover the input parameters. We repeated these tests for values of the nuisance parameters which were shifted from the nominal values, again recovering the injected spectrum. This test confirms the ability of the analysis to recover the underlying astrophysical flux model accurately.
We first perform a fit remaining blind to the astrophysical flux parameters, and ensure that the nuisance parameters are well behaved and within the allowed bounds imposed on them. This avoids any bias while performing the measurement. We followed a staged unblinding procedure, where we first performed a fit to only 10\% of our data, and checked agreement between data and simulation. The degree of agreement was computed using a $\chi^{2}$ fit to the log-likelihood distribution generated from pseudo-data, requiring a p-value $\geq$ 0.05 to verify that the histograms were compatible. Upon passing these criteria, we proceeded to fit the complete dataset. This system of checks helped ensure that our analysis was able to describe the observed data.

The bins used in both analyses are shown in Tab.~\ref{tab:Bins}.
\begin{table}[b]
\vspace{-1mm}
\caption {2D-Binning used in both analyses. Both the energy, and the zenith angle, $\theta$, are reconstructed quantities. The binning is optimized for the sensitivity and the resolution of the observable in the respective morphology.}
\begin{tabular}{c|cc|cc}
             & \multicolumn{2}{c|}{Cascades}                & \multicolumn{2}{c}{Tracks}                   \\ \cline{2-5} 
Analysis     & \multicolumn{2}{c|}{E (GeV) $|$ cos($\theta$)}
& \multicolumn{2}{c}{E (GeV) $|$ cos($\theta$)} \\ \hline  
MESE         
& \multicolumn{1}{c|}{\begin{tabular}[c]{@{}c@{}}($10^3-10^7$)\\ 22 bins\end{tabular}} & \begin{tabular}[c]{@{}c@{}}(-1, 1)\\ 10 bins\end{tabular} & \multicolumn{1}{c|}{\begin{tabular}[c]{@{}c@{}}($10^3-10^7$)\\ 13 bins\end{tabular}} & \begin{tabular}[c]{@{}c@{}}(-1, 1)\\ 10 bins\end{tabular}    \\ \hline
\begin{tabular}[c]{@{}c@{}}Combined \\ Fit\end{tabular} & \multicolumn{1}{c|}{\begin{tabular}[c]{@{}c@{}}($4\times10^2-10^7$)\\ 22 bins\end{tabular}} & \begin{tabular}[c]{@{}c@{}}(-1, 1)\\ 3 bins\end{tabular}  & \multicolumn{1}{c|}{\begin{tabular}[c]{@{}c@{}}($10^{2.5}-10^7$)\\ 45 bins\end{tabular}} & \begin{tabular}[c]{@{}c@{}}(-1, 0.09)\\ 34 bins\end{tabular}
\end{tabular}
\label{tab:Bins}
\end{table}
The samples use different binning based on their respective resolutions and the available statistics. For the zenith angle, the angular resolution between tracks and cascades strongly differs, resulting, e.g.,\ in only three not-equal-sized bins for the cascades in the CF.
In the case of the energy proxy, the samples inherently provide a large variation of correlations with the initial neutrino's energy.
Contained cascades reflect a calorimetric measurement of the total neutrino energy in case of charged current interactions, but not for neutral current interactions. In case of MESE starting events the interaction is contained in the detector but for tracks and neutral current interactions, part of the energy is not observed. In case of tracks for the CF, the initial neutrino interaction is usually far outside the detector boundaries and instead of the neutrino energy, only the muon energy within the detector volume can be estimated
from the observed energy loss.

\subsection{Analysis Framework}
\label{sec:nnmfit}
The procedure of likelihood maximization is computationally expensive, especially when dealing with large MC samples to compute bin expectations. We use a software toolkit, named \textsc{NNMFit}, to manage the computational challenge~\cite{NNMFit}. 
The \textsc{NNMFit} package propagates parameter variations to the likelihood and minimizes the negative log-likelihood for the chosen parameters.
Each data sample has a dedicated configuration, which gives information on the treatment of the binning and detector response uncertainties. Internally, the physics parameters of interest are shared between the expectations calculated for each data sample.

The per-bin expectation $\mu_i$ is the total sum of MC event weights in bin $i$. Any change in the weight of an MC event will change the expectation value in the given bin. The MC event weight in turn is a sum of all contributing flux components evaluated at the true energy and zenith of that event. 

We distinguish between fit parameters that can be implemented by directly re-weighting the events (such as flux normalizations and other components of the spectra), and parameters of the detector response that cannot be implemented by re-weighting, since they can affect the reconstruction of events.
The total expected counts in the i-th bin are given by: 
\begin{equation} \label{eq:binexpect}
\mu_{i}\,=\,\sum_j \mu_{i,j}^0 + \sum_{j,k} \frac{d \mu_{ij}^0}{d \xi_k} \cdot (\xi_k - \xi_k^0 ) 
\end{equation}
where $\mu_{i,j}^0 $ is the baseline expectation 
of the weighted flux component $j$. 
The second term in the equation represents the expectation attributed to the detector response, where 
$\xi_k$ is the fit parameter associated with a given detector systematic $k$ and $\xi_k^0 $ its baseline value. 
The details of this component's implementation are described in the next section.
The framework supports calculating analysis sensitivities based on MC using either pseudo-experiments, or the Asimov set (measured `data' is exactly the same as bin expectation, defined in~\cite{Cowan:2010js}).
\subsection{Treatment of Detector Systematics}
\label{sec:snowstorm}
We use the \textit{SnowStorm} method introduced in~\cite{SnowStormPaper} to include the effect of systematic uncertainties related to the detector response for both analyses presented here. Previous IceCube analyses used individual sets of simulations that included discrete variations of parameters related to the detector response to estimate their contribution to the overall systematic uncertainty. 
However, this approach is computationally expensive and does not provide a uniform  coverage of systematic variations of multiple parameters.
The SnowStorm approach overcomes this by generating an ensemble of events, where each event in the ensemble is simulated based on a varied combination of the parameters describing the systematic uncertainties on the detector response. The parameter values are chosen according to sampling distributions for each parameter derived from calibration measurements. 

For each analysis bin  and flux component we calculate the `nuisance gradient vector' 
which 
reflects the derivatives
in Eq.~\ref{eq:binexpect}
with respect to the respective nuisance parameter in the SnowStorm set.

Here, we ignore second-order corrections and correlations, under the assumption of linearity, considering only small variations in the detector-response parameters. The justification of this approach has been tested with simulation-based pseudo-experiments where the absence of biases of the reconstructed physics parameters was confirmed.
In order to avoid biases
related to the range of varied detector-response parameters, 
the bin expectations for the nominal detector response, 
$\mu_{i,j}^0$ are determined from a dedicated high statistics `baseline' simulation set without varied parameters. 
Thus, only the gradients depend on the range of variations of the detector response. 
Note that non-linearities in the calculations of the gradients would only affect the location of the global minimum for the fitted nuisance parameters, but not for the signal parameters.
The gradients can be recalculated iteratively if the fit moves away from the nominal values of the nuisance parameters. 
However, we have not included this in our analyses, as deviations from the nominal values were sufficiently small (below $1\,\sigma$).
The nuisance parameters that are treated with the SnowStorm method are five ice-related parameters and one parameter that accounts for the optical energy scale that is related to the DOM efficiency and the Cherenkov light yield uncertainty. Two of the ice parameters describe uncertainties in the absorption and scattering of light in the ice between the detector strings, usually referred to as `bulk ice'. Another parameter accounts for bulk ice anisotropy (used as a fit parameter in MESE only; the CF analysis does not include this parameter since the tracks are not affected by this systematic effect and the cascades sample in the fit uses only three cos($\theta_{\mathrm{reco}}$ bins) arising from the non-uniform layers of ice. These three bulk ice related parameters govern the propagation of the Cherenkov light that reaches the DOMs. Additional two parameters jointly describe the refrozen ice surrounding the string (referred to as 'hole ice'). The refrozen ice is filled with trapped air bubbles that modify the scattering properties of light in the local volume surrounding the strings. For more details on the implementation in the fit, see Sec.~\ref{sec:nuisanceparams}.

\subsection{Self Veto}
\begin{figure*}[tbh!]
\includegraphics[width=0.8\linewidth]{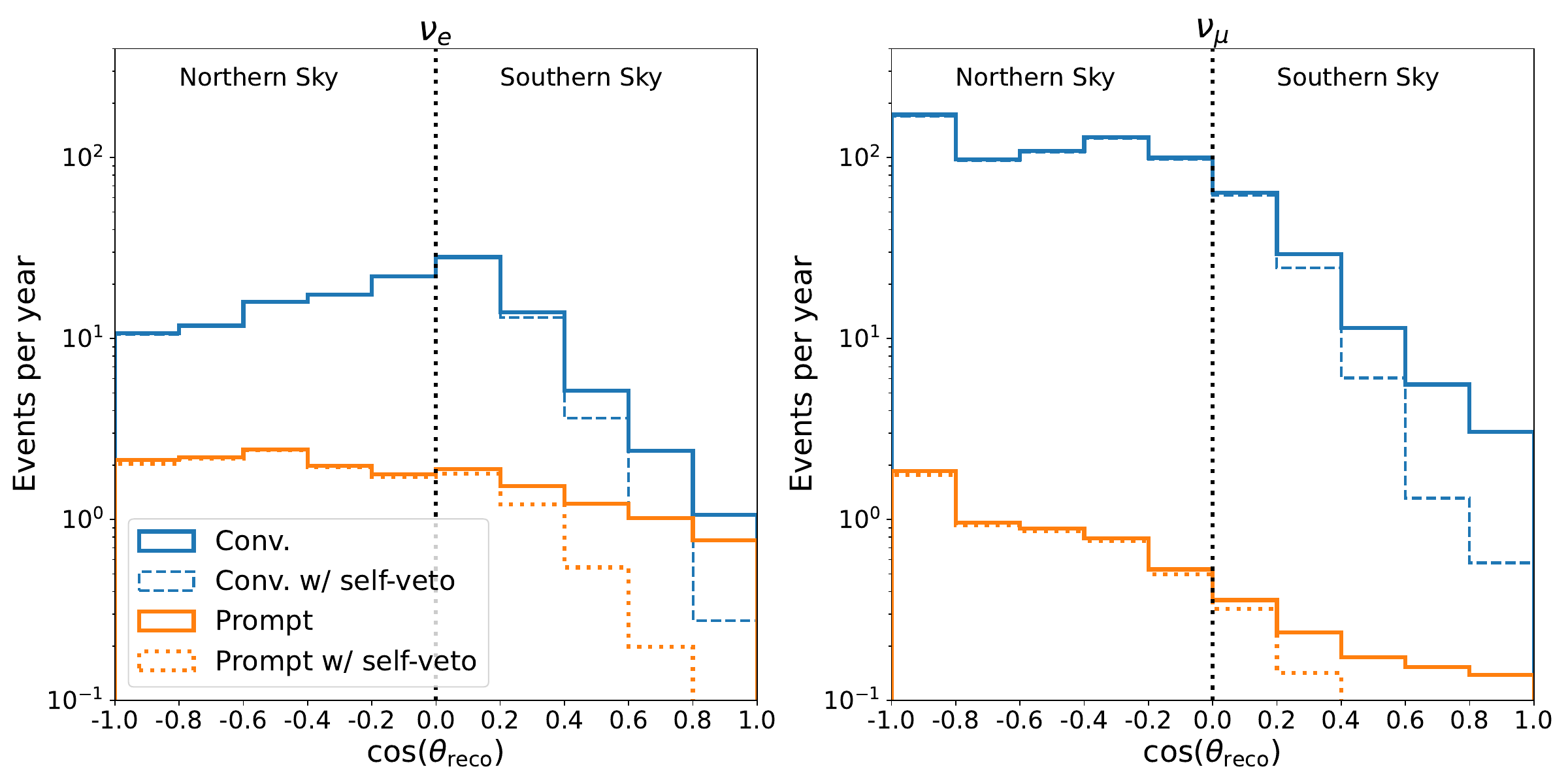}
\caption{\textbf{Atmospheric Self-Veto:} Atmospheric neutrinos from the Southern Hemisphere are often accompanied by muons from the same cosmic-ray air shower. These muons can cause the neutrino to be vetoed, resulting in a suppression of the atmospheric neutrino flux from the Southern sky. The effect of the self-veto on the final level atmospheric neutrino fluxes in the MESE event selection is depicted here for electron (left) and muon (right) neutrinos.
}
\label{fig:SelfVetoFluxes}
\end{figure*}
\begin{figure*}[tbh]
    \begin{minipage}[b]{0.32\linewidth}
     \centering
        \includegraphics[width=\linewidth]{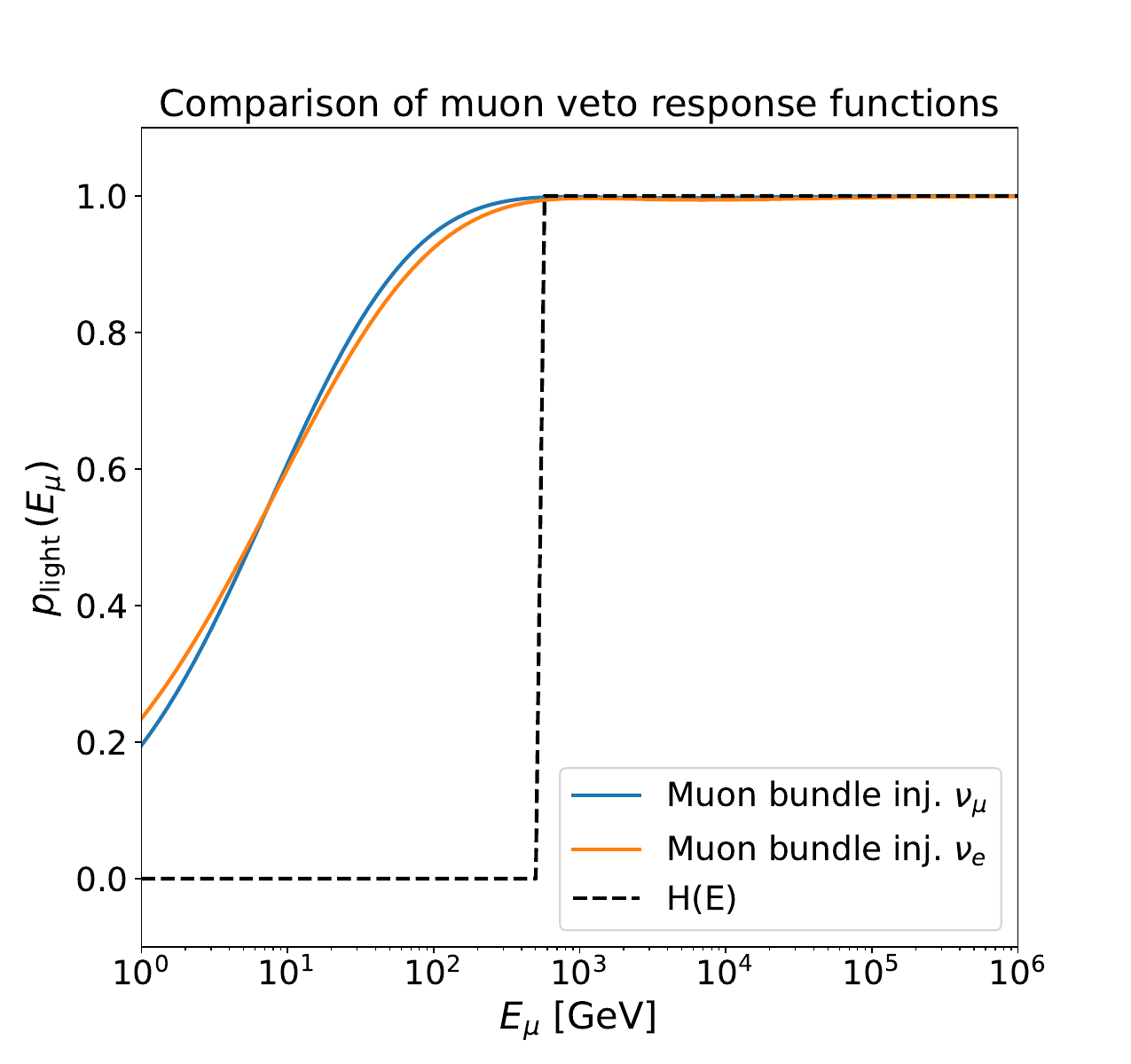}  
    \end{minipage}
    \begin{minipage}[b]{0.32\linewidth}
     \centering
        \includegraphics[width=\linewidth]{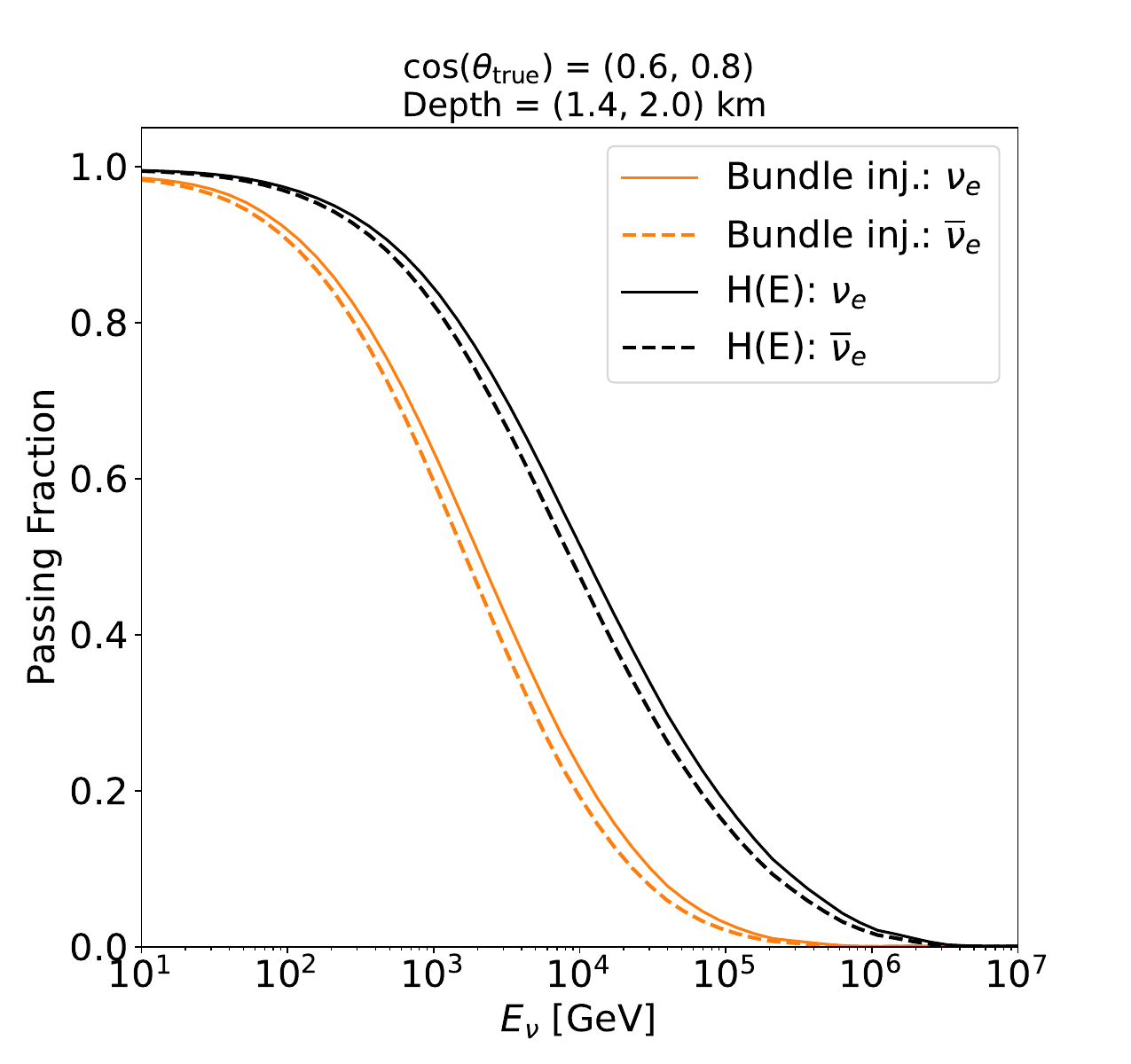}  
    \end{minipage}
     \begin{minipage}[b]{0.32\linewidth}
     \centering
        \includegraphics[width=\linewidth]{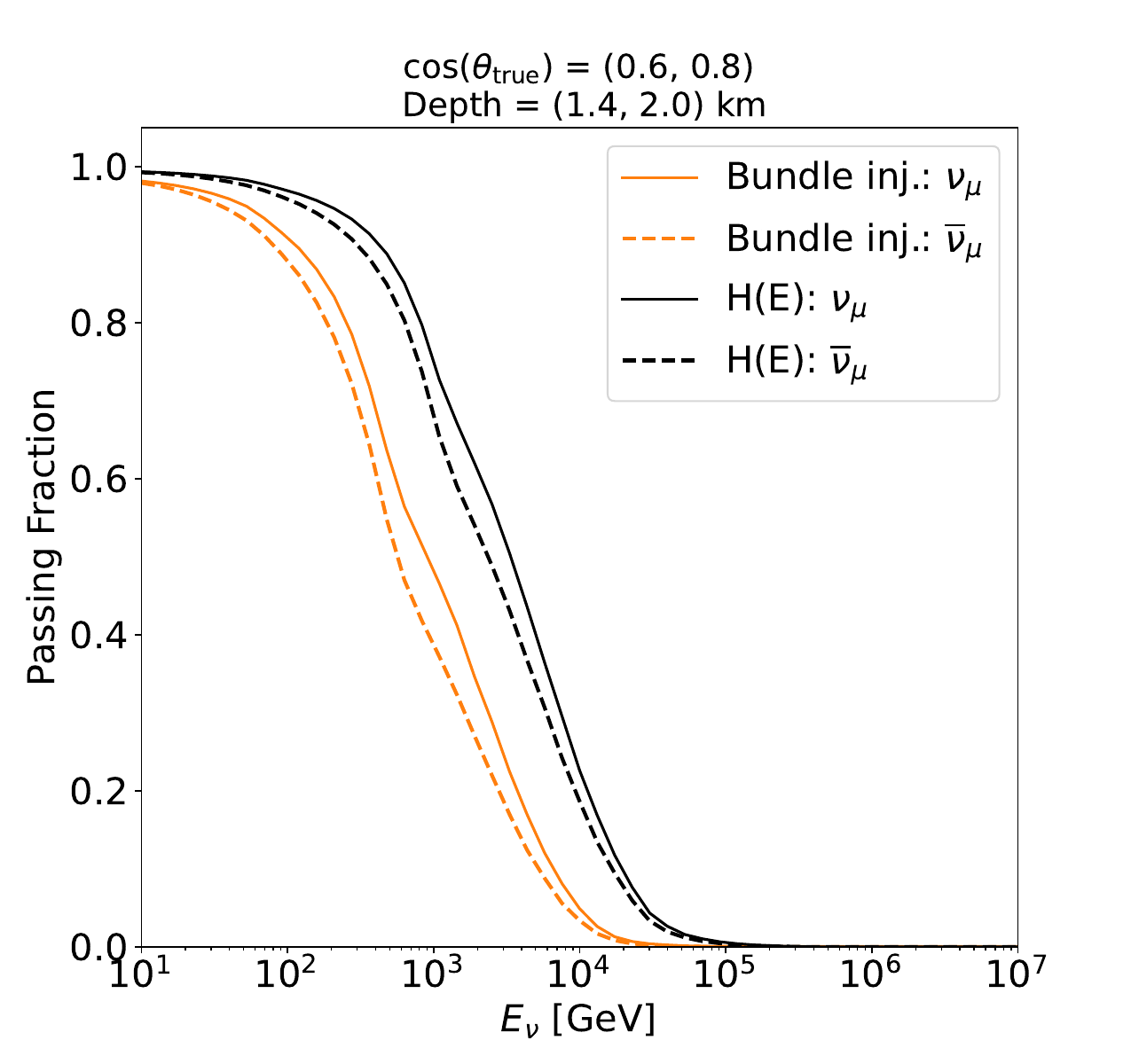}  
    \end{minipage}
\caption{\textbf{Self-Veto Passing Fraction Calculation:} The calculation of the passing fractions for the atmospheric neutrino self-veto in each analysis requires an estimation of the probability $p_{\mathrm{light}}$($E_{\mu}$) that a muon of a given energy ($E_{\mu}$), here measured at the detector boundary) deposits a sufficient amount of 
light in the detector to be vetoed. 
The plot on the left compares the muon bundle injection approach to the use of a Heaviside step function, $H(E)$, at \SI{500}{GeV} for the cascade sample of the CF analysis. Using $p_{\mathrm{light}}$ as an input, the \textsc{nuVeto} package \cite{nuVeto} is used to evaluate the passing fractions as a function of neutrino energy, flavour, cos($\theta_{\mathrm{true}}$) and interaction vertex depth. 
An example of the resulting passing fraction as a function of energy in the MESE analysis (orange) and the CF analysis (black) for a given depth and zenith range is illustrated for $\nu_e$ (solid) and $\bar{\nu}_e$ (dashed) in the middle panel and $\nu_\mu$ (solid) and $\bar{\nu}_{\mu}$ (dashed) in the right panel.
}
\label{fig:PlightPassingFractions}
\end{figure*}

Atmospheric neutrino self veto is the phenomenon where  atmospheric neutrinos are accompanied 
by muons originating from the same air shower and thus get vetoed by the event selection~\cite{Schonert:2008is, SelfVeto_AK_JVS}.
This results in an energy and zenith dependent suppression of the atmospheric neutrino component from the Southern sky after the event selection. 
This is illustrated in Fig.~\ref{fig:SelfVetoFluxes} for electron and muon neutrinos from air showers in the MESE analysis. 
The suppression depends on the specific selection cuts employed in each analysis. Therefore, the effects of self veto are specifically parameterized for each analysis. To account for this,
both MESE and CF analyses obtain these parameterizations from the \textsc{nuVeto} package~\cite{nuVeto}, which estimates the atmospheric neutrino passing fractions as a function of neutrino energy, zenith, and interaction depth. The neutrino passing fraction is defined as the probability that a given neutrino passes the cuts and is retained in the sample because no accompanying muon triggers the veto. The package requires $p_{\mathrm{light}}$ as an input, which is the probability that a muon of a given energy, zenith, and interaction depth deposits a sufficient amount of light within the detector in order to trigger the veto. 
The $p_{\mathrm{light}}$ function therefore varies from 0 for dim low-energy muons to 1 for high-energy muons which deposit more light in the detector. 
The CF analysis approximates this behavior with a Heaviside step function at a given cutoff energy, $H(E)$, which is a conservative estimate of this behavior. The MESE analysis, on the other hand, determines these functions by using simulation.
To this end, muon bundles, with energies and multiplicities derived from \textsc{CORSIKA} simulations of CR air showers, are injected into simulated neutrino events 
that pass to the final level of the event selection. We determine if the inclusion of these muons will trigger the veto or not.
 By comparing the number of events before ($N_{\mathrm{b}}$) and after injecting the muon ($N_{\mathrm{a}}$), we can estimate the probability as:
 \begin{equation}
     p_{\mathrm{light}}\,=\, 1\,-\,\frac{N_{\mathrm{a}}}{N_{\mathrm{b}}}.
 \end{equation}
We then parameterize $p_{\mathrm{light}}$ as a function of energy for a given zenith angle and depth, with a combination of a Gaussian and a sigmoid function. 
The respective $p_{\mathrm{light}}$ curves for MESE and CF are then processed through \textsc{nuVeto}, to calculate the passing fractions as a function of true energy, flavor, cos($\theta_{\mathrm{true}}$), and detector depth. Fig.~\ref{fig:PlightPassingFractions} shows the $p_{\mathrm{light}}$ curves and passing fractions of the two analyses. In order to account for the uncertainty of $H(E)$, the respective threshold is not fixed but included as one of the free nuisance parameters in the CF analysis. A description of all nuisance parameters is given in Sec.~\ref{sec:nuisanceparams}.

\subsection{Treatment of low statistics backgrounds}

In cases of efficient background rejection, 
only a small number of MC events for the background components survive the final selection level. 
When distributing these events 
to the analysis bins, these background estimations are subject to strong statistical fluctuations, even if the overall statistics is sufficient for an estimation of the total background rate.
In the forward folding 
of the summed expectations from all components in the analysis, fluctuations in the assumed background rate will directly translate into a bias of the estimated signal in that bin.

This is the case for the cosmic-ray muon template for the tracks sample used in the CF analysis and the \textsc{MuonGun} background in the MESE selection.
Both components can be
reasonably assumed to follow a smooth distribution with respect to the analysis observables: the reconstructed energy and cos($\theta_{\mathrm{reco}}$).
Hence, we apply an adaptive kernel density estimator (KDE) to generate a smooth template in energy and cos($\theta_{\mathrm{reco}}$) (described in App.~B in \cite{Schoenen:696221}). 
We use the same method of generating the muon template as in~\citep{NorthernTracks}. The template is generated by bootstrapping the \textsc{MuonGun} MC distribution at the final level of the selection, resulting in an ensemble of KDE-based templates. The ensemble of bootstrapped templates  are then used to obtain the median template and its statistical variance to be used as bin uncertainty in the modified likelihood described in Eq.~\ref{eq:likelihood}. 
The template is then included in the fit with a free overall normalization parameter, denoted as $\phi_{\mathrm{muon}}$ in Tab.~\ref{tab:systs_MESE} and $\phi_{\mathrm{muon\,template}}$ in Tab.~\ref{tab:systs_CombinedFit}, to account for the muon flux. We additionally validated the atmospheric muon flux predicted by \textsc{MuonGun} by checking the consistency between MC and data in the background region with the MESE sample. The rates were checked for 500\,GeV to 1\,TeV events on the events that pass the down-going track veto, since this is outside the signal region and is muon dominated. We define a Gaussian prior for the atmospheric muon flux in the MESE analysis based on the difference in scale between MC and data for the most vertical bins (see Tab.~\ref{tab:systs_MESE}).

\subsection{Astrophysical Neutrino Spectral Models}\label{sec:AstrophysicalFitModels}

We test several spectral models for the astrophysical neutrino flux with both analyses, where we fit for the total flux, $\Phi_{\nu+\bar{\nu}}$, measured in units of 
$ \SI{e-18}{GeV^{-1}  cm^{-2}  s^{-1} sr^{-1}} $. 
In these tests we assume a total neutrino and antineutrino astrophysical flavor ratio of \mbox{$\nu_e:\nu_{\mu}:\nu_{\tau} = 1:1:1$}, unless specified otherwise. 
For each tested model we fit the nuisance parameters that reflect systematic uncertainties along with the physics parameters that describe the given astrophysical model. The simplest model for non-thermal emission is a single power law. In reality, we see structure beyond the single power law in the spectra over many decades in energy, and so we test the other models as modifications to this flux model.
The model hypotheses tested in both analyses are: \\
\begin{enumerate}
    \item single power law (SPL): \\
\begin{center}\vspace{-5mm}
$\Phi_{\nu+\bar{\nu}} =\phi_0\,  
\left(\frac{E_{\nu}}{100\,\mathrm{TeV}}\right)^{-\gamma}$,\end{center}
with the  normalization $\phi_0 $ and spectral index $\gamma $ as free parameters. $E_{\nu}$ is the energy of the neutrino.

\item broken power law (BPL): \\
\begin{center}\vspace{-5mm}
$\Phi_{\nu+\bar{\nu}} =\phi_{0,\mathrm{broken}}\,\left\{
\begin{array}{c}
      \left(\frac{E_{\nu}}{E_{\mathrm{break}}}\right)^{-\gamma_1}\, (E_{\nu} < E_{\mathrm{break}})\\
      \left(\frac{E_{\nu}}{E_{\mathrm{break}}}\right)^{-\gamma_2}\, (E_{\nu} > E_{\mathrm{break}})
      \end{array}\right.
$,\\
\end{center}
where 
    \begin{center}
$\phi_{0,\mathrm{broken}}=\phi_{0}\left\{
\begin{array}{c}
       \left(\frac{E_{\mathrm{break}}}{100\,\mathrm{TeV}}\right)^{-\gamma_1}\, (E_{\mathrm{break}} > 100\,\mathrm{TeV})\\
       \left(\frac{E_{\mathrm{break}}}{100\,\mathrm{TeV}}\right)^{-\gamma_2}\, (E_{\mathrm{break}} \leq 100\,\mathrm{TeV})
    \end{array}\right.$,
    \end{center}

with the  normalization $\phi_0 $, spectral indices $\gamma_1, \gamma_2$ and break energy $E_{\mathrm{break}}$ as free parameters.

\item log parabola (LP): \\
\begin{center}\vspace{-5mm}
$\Phi_{\nu+\bar{\nu}} =\phi_0\, \left(\frac{E_{\nu}}{100\,\mathrm{TeV}}\right)^{-\alpha_\mathrm{LP}-\beta_\mathrm{LP}\log_{10}\left(\frac{E_{\nu}}{100\,\mathrm{TeV}}\right)
}$, \end{center}
where the  normalization $\phi_0 $, spectral index $\alpha_{\rm{LP}}$ and the curvature parameter $\beta_{\rm{LP}}$ are free parameters.

\item and single power law with an exponential cutoff (SPE): \\
\begin{center}\vspace{-5mm}
$\Phi_{\nu+\bar{\nu}} =\phi_0\, \left(\frac{E_{\nu}}{100\,\mathrm{TeV}}\right)^{-\gamma} e^{\frac{-E_{\nu}}{E_\mathrm{cutoff}}}$,\end{center}
where the  normalization $\phi_0 $, spectral index $\gamma$ and the cut-off energy $E_{\mathrm{cutoff}}$ are free parameters.
\end{enumerate}

In the MESE analysis, we also tested three additional models with specific predictions within the sensitive energy range. The models were chosen before unblinding the data, based on previous measurements from IceCube with datasets including starting events. These are:
\begin{enumerate}\addtocounter{enumi}{4}
 \item a single power law with an additional Gaussian bump (SPB): \\

 \begin{center}\vspace{-5mm}
$\Phi_{\nu+\bar{\nu}} =\phi_0\,
\left(\frac{E_{\nu}}{100\,\mathrm{TeV}}\right)^{-\gamma}$ + $\phi_{\mathrm{bump}}\,
\left(e^{-\frac{(E-E_{\mathrm{bump}})^2}{2\sigma_{\mathrm{bump}}^2}}\right)$,\end{center}
where the  normalization $\phi_0 $, spectral index $\gamma$, bump normalization $\phi_{\mathrm{bump}}$, bump energy $E_{\mathrm{bump}}$, and the width of the bump $\sigma_{\mathrm{bump}}$ are free parameters.

\item a single power law and an additional a template flux from AGN cores~\cite{SteckerAGN} (SPL with AGN), where $\phi_0 $, $\gamma$ and the template normalization are free parameters.

\item and a single power law along with model predictions from BLLacs~\cite{BLLac} (SPL with BLLac), where $\phi_0 $, $\gamma$ and the template normalization are free parameters.
\end{enumerate}

The CF analysis also tested some additional flux models. These additional models were chosen based on previous IceCube studies of the tracks-only sample. These are:
\begin{enumerate}\addtocounter{enumi}{7}
 \item a two-component flux model, to incorporate the effect of additional source populations: 
\begin{multline*}
\Phi_{\nu+\bar{\nu}} = \phi_0 \times\\\left[(1-\alpha) \left(\frac{E_{\nu}}{100\,\mathrm{TeV}}\right)^{-\gamma} +  \alpha\left(\frac{E_{\nu}}{100\,\mathrm{TeV}}\right)^{-\gamma+\Delta} \right],
\end{multline*}
where the  normalization $\phi_0 $, mixture parameter $\alpha$, spectral index $\gamma$ and modifier $\Delta$ are free parameters.

\item an extension of the single power law flux model with additional muon dampening at the source~\cite{kashti2005astrophysical}, which includes a suppression in the $\nu_e,\nu_\mu$ flux due to dampened muon decay above a certain critical energy $E_{\mu\mathrm{crit}}$. The damped flux is given by
\begin{multline*}
\Phi_{\nu+\bar{\nu},\mathrm{dampened}} = \phi_{\mathrm{undampened}} (E_\nu) \times \\
\left(1+\left(\frac{E}{0.3\,E_{\mu\mathrm{crit}}}\right)^\epsilon \right)^{-\Delta\gamma/\epsilon},
\end{multline*}
with the additional free parameter $E_{\mu\mathrm{crit}}$. $\phi_{\mathrm{undampened}}$ is the flux without damping, 
$\Delta\gamma=2$ represents the change in the spectrum across 2 orders of magnitude in energy and $\epsilon$, the smoothing parameter, is chosen as 5 following~\citep{2022JInst..1708009V,IceCube-Gen2:2020qha}.

\item and a broken power law with independent flavor normalizations, to model the effects of varying source flavor compositions~\cite{bustamante2019inferring}. These are parameterized by the scaling factors $s_{\nu_\mathrm{e}}$ and $s_{\nu_\tau}$. The relative fractions of the various neutrino flavors are thus given by $f_{\nu_\mathrm{e}},\,f_{\nu_{\mu}},$ and $f_{\nu_{\tau}}$ where
\begin{align*}
f_{\nu_\mathrm{e}} &= \frac{s_{\nu_\mathrm{e}}}{1+s_{\nu_\mathrm{e}}+s_{\nu_\tau}}\\
f_{\nu_\mu} &= \frac{1}{1+s_{\nu_\mathrm{e}}+s_{\nu_\tau}} \\
f_{\nu_\tau} &= \frac{s_{\nu_{\tau}}}{1+s_{\nu_\mathrm{e}}+s_{\nu_\tau}}. 
\end{align*}
\end{enumerate}

In addition to the fits of predefined spectral models, we aim in both analyses for a determination of the energy distribution with minimal model dependence.
For this, we fit 
the normalization of the astrophysical neutrino flux independently in 13 energy segments of the total spectrum (see Tab.~\ref{tab:segmented_fit_results}), assuming a power-law spectrum with index 2 in each segment. This fit is denoted as ``segmented'' fit below.

\subsection{Nuisance parameters}
\label{sec:nuisanceparams}
Systematic uncertainties are modeled with nuisance parameters which are optimized concurrently with the signal parameters.
A summary of all nuisance parameters used in the MESE fit is given in Tab.~\ref{tab:systs_MESE}, with the corresponding parameters for the CF analysis in Tab.~\ref{tab:systs_CombinedFit}.

The fit includes parameters that describe the conventional and prompt atmospheric neutrino flux normalizations, and a normalization parameter for the muon flux, as nuisance parameters. MESE analysis applies a Gaussian prior to the muon normalization. This prior is derived from pre-final level data, which is dominated by muons. 
Comparing this data to MC in the down-going region allows us to determine the corresponding prior to be applied to the muon normalization. This is done at reconstructed-event energies of \SI{500}{GeV}--\SI{1}{TeV}, which is outside the energy range used for determining the astrophysical neutrino spectrum.  
In order to allow for 
variations from the baseline assumption for conventional atmospheric fluxes (see Sec.~\ref{sec:detandsim}), we introduce additional nuisance parameters. 
The variations arising from uncertainties in the hadronic interaction model may lead to modifications in the $\pi/K$ ratio in cosmic-ray showers, following the work in~\cite{barr_uncertainties_2006}. These are modelled using nuisance parameters which modify the expected hadronic yield in a given energy range. 
We use four of these 
parameters (\textit{h, w, y, z}, described in Tab.~\ref{tab:systs_MESE} and Tab.~\ref{tab:systs_CombinedFit}), which have been identified to have a relevant effect in the considered energy range of this measurement. An 
additional parameter, $\Delta\gamma$, that allows for variations in the 
spectral index of the primary cosmic-ray spectrum is 
also included. CR composition uncertainties are included in the fit  using a parameter that interpolates between the H4a and the GST~\cite{gaisser_cosmic_2013} predictions of the cosmic-ray composition. 
We do not explicitly account for uncertainties in the prompt atmospheric flux with additional nuisance parameters beyond the normalization. As the prompt neutrino flux is subdominant in the energy range under consideration, any uncertainty on the shape of the prompt neutrino spectrum is expected to be absorbed by the nuisance parameters already included in the fit. 
 Both analyses include a self-veto parameter as a systematic uncertainty. This parameter allows for modifications of the passing fractions from its baseline assumption. The allowed modifications are derived from the uncertainty bands of the passing-fraction curve.
 The parameter $\epsilon$ describing each realization of the passing-fraction curve is treated as a nuisance parameter within the fit. Any variation in $\epsilon$ 
 can be interpreted as a variation in the threshold energy of $H(E)$ for the CF analysis. Similarly for the MESE analysis, this corresponds to a variation in $p_{\mathrm{light}}$, which in turn changes the passing-fraction curve.
 A Gaussian prior is applied on $\epsilon$ for the MESE fit, while CF analysis does not apply a prior.

We also include nuisance parameters that describe the detector response, and the effect of the ice. They are: \textit{ice absorption} and \textit{ice scattering} parameters, that modify the effect of light propagation through bulk ice by globally scaling the depth-dependent absorption and scattering length, and two \textit{hole ice} parameters that modify the angular acceptance function of the DOMs. 
The MESE analysis additionally includes a parameter that allows for variations from the nominal model of anisotropy of light scattering in the bulk of the ice. The CF analysis does not include this parameter since the tracks are not affected by this systematic effect and the cascades sample in the fit uses only 3 bins in cos($\theta_{\mathrm{reco}}$) . For more details of these parameters see Sec.~\ref{sec:snowstorm}.
All the ice parameters are highly correlated to each other in the fit. An additional nuisance parameter that accounts for the uncertainties in the optical efficiency, that is the production of Cherenkov photons and the efficiency of the DOMs to detect that light, is included in both analyses. This parameter absorbs all effects that  
influence the overall assumed energy scale of the Cherenkov detector.
The CF analysis utilizes the high statistics of the track sample to limit the impact of these systematic parameters implicitly in the analysis. MESE uses Gaussian priors on the ice absorption and scattering, informed from calibration runs of the detector, to constrain the fit. 

Finally, the MESE analysis includes a parameter to absorb the variations arising from the imperfect understanding of the inelasticity of neutrino interactions, following the parameterization in~\cite{IceCube:2018pgc}. This nuisance parameter absorbs phenomena that can impact the inelasity,eg. final state radiation described in~\cite{FSR}. The CF analysis included this parameter in a cross-check study (see Sec.~\ref{sec:crosschecks}).
The standard cross sections used in the analysis are derived from CSMS~\cite{CSMS}. We allow for variations of the mean inelasticity by up to \SI{20}{\percent} in the MESE analysis and include a scaling factor that describes the relative value of the mean inelasticity parameter with respect to the baseline from CSMS as a nuisance parameter in the fit. This is defined as a single parameter for the entire event sample, over the entire energy range. We additionally apply a Gaussian prior with a width of \SI{10}{\percent}  on this mean inelasticity scaling parameter during the fit.
\section{Results}
\label{sec:results}
\begin{figure*}[t!]
    \begin{minipage}[b]{0.49\linewidth}
     \centering
        \includegraphics[width=\linewidth]{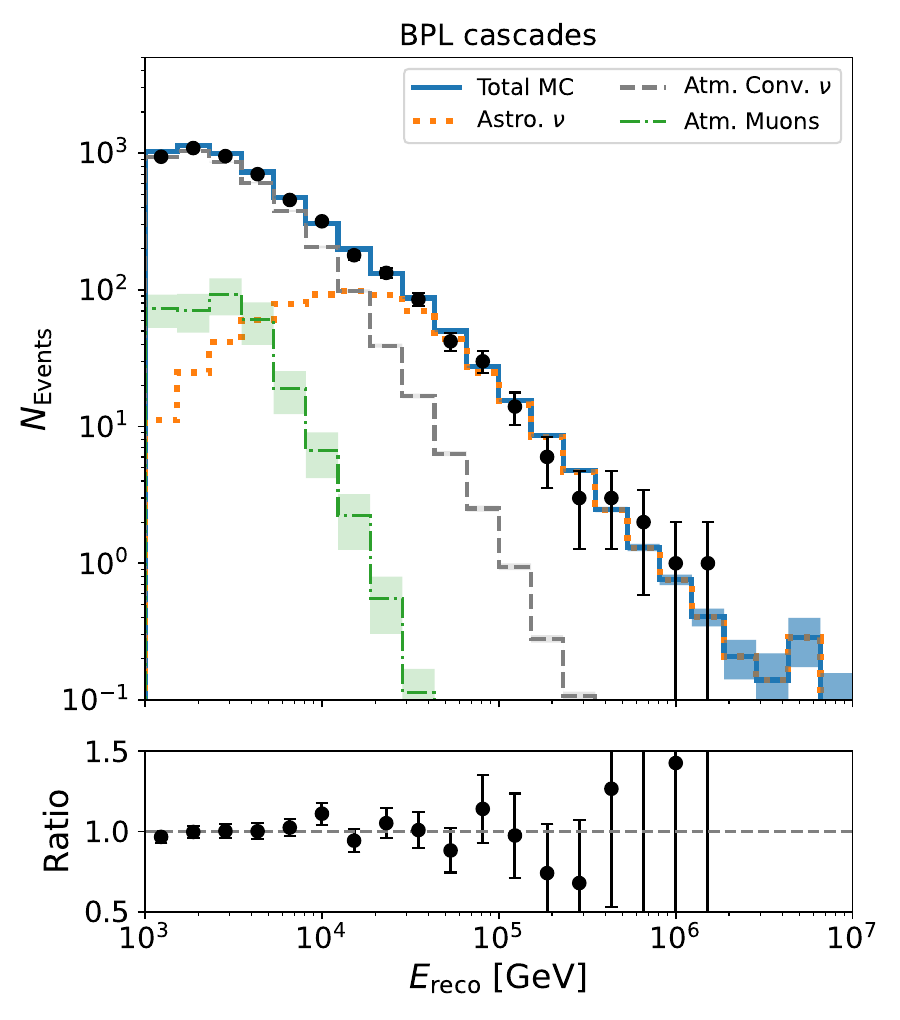}  
    \end{minipage}
    \begin{minipage}[b]{0.49\linewidth}
     \centering
        \includegraphics[width=\linewidth]{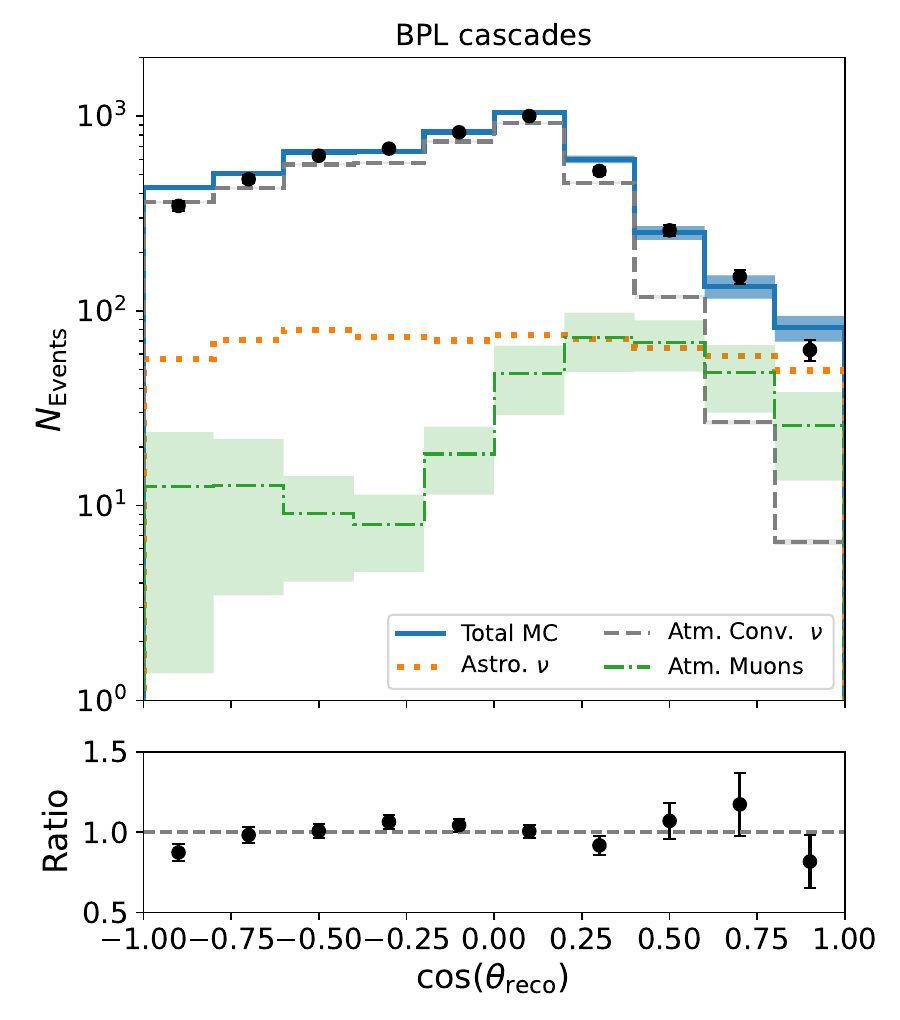}  
    \end{minipage}
     \begin{minipage}[b]{0.49\linewidth}
     \centering
        \includegraphics[width=\linewidth]{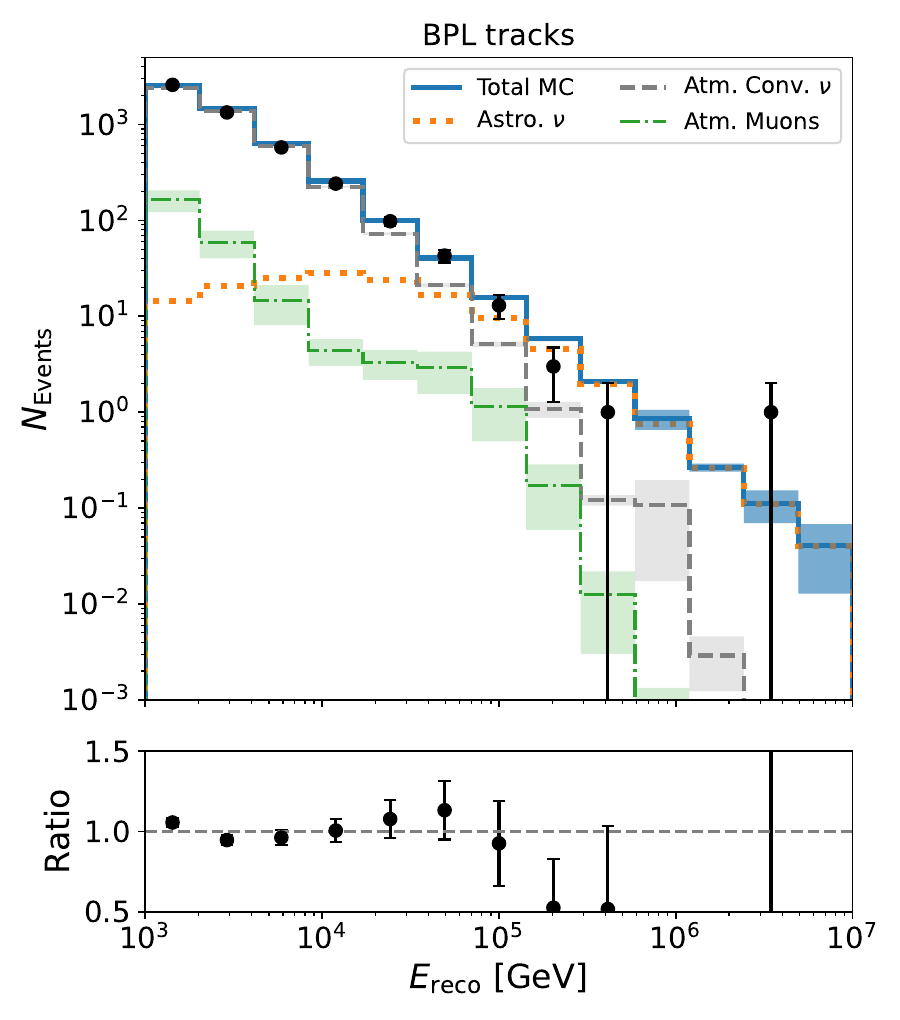}  
    \end{minipage}
    \begin{minipage}[b]{0.49\linewidth}
     \centering
        \includegraphics[width=\linewidth]{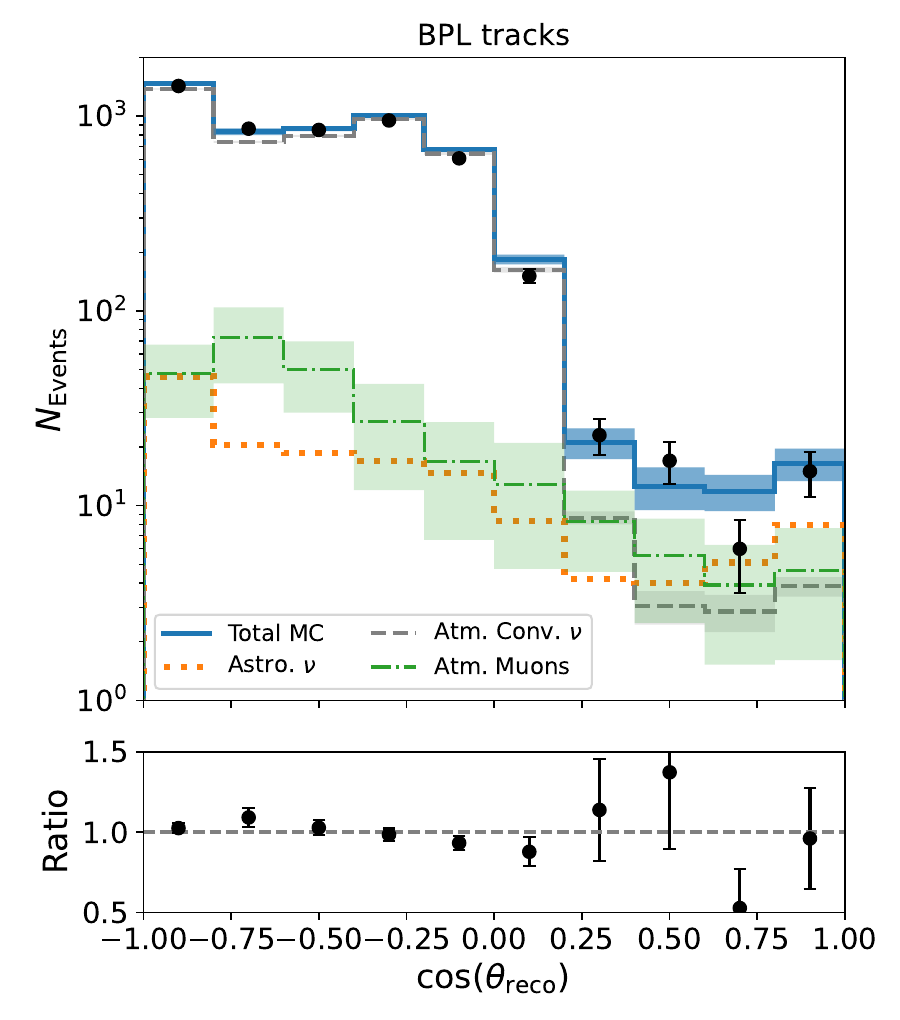}  
    \end{minipage}
\caption{\textbf{MESE Data \& MC:} Comparison of data and MC simulation for the best fit BPL spectral model. Displayed are reconstructed cascade energy (upper left), cos($\theta_{\rm{reco}}$)  (lower left), reconstructed track energy (upper right), and cos($\theta_{\rm{reco}}$) (lower right).  A DNN is used to classify tracks and cascades, and we use separate reconstructions for them. The atmospheric prompt flux normalization is a free parameter which fits to zero and is therefore not shown. The shaded regions show the MC error for the respective component. The bottom panel of each plot shows the ratio data/MC.
}
\label{fig:MESEBPLDataMC}
\end{figure*}

\begin{figure*}[t!]
    \begin{minipage}[b]{0.49\linewidth}
     \centering
        \includegraphics[width=\linewidth]{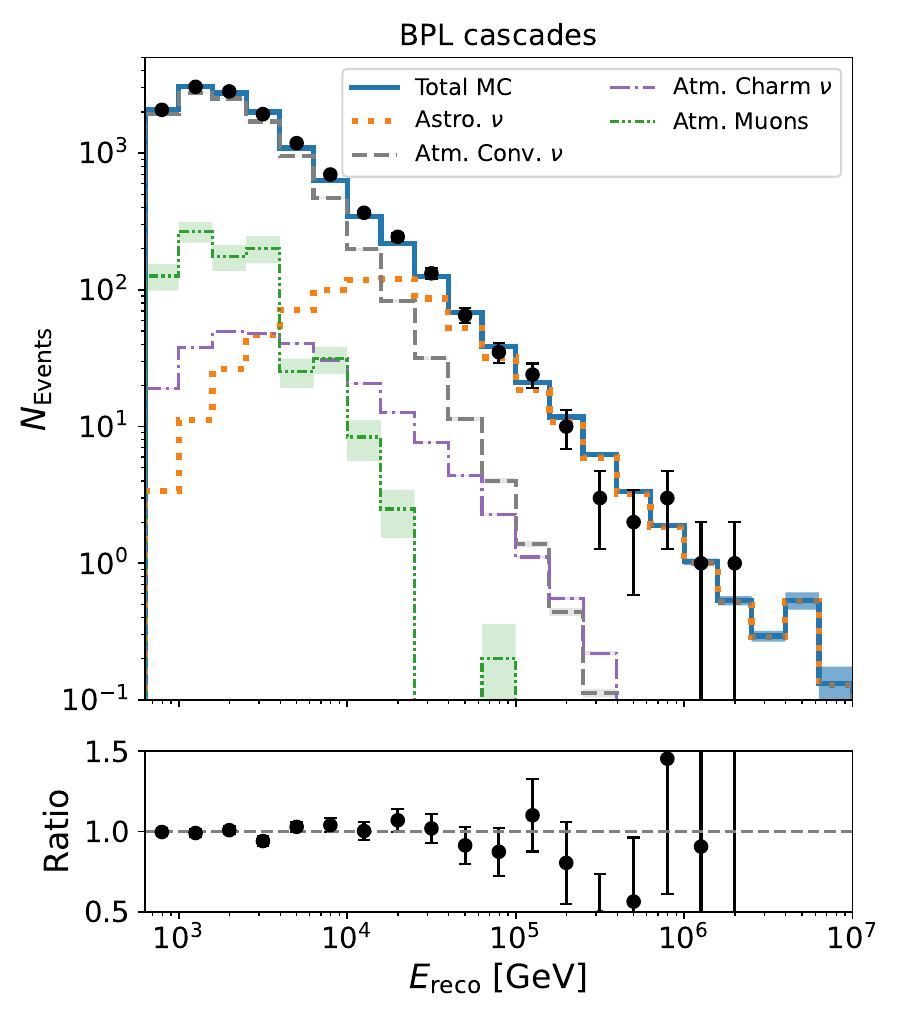}  
    \end{minipage}
    \begin{minipage}[b]{0.49\linewidth}
     \centering
        \includegraphics[width=\linewidth]{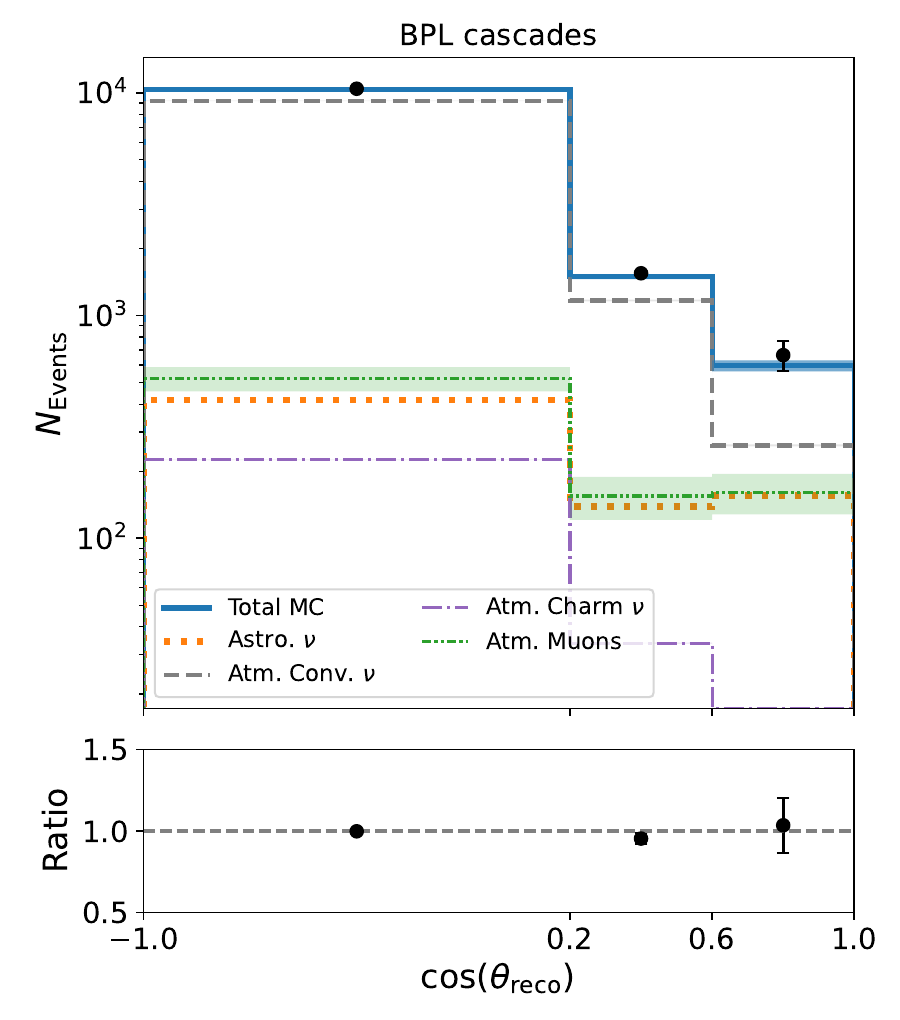}  
    \end{minipage}
     \begin{minipage}[b]{0.49\linewidth}
     \centering
        \includegraphics[width=\linewidth]{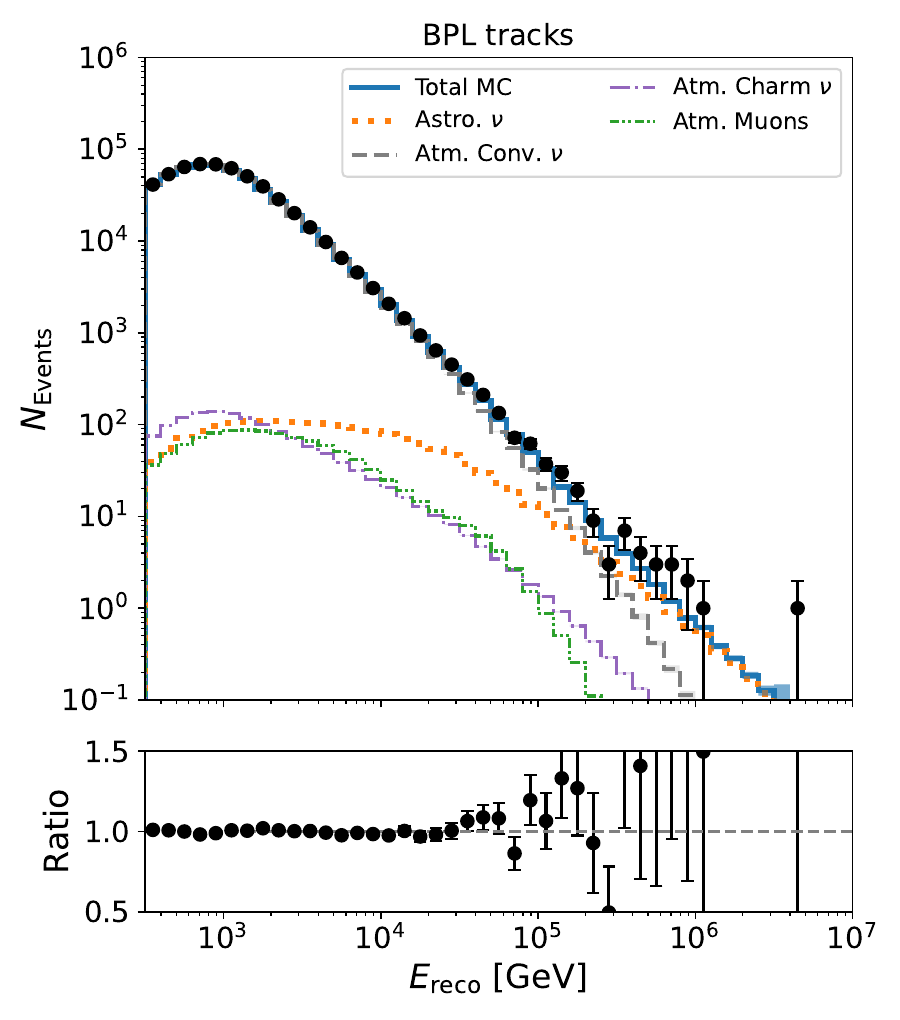}  
    \end{minipage}
    \begin{minipage}[b]{0.49\linewidth}
     \centering
        \includegraphics[width=\linewidth]{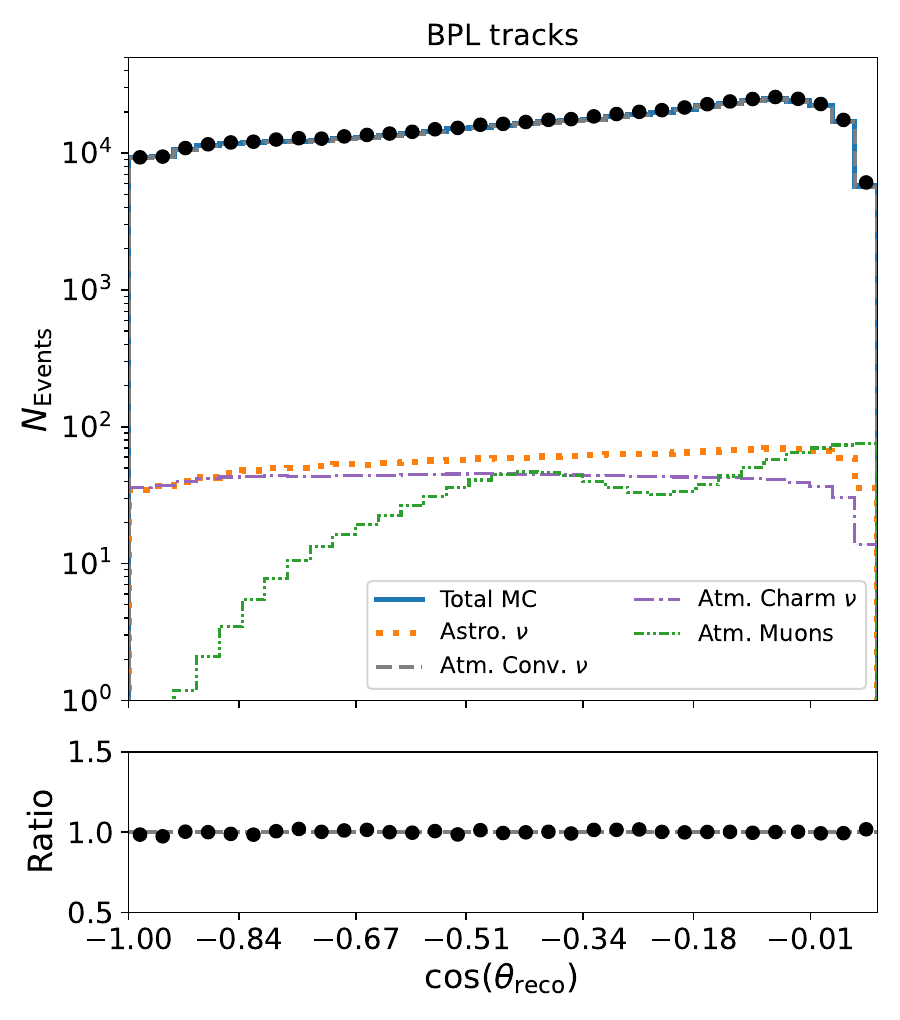}  
    \end{minipage}
\caption{\textbf{CF Data \& MC:} Comparison of data and MC simulation for the best fit BPL spectral model. Displayed are reconstructed cascade energy (upper left) , cos($\theta_{\rm{reco}}$)  (lower left), reconstructed track energy (upper right), and cos($\theta_{\rm{reco}}$) (lower right). The shaded regions show the MC error for the respective component. The bottom panel of each plot shows the ratio data/MC.
}
\label{fig:CombinedFitBPLDataMC}
\end{figure*}
\subsection{Spectral Measurements}
The results for the models tested in both CF and MESE analyses are summarized in Tab.~\ref{tab:main-results}.
The BPL model and LP model are  statistically preferred with a $p$-value equivalent to 4\,$\sigma$ level with respect to the SPL model, while the preference for the SPE model is not statistically significant (see Tab.~\ref{tab:main-results}). In this analysis, we calculate the test statistic as the ratio of the maximized likelihoods, with $\mathrm{TS}\,=\,-2 (\mathrm{ln} \mathcal{L}_{\mathcal{H}_1(\eta)}-\mathrm{ln} \mathcal{L}_{\mathcal{H}_0(\eta)})$, where $\mathcal{H}_0(\eta)$ represents the SPL hypothesis and $\mathcal{H}_1(\eta)$ the other spectral models. 
The largest significance is found for the BPL model, with an improvement of $\mathrm{TS}\,=\,-2 \Delta \mathrm{ln} \mathcal{L}\,=\,27.3$ for the MESE analysis when compared to the SPL model. The obtained TS for the CF analysis is 24.7. We also obtain an improvement of the likelihood when we compare LP model to an SPL spectrum ($TS = 18.8$ for MESE and $16.4$ for CF). 

\begin{table*}[htbp]
\caption{Results for the spectral models tested in both analyses. The uncertainties are derived from 1D profile likelihood scans, assuming Wilks' theorem~\cite{Wilks} applies. We show the preference over the single power-law hypothesis in terms of $-2 \Delta \mathrm{ln} \mathcal{L}$. The sensitive energy range of each model is shown for both analyses.
The flux is measured in units of $\SI{e-18}{GeV^{-1} cm^{-2} s^{-1} sr^{-1}}$. All flux normalizations are at \SI{100}{TeV}.}
\label{tab:main-results}
\renewcommand{\arraystretch}{1.5}
\centering
\small
\begin{tabularx}{\textwidth}{c|c|c|c|c}
 & \multicolumn{4}{c}{Astrophysical model} \\ \cline{2-5} 
Analysis & \makecell[tc]{SPL} & 
\makecell[tc]{SPE
} & 
\makecell[tc]{BPL} & 
\makecell[tc]{LP}
 \\ 
\cline{1-5} 
MESE & 
\begin{tabular}[t]{@{} l l @{}}
    $\phi_0$ & $= 2.13^{+0.18}_{-0.17}$ \\
    $\gamma$ & $= 2.548^{+0.039}_{-0.041}$
\end{tabular} & 
\begin{tabular}[t]{@{} l l @{}} 
    $\phi_0$ & $= 3.9^{+1.2}_{-1.2}$ \\
    $\gamma$ & $= 2.16^{+0.11}_{-0.16}$ \\
    $\log_{10}(\frac{\mathrm{E}_\mathrm{cutoff}}{\mathrm{GeV}})$ & $= 5.52^{+0.39}_{-0.35}$ \\
\end{tabular} & 
\begin{tabular}[t]{@{} l l @{}}
    $\phi_0$ & $= 2.28^{+0.22}_{-0.20}$ \\
    $\gamma_1$ & $= 1.72^{+0.26}_{-0.35}$ \\
    $\gamma_2$ & $= 2.839^{+0.11}_{-0.091}$ \\
    $\log_{10}(\frac{\mathrm{E}_\mathrm{break}}{\mathrm{GeV}})$ & $= 4.524^{+0.097}_{-0.087}$ \\
\end{tabular} & 
\begin{tabular}[t]{@{} l l @{}}
    $\phi_0$ & $= 2.58^{+0.26}_{-0.26}$ \\
    $\alpha_\mathrm{LP}$ & $= 2.668^{+0.12}_{-0.061}$ \\
    $\beta_\mathrm{LP}$ & $= 0.359^{+0.11}_{-0.082}$ \\
\end{tabular} \\ \cline{2-5} 
 & \begin{tabular}[t]{@{} c @{}}
\\ 
\\ $E$: \SI{5.3}{TeV} - \SI{7.5}{PeV}
\end{tabular}
 & 
\begin{tabular}[t]{@{} c @{}}
$-2 \Delta \mathrm{ln} \mathcal{L}$ $= 1.8$ 
\\ $p= 0.18$ ($0.9\,\sigma$)
\\ $E$: \SI{5.3}{TeV} - \SI{6.5}{PeV}
\end{tabular} & 
\begin{tabular}[t]{@{} c @{}}
$-2 \Delta \mathrm{ln} \mathcal{L}$ $= 27.3$
\\ $p= 1.2\cdot 10^{-6}  $ ($4.7\,\sigma$)
\\ $E$: \SI{6.1}{TeV} - \SI{7.5}{PeV}
\end{tabular} & 
\begin{tabular}[t]{@{} c @{}}
$-2 \Delta \mathrm{ln} \mathcal{L}$ $= 18.84$
\\ $p= 1.42\cdot 10^{-5}  $ ($4.2\,\sigma$)
\\ $E$: \SI{6.1}{TeV} - \SI{6.5}{PeV}
\end{tabular}\\ \cline{1-5} 
CF & 
\begin{tabular}[t]{@{} l l @{}}
    $\phi_0$ & $= 1.800^{+0.13}_{-0.16}$ \\
    $\gamma$ & $= 2.52^{+0.036}_{-0.038}$
\end{tabular} & 
\begin{tabular}[t]{@{} l l @{}}
    $\phi_0$ & $= 2.20^{+0.30}_{-0.25}$ \\
    $\gamma$ & $= 2.386^{+0.081}_{-0.090}$ \\
    $\log_{10}(\frac{\mathrm{E}_\mathrm{cutoff}}{\mathrm{GeV}})$ & $= 6.15^{+0.37}_{-0.24}$ \\
\end{tabular} & 
\begin{tabular}[t]{@{} l l @{}}
    $\phi_0$ & $= 1.77^{+0.19}_{-0.18}$ \\
    $\gamma_1$ & $= 1.31^{+0.51}_{-1.30}$ \\
    $\gamma_2$ & $= 2.735^{+0.067}_{-0.075}$ \\
    $\log_{10}(\frac{\mathrm{E}_\mathrm{break}}{\mathrm{GeV}})$ & $= 4.39^{+0.1}_{-0.1}$ \\
\end{tabular} & 
\begin{tabular}[t]{@{} l l @{}} 
    $\phi_0$ & $= 2.13^{+0.16}_{-0.19}$ \\
    $\alpha_\mathrm{LP}$ & $= 2.572^{+0.062}_{-0.053}$ \\
    $\beta_\mathrm{LP}$ & $= 0.228^{+0.098}_{-0.067}$ \\
\end{tabular}\\ \cline{2-5}
 &\begin{tabular}[t]{@{} c @{}}

\\
\\ $E$: \SI{2.4}{TeV} - \SI{6.4}{PeV}
\end{tabular}
 & 
\begin{tabular}[t]{@{} c @{}}
$-2 \Delta \mathrm{ln} \mathcal{L}$ $= 7.5$
\\ $p= 6.17\cdot 10^{-3}  $ ($2.5\,\sigma$) 
\\ $E$: \SI{3.8}{TeV} - \SI{1.8}{PeV}
\end{tabular} & 
\begin{tabular}[t]{@{} c @{}}
$-2 \Delta \mathrm{ln} \mathcal{L}$ $= 24.7$
\\ $p= 4.33 \cdot 10^{-6}  $ ($4.4\,\sigma$)
\\ $E$: \SI{13.7}{TeV} - \SI{4.7}{PeV}
\end{tabular} & 
\begin{tabular}[t]{@{} c @{}}
$-2 \Delta \mathrm{ln} \mathcal{L}$ $= 16.4$
\\ $p= 5.13\cdot 10^{-5}  $ ($3.9\,\sigma$) 
\\ $E$: \SI{7.5}{TeV} - \SI{2.2}{PeV}
\end{tabular} 
\end{tabularx}
\end{table*}

We note that the BPL model has more degrees of freedom than LP, and that these models are not a parametric family of hypotheses.
Therefore, unlike the above tests of nested hypotheses 
Wilks' theorem cannot be applied for comparing the likelihood difference between the two model fits directly. Hence, we run pseudo experiments to obtain a distribution of $\Delta\mathrm{ln}\mathcal{L}$ values, which can then be compared to the value obtained from data. The pseudo experiments are prepared by injecting the best fit obtained for the LP model for each analysis. Each pseudo experiment is then fitted with both, the BPL and the LP models from which the TS ($-2\,\Delta\mathrm{ln}\mathcal{L}\,=\,-2\mathrm{ln}\mathcal{L}_{\mathrm{BPL}}-(-2\mathrm{ln}\mathcal{L}_{\mathrm{LP}})$) is calculated. Based on this test statistic and the pseudo experiments, we derive $p$-values of 0.008 for MESE and 0.018 for the CF analysis, for the observed preference of the BPL model under the assumption that the true spectral model is the best-fit LP spectrum. We therefore consider the BPL model as the spectral parameterization that best fits the data among the ones tested for both analyses.

The MESE analysis has also probed additional spectral shapes not reported in~\cite{IceCube:2025tgp}. These models and their results are shown in Tab. \ref{tab:mese_results}. We note that 
the assumed spectral shape of SPB, fit only for the MESE analysis, also provides a significant deviation from an SPL with a TS of 22.3. We again determine the chances of mis-identifying a true SPB model as a BPL model from pseudo experiments. Here, we inject the SPB best fit and calculate TS = $-2\,\Delta\mathrm{ln}\mathcal{L}\,=\,-2\mathrm{ln}\mathcal{L}_{\mathrm{BPL}}-(-2\mathrm{ln}\mathcal{L}_{\rm{SPB}})$ to compare the two model fits. Obtaining a p-value of 0.09 for this test, we are not able to distinguish this model from the best-fit model of a BPL. 

Similarly, the CF analysis tested for additional models that can describe the shape of the astrophysical spectrum.  Of particular interest is the BPL with independent flavor normalizations that demonstrates a higher deviation from the SPL case when compared to the standard BPL model, with a TS of 28.7. The results from these models are shown in Tab.~\ref{tab:CF_results}.

A comparison of the data and the simulation reflecting the best-fit BPL model for the MESE analysis is shown in Fig. \ref{fig:MESEBPLDataMC} and for the CF analysis in Fig.~\ref{fig:CombinedFitBPLDataMC}. 
The figures show the 1D-projections of cos($\theta_{\rm{reco}}$) and the reconstructed energy of the 2D-histograms used in the analysis. Data and best fit MC are compatible within 1 to 2 $\sigma$ in the analysis bins of all four histograms, which can also be seen in the panels showing the 1D-projections of the ratios. The CF tracks sample exhibits small, percent-level deviations between data and MC at energies below 10 TeV, which do not impact the physics results ( see the checks performed in Sec.~\ref{sec:bkg_data_mc}).

It is important to note that the physical properties of the shown data are very different between the four samples: tracks and cascades, and the two analyses, as explained in Sec.~\ref{sec:methods}. Hence, direct comparison of the reconstructed quantities should be done with caution. 

The MESE analysis is unable to measure a significant flux of prompt neutrinos and sets an upper limit on the prompt normalization as \textbf{$0.5$} in units of $\SI{e-18}{GeV^{-1} cm^{-2} s^{-1} sr^{-1}}$. This is hence not shown in the plots. The CF analysis, on the other hand, fits a non-zero prompt flux of \textbf{$1.04^{1.2}_{-1.1}$} in units of $\SI{e-18}{GeV^{-1} cm^{-2} s^{-1} sr^{-1}}$, consistent with a zero prompt flux within a confidence level of 1~$\sigma$. 
Predictions of the prompt neutrino flux from hadronic interaction models (e.g. Sibyll 2.3c~\cite{riehn_hadronic_2018}) indicate an order of magnitude higher flux of muon neutrinos from charmed mesons when compared to the prompt flux of electron or tau neutrinos. Therefore, the high statistics sample of tracks in the CF analysis can drive a non-zero prompt flux in the best fit, while the cascades-dominated MESE sample fits a zero prompt flux. For checks on the prompt flux, see Sec.~\ref{sec:prompt}.
The cos($\theta_{\rm{reco}}$) distributions of tracks and cascades both show the effect of the suppression of the atmospheric flux in the Southern sky due to the atmospheric self veto. The total contribution from the muon and astrophysical flux component is greater towards vertically down-going zenith angles than that from the conventional atmospheric neutrino flux, due to the suppression by the self veto. 

\begin{table}[h!]
\caption{Results for the additional spectral models tested in the MESE analysis. The uncertainties are derived from 1D profile likelihood scans, assuming Wilks' theorem. We show the preference over the single power-law hypothesis in terms of $-2 \Delta \mathrm{ln} \mathcal{L}$.
The flux is measured in units of $\SI{-18}{GeV^{-1} cm^{-2} s^{-1} sr^{-1}}$ and the normalization is determined at an energy of \SI{100}{TeV}.}
    \label{tab:mese_results}
    \renewcommand{\arraystretch}{1.5}
\centering
\begin{tabular}{c|c}
Astrophysical Model & Results \\ \hline 
SPB          &   \multicolumn{1}{c}{$\begin{array}{ccl}
          \phi_0 \, &=&1.42^{+0.21}_{-0.20} \\
          \gamma&=&2.512^{+0.059}_{-0.067} \\
          \log_{10}(\frac{E_\mathrm{bump}}{\mathrm{GeV}})&=& 4.30^{+0.13}_{} \\
          \log_{10}(\frac{\sigma_{\mathrm{bump}}}{\mathrm{GeV}})&=&4.421^{+0.097}_{-0.15}\\
          \phi_{\mathrm{bump}}\,/\,C\,&=&24.4^{+13}_{-8.4}\\
        -2 \Delta \mathrm{ln} \mathcal{L}&=&22.3
       \\ p= 5.65\cdot 10^{-5} && (3.9\, \sigma) 
         \end{array}$}      \\ \hline
SPL with BLLac         &    \multicolumn{1}{c}{$\begin{array}{ccl}
          \phi_0 \, &=&2.13^{+0.18}_{-0.17} \\
          \gamma&=&2.548^{+0.039}_{-0.041} \\
          \phi_{\mathrm{model}}&=&0^{+0.082} \\
        -2 \Delta \mathrm{ln} \mathcal{L}&=&0
               \\ p= 1 && (0\, \sigma) 
          \end{array}$}     \\ \hline
SPL with AGN           &   \multicolumn{1}{c} {$\begin{array}{ccl}
          \phi_0 \, &=&2.13^{+0.18}_{-0.17} \\
          \gamma&=&2.548^{+0.039}_{-0.041} \\
          \phi_{\mathrm{model}}&=&0^{+0.0025} \\
        -2 \Delta \mathrm{ln} \mathcal{L}&=&0
                       \\ p= 1 && (0\, \sigma) 
          \end{array}$}     
\end{tabular}
\end{table}
\begin{table}[h!]
\caption{Results for the additional spectral models tested in the CF analysis. The uncertainties are derived from 1D profile likelihood scans, assuming Wilks' theorem applies. We show the preference over the single power-law hypothesis in terms of $-2 \Delta \mathrm{ln} \mathcal{L}$.
The flux is measured in units of $\SI{-18}{GeV^{-1} cm^{-2} s^{-1} sr^{-1}}$ and the normalization is determined at an energy of \SI{100}{TeV}.}
    \label{tab:CF_results}
    \renewcommand{\arraystretch}{1.5}
\centering
\begin{minipage}{0.5\textwidth}
\begin{tabular}{c|c}
Astrophysical Model & Results \\ \hline
Two-Component Flux\footnote[2]{The two component flux model can identify a hardening in the spectrum, and when $\alpha$ is either zero or one, this reduces to an SPL. At the best fit value of $\alpha=0$, the likelihood space for $\Delta$ is flat.}       &   \multicolumn{1}{c}{$\begin{array}{ccl}
          \phi_0 \, &=&1.80^{+0.12}_{-0.16} \\
          \gamma&=&2.52^{}_{-0.02} \\
          \Delta&=&0.078\\
          \alpha&=&0.0^{}_{} \\
        -2 \Delta \mathrm{ln} \mathcal{L}&=&0
        \\ p= 1 && (0\, \sigma)
          \end{array}$}      \\ \hline
BPL, independent flavor         &    \multicolumn{1}{c}{$\begin{array}{ccl}
          \phi_0 \, &=&1.76^{+0.35}_{-0.29} \\
          \gamma_1&=&1.20^{+0.58}_{-1.50} \\
          \gamma_2&=&2.713^{+0.082}_{-0.086} \\
          \log_{10}(\frac{\mathrm{E}_\mathrm{break}}{\mathrm{GeV}})&=&4.44^{+0.12}_{-0.10} \\
          s_{\mathrm{e}}&=&0.0^{+0.24}_{} \\
          s_{\tau}&=&2.62^{+0.65}_{-0.66} \\
        -2 \Delta \mathrm{ln} \mathcal{L}&=&28.7
        \\ p= 8.99\cdot 10^{-6} && (4.3\, \sigma)
          \end{array}$}     \\ \hline
Muon Dampening           &   \multicolumn{1}{c}{$\begin{array}{ccl}
          \phi_0 \, &=&120^{+66}_{-36} \\
          \gamma&=&0.764^{+0.060}_{-0.079} \\
          \log_{10}(\frac{\mathrm{E}_{\mu,\mathrm{crit}}}{\mathrm{GeV}})&=&3.677^{+0.077}_{-0.069} \\
        -2 \Delta \mathrm{ln} \mathcal{L}&=&24.5
        \\ p= 7.43\cdot 10^{-7} && (4.8\, \sigma)
          \end{array}$}     
\end{tabular}
\end{minipage}
\end{table}
\begin{figure*}[t!]
    \begin{minipage}[b]{0.49\linewidth}
     \centering
        \includegraphics[width=\linewidth]{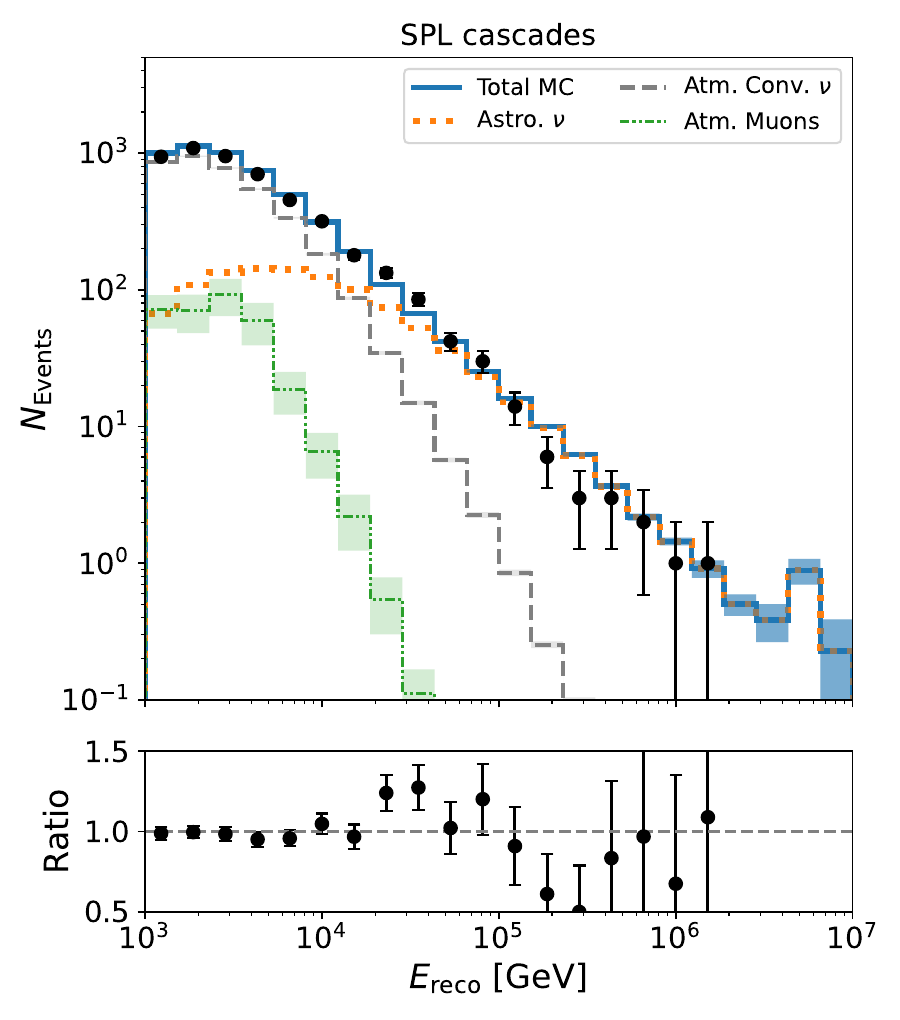}  
    \end{minipage}
    \begin{minipage}[b]{0.49\linewidth}
     \centering
        \includegraphics[width=\linewidth]{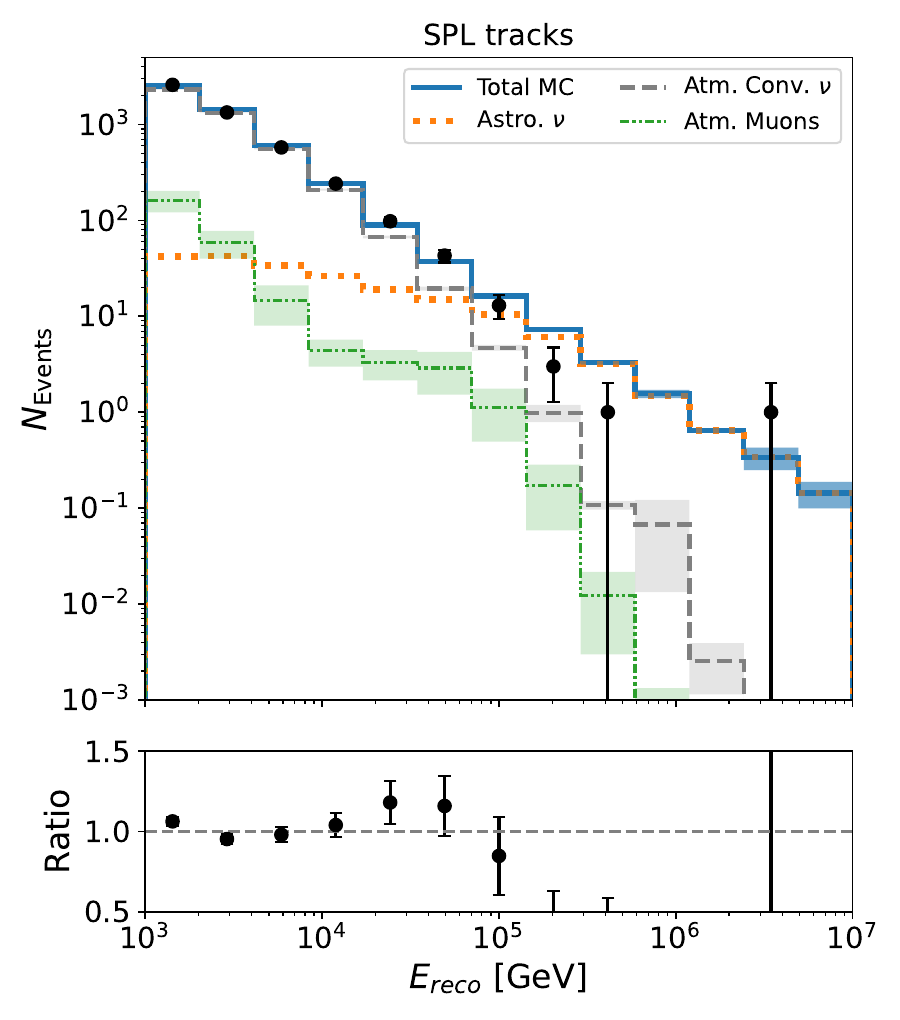}  
    \end{minipage}
\caption{\textbf{MESE SPL Data \& MC:} Comparison of Data and MC simulation for the best-fit SPL model. Shown are reconstructed cascade energy (left), and reconstructed track energy (right). The atmospheric prompt flux normalization is a free parameter which fits to zero. The excess of data events at energies close to \SI{30}{TeV} and the deficit of events above \SI{100}{TeV} drive the preference for a BPL spectral model. The excess disappears for a fit with a BPL model (cf. Fig.~\ref{fig:MESEBPLDataMC}).
}
\label{fig:MESE_SPLDataMC}
\end{figure*}
\begin{figure*}[t!]
    \begin{minipage}[b]{0.49\linewidth}
     \centering
        \includegraphics[width=\linewidth]{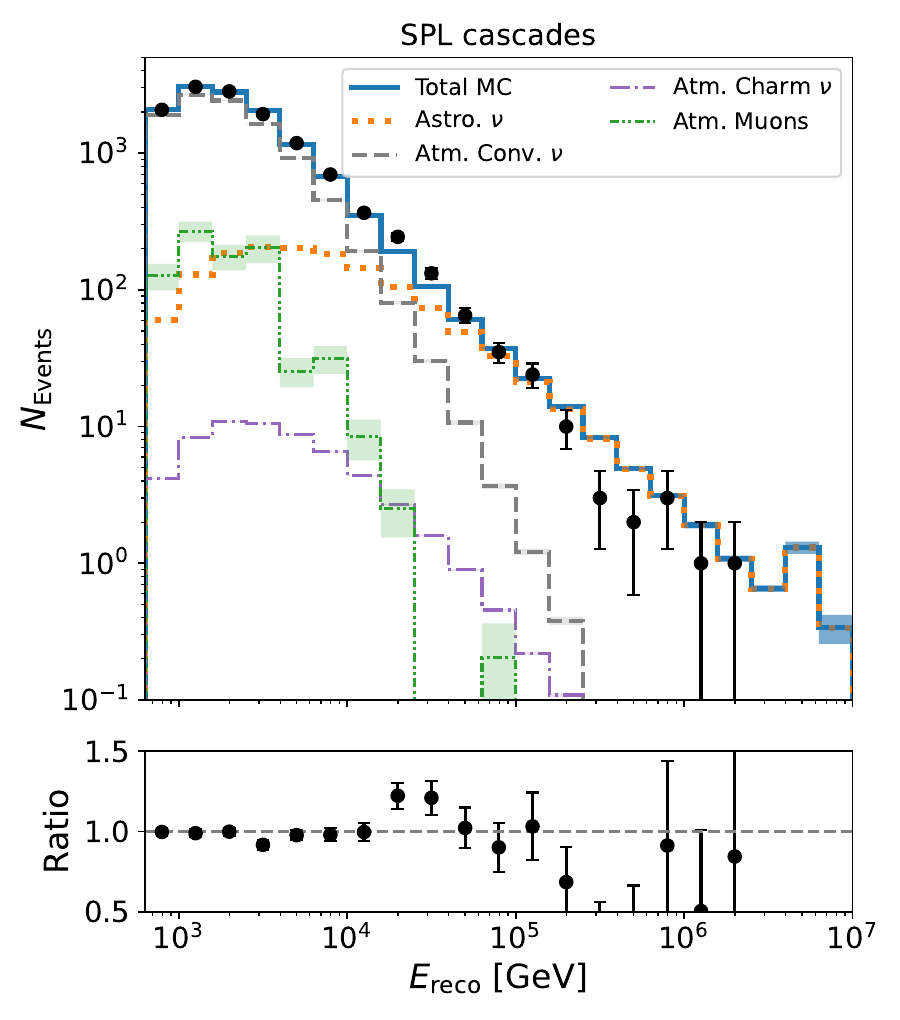}  
    \end{minipage}
    \begin{minipage}[b]{0.49\linewidth}
     \centering
        \includegraphics[width=\linewidth]{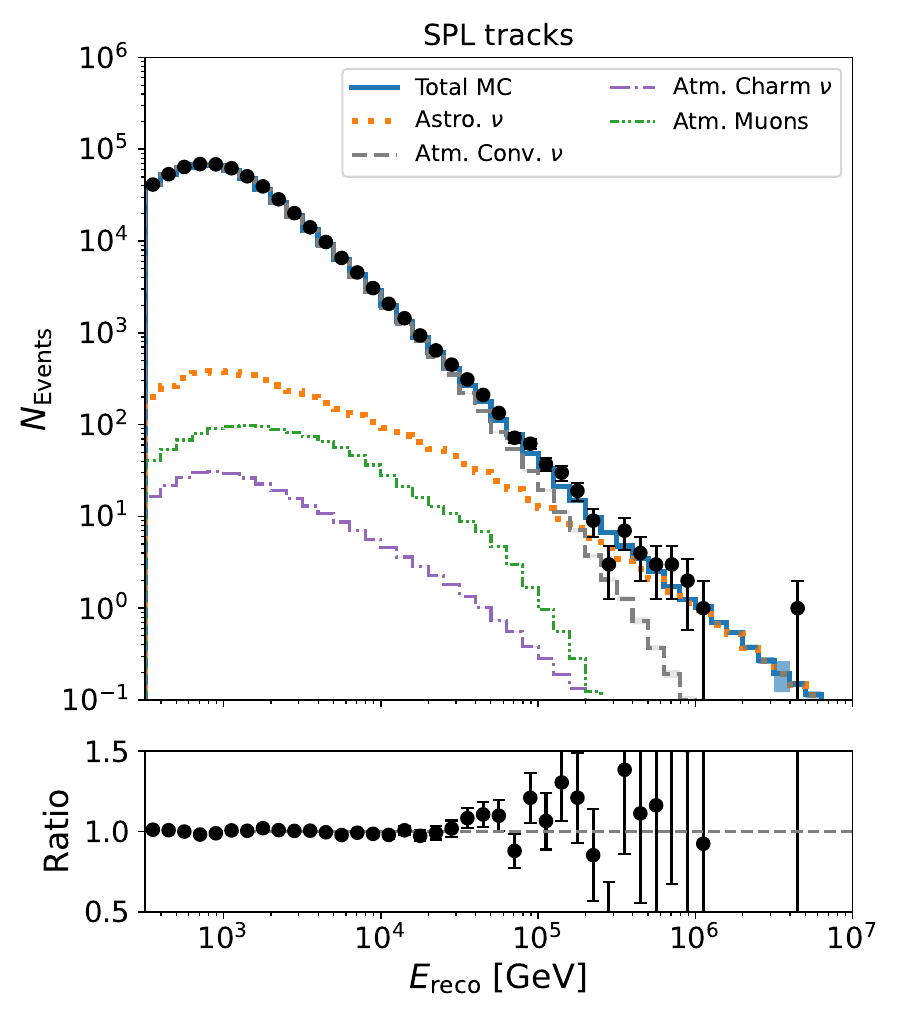}  
    \end{minipage}
\caption{\textbf{CF SPL Data \& MC:} Comparison of Data and MC for the best-fit SPL model for the CF analysis. Shown are reconstructed cascade energy (left), and reconstructed track energy (right). The excess of cascade-data events at energies close to \SI{30}{TeV} and the deficit of events above \SI{100}{TeV} drive the preference for a BPL spectral model.
}
\label{fig:CF_SPLDataMC}
\end{figure*}
Two features drive the deviation from the SPL in both analyses, an excess at $\sim \SI{30}{TeV}$  and a deficit at a few hundred TeV, when compared to the baseline SPL model. 
Fig. \ref{fig:MESE_SPLDataMC} shows the distribution of the reconstructed energy proxy for cascades and tracks in the MESE analysis, where simulations for the best-fit SPL model are compared to observed data. There is a visible excess of data compared to this simulation around \SI{30}{TeV} energy. 
This is more prominent in the cascades channel, and is also visible within a $1$ to $2\,\sigma$ deviation for the MESE tracks sample. The same features are also seen in the reconstructed energy proxy distributions of the CF analysis when compared to the best-fit SPL model, shown in Fig.~\ref{fig:CF_SPLDataMC}.
Another noticeable feature in the energy distributions is the dip at $\mathcal{O}$(\SI{100}{TeV}) seen in the cascades dataset of both analyses and in the starting-tracks sample of MESE. The poorer energy resolution of the through-going tracks sample renders it difficult to measure the possible dip with this channel. However, this dip is not yet statistically significant, and more data is needed to confirm or reject its existence.

\begin{figure*}[t!]
    \begin{minipage}[b]{0.8\linewidth}
\includegraphics[width=1\linewidth]{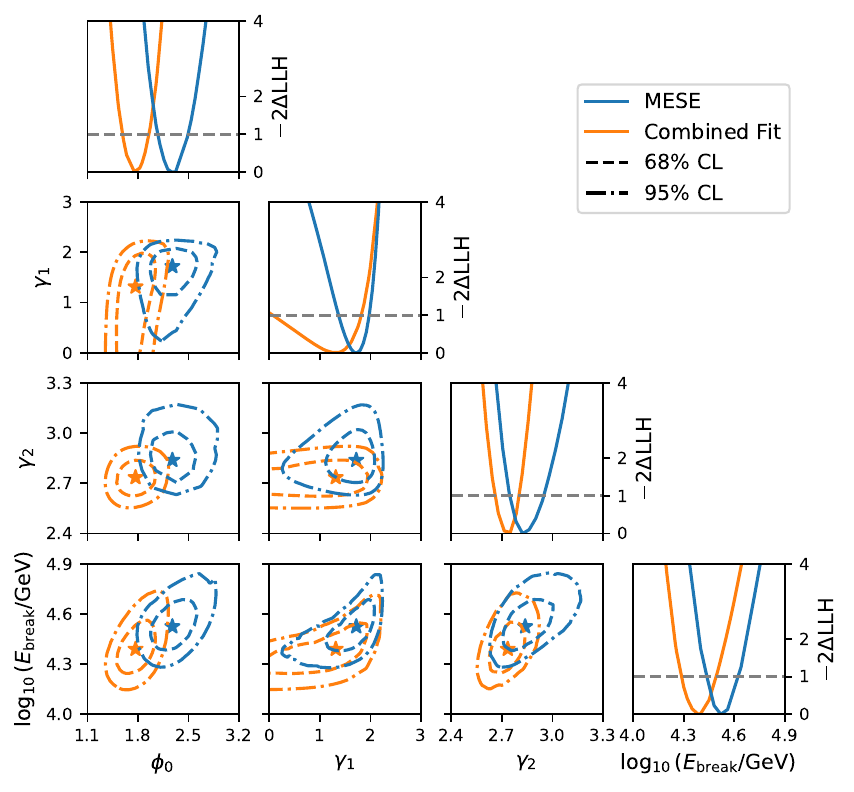}
\end{minipage}
\caption{\textbf{BPL spectral parameters:} Two-dimensional profile likelihood scans of all physics parameters in the BPL model fit. The star markers indicate the best fit parameter values with $\phi_0  =2.28$, $\gamma_1 = 1.72$, $\gamma_2 = 2.84$,  and log$_{10}$($\mathrm{E}_{\mathrm{break}}/\si{GeV}) = 4.52 $ for MESE, and  $\phi_0 =1.77$, $\gamma_1 = 1.31$, $\gamma_2 = 2.74$,  and log$_{10}$($\mathrm{E}_{\mathrm{break}}$/\si{GeV}) = 4.39 for CF.  The contours represent the \SI{68}{\percent} and \SI{95}{\percent} confidence regions for the parameters based on Wilks’ theorem. 
}
\label{fig:BPLContours}
\end{figure*}

Fig. \ref{fig:BPLContours} shows the 2D-profile likelihood scans of the parameters of the BPL spectral model ($\phi_0$: the astrophysical flux normalization at \SI{100}{TeV}, $\gamma_1$: the low-energy spectral index, $\gamma_2$: the high-energy spectral index, and $E_{\mathrm{break}}$: the break energy) for the two analyses presented in this paper. 
The figure shows the relative strengths of each analysis in measuring each component. The CF analysis shows better constraints in the measurement of the flux normalization and $\gamma_2$. This can be accounted for by the better constraints on the atmospheric neutrino flux derived from the through-going tracks sample. The MESE sample, on the other hand, demonstrates better performance in constraining $\gamma_1$. This can be attributed to the lower limit on the sensitive energy range for MESE. The sensitive energy range for MESE extends from \SI{5}{TeV} to \SI{7.5}{PeV}, while that of the CF analysis extends from \SI{13}{ TeV} to \SI{10}{PeV}. The sensitive energy range is calculated by comparing the per-energy bin likelihoods  ($\mathcal{L}_i$ in Eq.~\ref{eq:likelihood}) of a background-only hypothesis (no astrophysical flux) to the signal hypothesis with the astrophysical flux component included. The difference in $\mathcal{L}_i$ with both hypotheses shows the energy bins which have the most power to distinguish these two hypotheses. This distribution is integrated to yield a cumulative distribution, and the sensitive energy range is defined as the interval covering the \SIrange{5}{95}{\percent}  region of this cumulative distribution. 
The MESE dataset's sensitive energy range extends to a lower value than that of CF. This may be attributed to the finer zenith binning of MESE cascades, which in turn can give better constraints, particularly in the Southern Hemisphere, where the suppression of the atmospheric flux by the self-veto effect becomes important. It was also seen from Asimov studies that a softer spectral index (similar to the MESE results) gives stronger constraints on the $\gamma_1$ uncertainties than a harder index at lower energies (similar to CF results). It is difficult to disentangle this effect from the zenith-binning effect, both of which can contribute to the observed differences in the constraints on $\gamma_1$ from the two analyses reported here.

\begin{figure*}[t!]
    \begin{minipage}[b]{0.7\linewidth}
\includegraphics[width=1\linewidth]{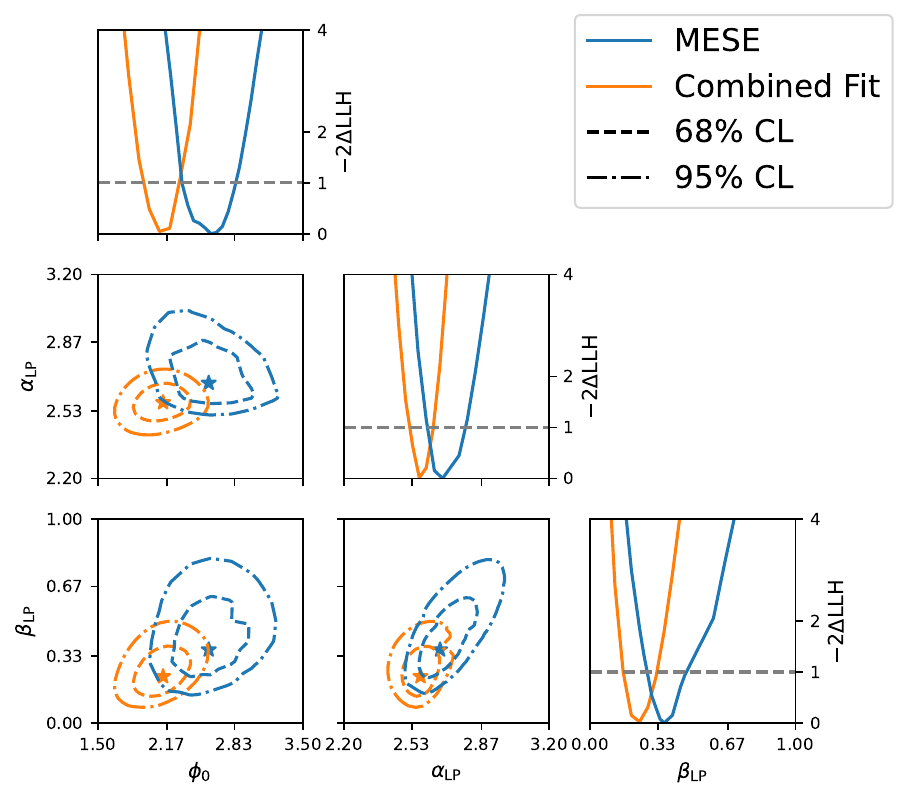}
\end{minipage}
\caption{\textbf{LP physics parameters:} Two-dimensional profile likelihood scans of all physics parameters in the LP model fit. The star markers indicate the best fit parameter values with $\phi_0 =2.58$, $\alpha_{\mathrm{LP}} = 2.67$, and $\beta_{\mathrm{LP}} = 0.36$ for MESE, and  $\phi_0 =2.13$, $\alpha_{\mathrm{LP}} = 2.57$, and $\beta_{\mathrm{LP}} = 0.23$ for CF.  The contours represent the \SI{68}{\percent} and \SI{95}{\percent} confidence regions based on Wilks’ theorem. 
}
\label{fig:LPContours}
\end{figure*}
Figure \ref{fig:LPContours} compares the 2D-profile likelihood scans for the parameters of the LP model ($\phi_{\mathrm{astro}}$: the astrophysical flux normalization at \SI{100}{TeV}, $\alpha$: the parameter describing the spectral index, and  $\beta$: the curvature parameter) for both analyses. We note that both $\alpha$ and $\beta$ are better constrained by the CF.

In addition to the fits of predefined spectral models, the normalization of the astrophysical neutrino flux in 13 independent energy bands (3 bins per decade and an additional bin at the highest energy, covering one decade) was fitted for both analyses, assuming a power-law spectrum with index 2 in each neutrino energy band. This fit is denoted as ``segmented'' fit below. 
 For bins where the best fit flux normalization is 0, we report the upper limits at \SI{68}{\percent} confidence level (C.L.). The best fit normalizations are provided in section Appendix \ref{sec:SegmentedFluxParameters}.  
\begin{figure*}[tbh!]
    \begin{minipage}[b]{0.49\linewidth}
     \centering
        \includegraphics[width=\linewidth]{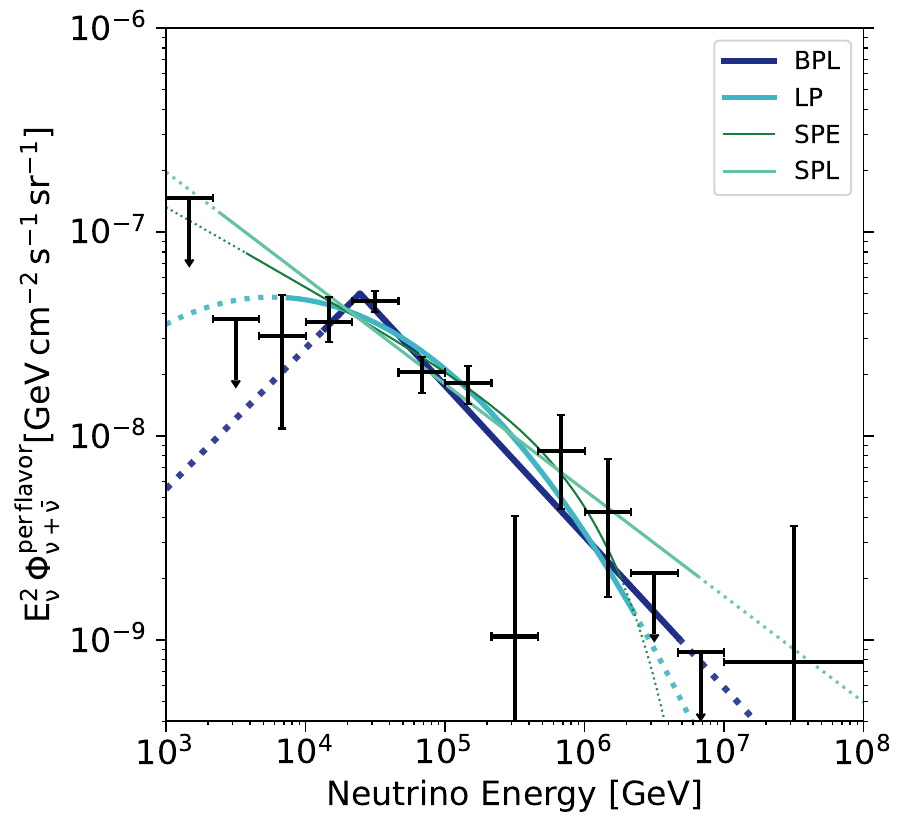}  
    \end{minipage}
    \begin{minipage}[b]{0.49\linewidth}
     \centering
        \includegraphics[width=\linewidth]{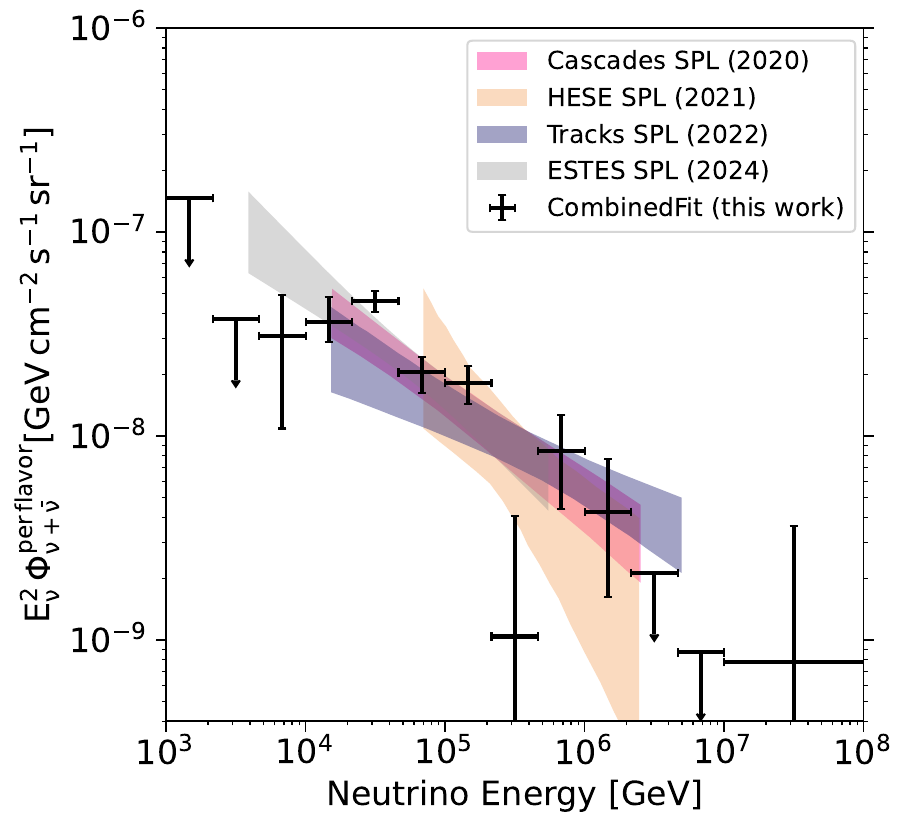}  
    \end{minipage}
\caption{\textbf{CF Segmented Flux:} Results of a fit of the astrophysical neutrino flux in independent energy bands. \textit{Left:} The results are compared to the models fitted in the analysis. The legend is ordered according to the likelihood obtained for each model. The solid lines show the energy range where the dataset is sensitive to the respective model, and the dotted lines show the energy range over which the fit is performed. \textit{Right:} The segmented fit is compared to previous measurements from IceCube, all under the assumption of a SPL.
}
\label{fig:CF_seg}
\end{figure*}

\begin{figure*}[tbh!]
    \begin{minipage}[b]{0.49\linewidth}
     \centering
        \includegraphics[width=\linewidth]{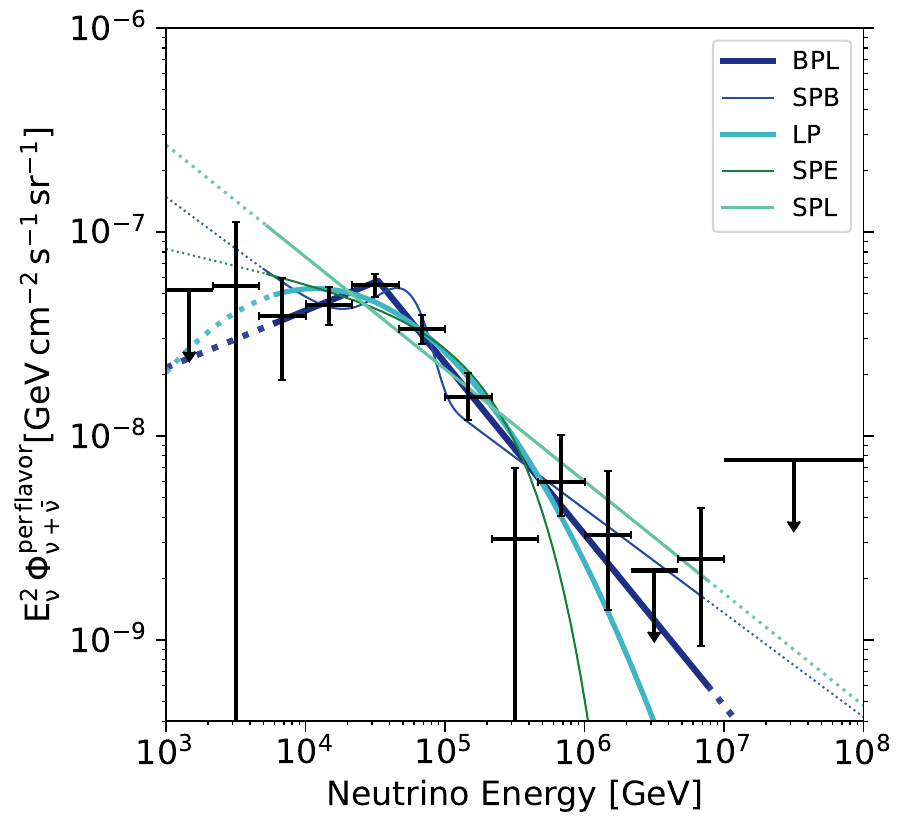}  
    \end{minipage}
    \begin{minipage}[b]{0.49\linewidth}
     \centering
        \includegraphics[width=\linewidth]{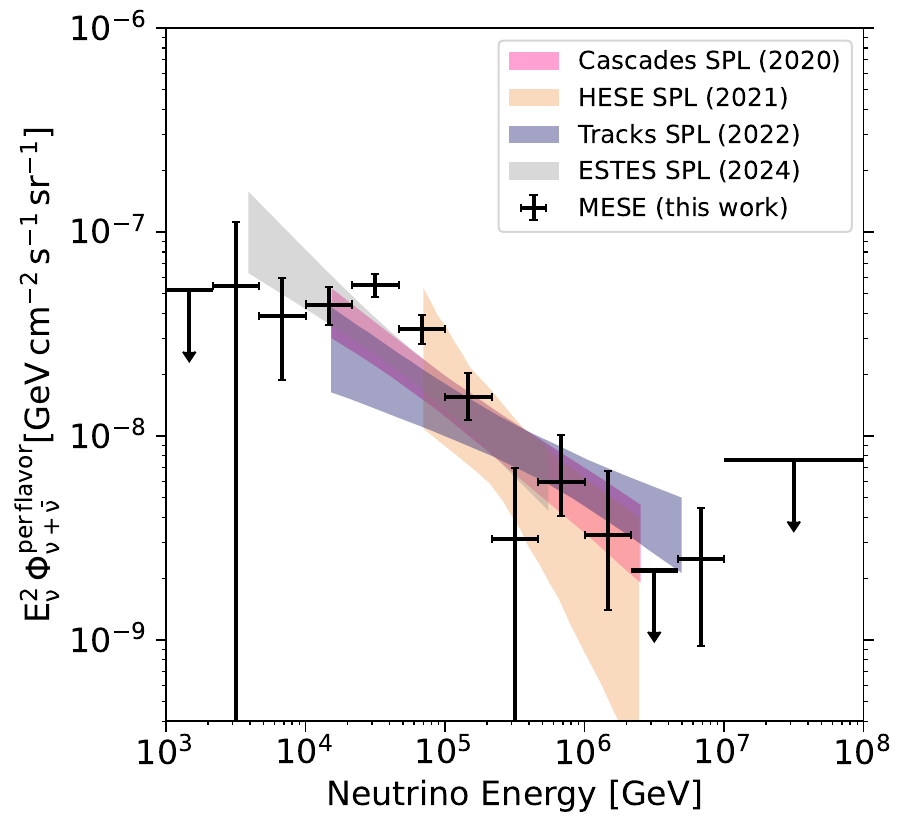}  
    \end{minipage}
\caption{\textbf{MESE Segmented Flux:} Results of a fit of the astrophysical neutrino flux in independent energy bands. \textit{Left:} The results are compared to the models fitted in the analysis. The legend is ordered according to the likelihood obtained for each model. The solid lines show the energy range where the dataset is sensitive to the respective model, and the dotted lines show the energy range over which the fit is performed. An additional fit of SPL + Gaussian bump is performed for the MESE analysis. \textit{Right:} The segmented fit is compared to previous measurements from IceCube, all under the assumption of an SPL. 
}
\label{fig:MESE_seg}
\end{figure*}
Figures \ref{fig:CF_seg} and \ref{fig:MESE_seg} show the best fits for the astrophysical neutrino flux obtained with this method for the CF analysis and the MESE analysis respectively. The best fit differential neutrino flux is compared to the astrophysical flux models tested previously for each analysis. The SPL~with~BLLac and SPL~with~AGN models tested with the MESE dataset are not shown here since the best-fit value for the non-power-law model component is zero in both cases, resulting in curves that look identical to the SPL fit. 
The fits obtained in the CF and MESE analyses agree well with each other, as demonstrated in~\cite{IceCube:2025tgp}.
A comparison with previous measurements from IceCube is also shown in these figures, where the \SI{68}{\percent} C.L. envelopes obtained from previous analyses are shown for the SPL model. 
The ESTES analysis~\cite{ESTES}, which utilizes starting tracks to measure the neutrino spectrum, also extends to the few TeV energy scale and is consistent with a single power law hypothesis in the entire energy range (see Section \ref{sec:estes_crosschecks}
 for a comparison of the segmented fit results). The ESTES analysis used a different treatment of systematic uncertainties and a different parameterization of the atmospheric and self-veto uncertainties when compared to the MESE and the CF analyses. 
 
\subsection{The Highest Energy Event}
\label{sec:highestenergyevent}
\begin{figure}[ht!]
    \begin{minipage}[b]{1\linewidth}
        \includegraphics[width=0.95\linewidth]{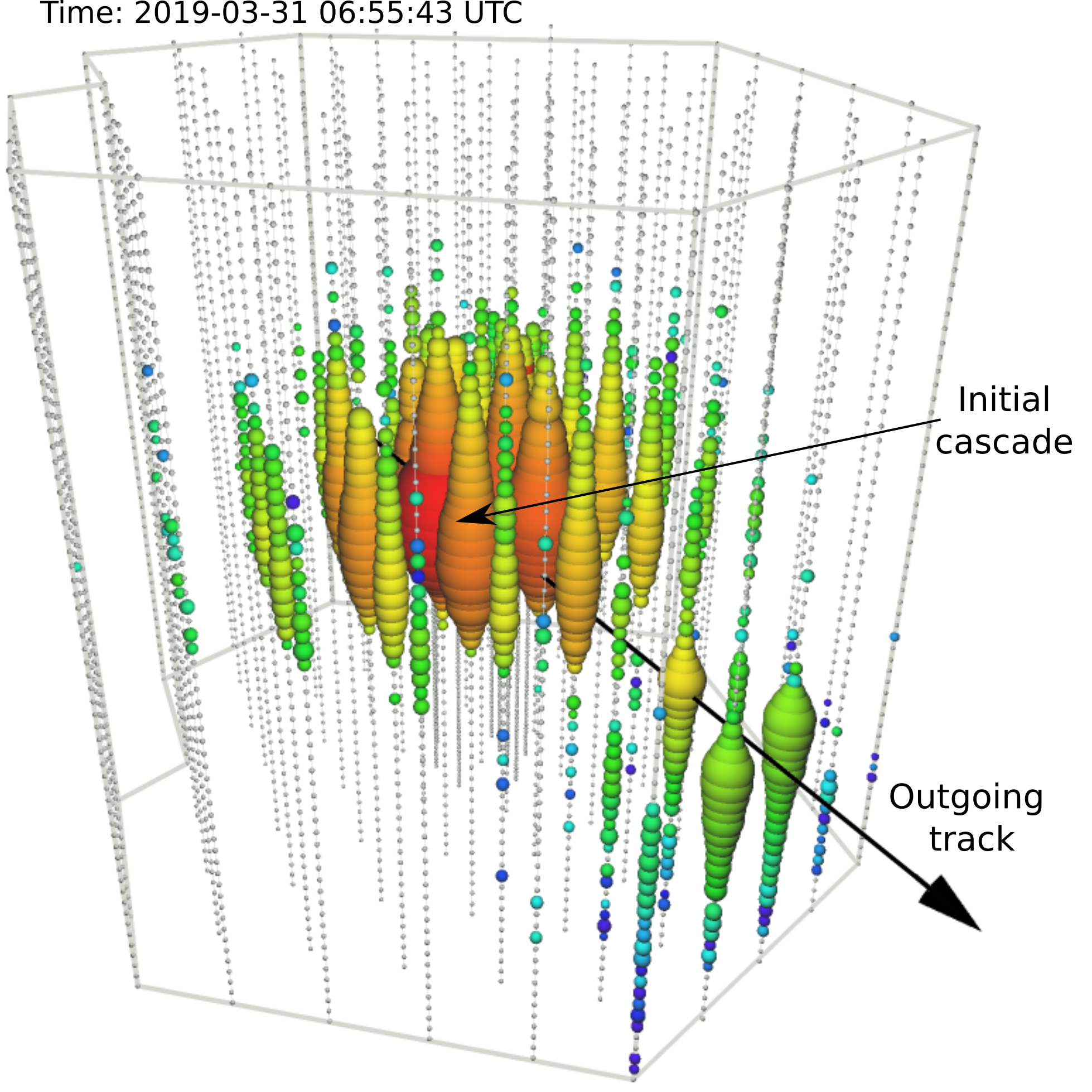}  
    \end{minipage}
    \begin{minipage}[b]{1\linewidth}
        \includegraphics[width=0.9\linewidth]{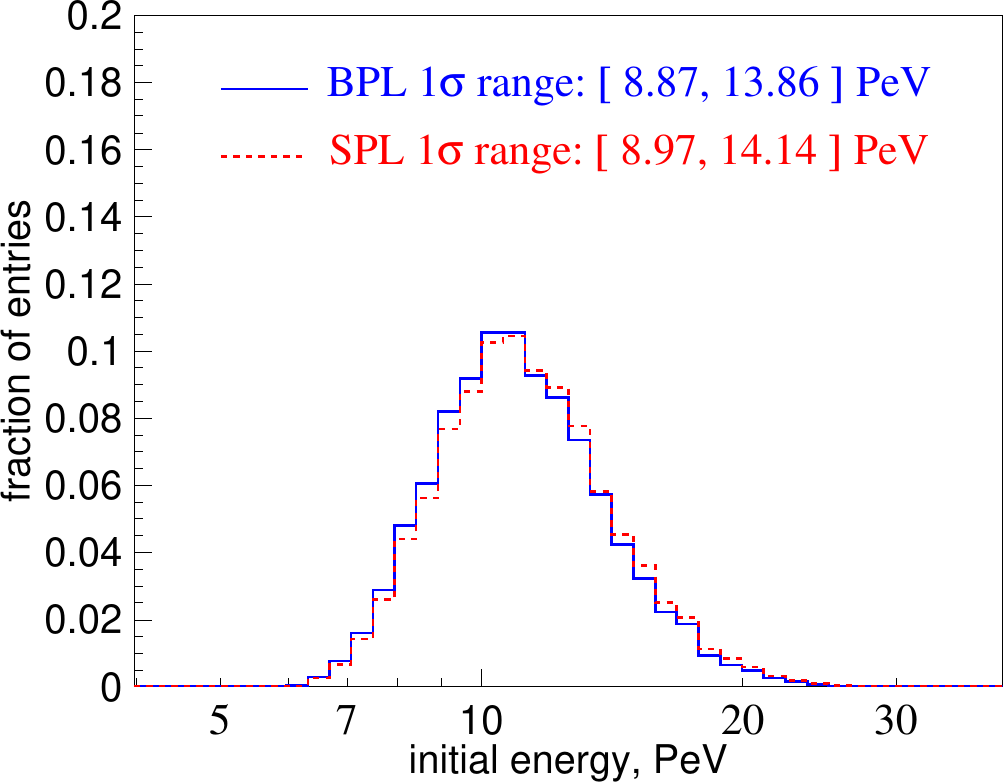}  
    \end{minipage}
    \caption{\textbf{IceCube's Highest Energy Neutrino Event:} An event view of the highest neutrino energy event recorded to date by IceCube is shown (top). The size of the spheres indicates the number of photoelectrons recorded by the respective DOMs. The earliest deposits are shown in red, with later deposits depicted first in yellow, then green, with the last deposits in blue. The duration of the event is \SI{22596}{ns}. This event stands out as an exceptional instance of a starting track, where the initial muon-neutrino interaction occurs inside the detector, creating a cascade feature, and the outgoing muon lepton leaves a track signature in the detector volume. The cascade and track portions of the event are highlighted. 
    The reconstructed neutrino energy distribution of the highest neutrino energy event recorded to date during IceCube’s data taking is shown (bottom). The distribution is derived from multiple simulations of the neutrino interaction that could have occurred to produce the observed event. The median neutrino energy is \SI{11.4}{PeV} assuming the best fit broken power law spectral model for the astrophysical neutrino flux.
    }
    \label{fig:highest_energy_event}
\end{figure}

The MESE dataset contains the highest-energy event recorded by IceCube as of August 2024. 
The event was observed on March 31, 2019 (issued as a public alert~\cite{2019GCN.24028....1I}), and is identified as a down-going starting track. An event view is shown in Fig.~\ref{fig:highest_energy_event}.
The event is tagged as a HESE event within the event selection. It has a reconstructed deposited energy of $\sim  \SI{3.7}{PeV}$ using the reconstruction algorithms introduced above for the MESE analysis. 
The event was also part of the 12-year HESE dataset which has been released publicly~\cite{HESE12}, in addition to a sample of Extremely High Energy (EHE) neutrino events observed by IceCube~\cite{EHE_PRL}.
A further detailed reconstruction of the event was conducted using the \textsc{DirectFit} algorithm\footnote[1]{
\textsc{DirectFit} operates by simulating several iterations similar to the given event ($\sim 10000$ for determining the solution and another $\sim 10000$ to calculate the uncertainties), exploring its allowed parameter space, and further propagating the event through ice to obtain the deposited light in the DOMs. This procedure enables accurate modeling of the reconstructed event, however at very high computational costs, and is therefore performed only for the highest energy event reported in this paper. 
All events released in~\cite{HESE12} and additional interesting events such as the multi-PeV track-like event in~\cite{IceCube:2016umi} and~\cite{IceCube:2021rpz} have also been reconstructed using \textsc{DirectFit}.}~\citep{HESE12,Chirkin:2013avz}.
\textsc{DirectFit} uses the most recent ice models, including details of ice layer undulations and ice model anisotropy~\cite{IceModelFTP}, 
not otherwise used in the standard simulations generated for the measurement of the spectrum presented in this paper. 
The visible energy of the event was reconstructed using these updates as a part of the HESE data release \cite{HESE12}. The initial cascade's visible energy is estimated as \SI{4.4}{PeV} (which can be approximated as the equivalent electromagnetic (EM) loss, and scales up to \SI{4.6}{PeV}  for a hadronic cascade). The muon energy is estimated by studying its energy losses along the path in the detector.
This energy loss is found to be $\mathrm{d}E/\mathrm{d}x \simeq 
\SI{1.125}{TeV \, m^{-1}}$ after \SI{400}{m} of traveling~\cite{Chirkin:2004hz}, which is the distance of the track to the edge of the detector. The energy loss for an equivalent EM cascade for the whole event is therefore \SI{4.8}{PeV} (scaled to \SI{5}{PeV} if the first cascade is hadronic).
We draw samples from an energy distribution from $\sim \SI{5}{PeV} $ to \SI{100}{PeV}, assuming a neutrino spectrum  that follows a broken power law with the parameters obtained as our best fit, and perform multiple repeated simulations. 
The lower limit of this sampled energy distribution arises from the reconstructed energy loss.
These simulations are performed under the constraint that the event deposits \SI{440}{TeV} in these \SI{400}{m}, obtained from the summed energy deposit along the outgoing track.
The muon energy distribution is further constrained by the inelasticity relationship (neutrino cross section), assuming the first interaction is a hadronic cascade with an energy of \SI{4.6}{PeV}.
This results in an initial muon energy between \SIrange{4.3}{9.3}{PeV}, which corresponds to a total neutrino energy of \SIrange{8.9}{13.9}{PeV} (\SI{68}{\percent} C.L.). If we assume that the spectrum, instead, follows a SPL according to the best-fit values from MESE, the estimated neutrino energy lies within the interval $9.0-14.1$~PeV.
The event reported here is lower in energy than the ultra-high energy event reported by KM3NeT in~\cite{KM3NeT:2025npi}. It is worth noting that the KM3NeT observation of the $\mathcal{O}$(100 PeV) event is in $\sim 3\,\sigma$ tension with the limits reported by IceCube in~\cite{EHE_PRL}.

Fig.~\ref{fig:highest_energy_event} shows a visualization of this event, as well as the posterior (reconstructed) energy distribution.
\section{Robustness}
\label{sec:crosschecks}

\begin{figure*}[t!]
    \begin{minipage}[b]{0.49\linewidth}
     \centering
        \includegraphics[width=\linewidth]{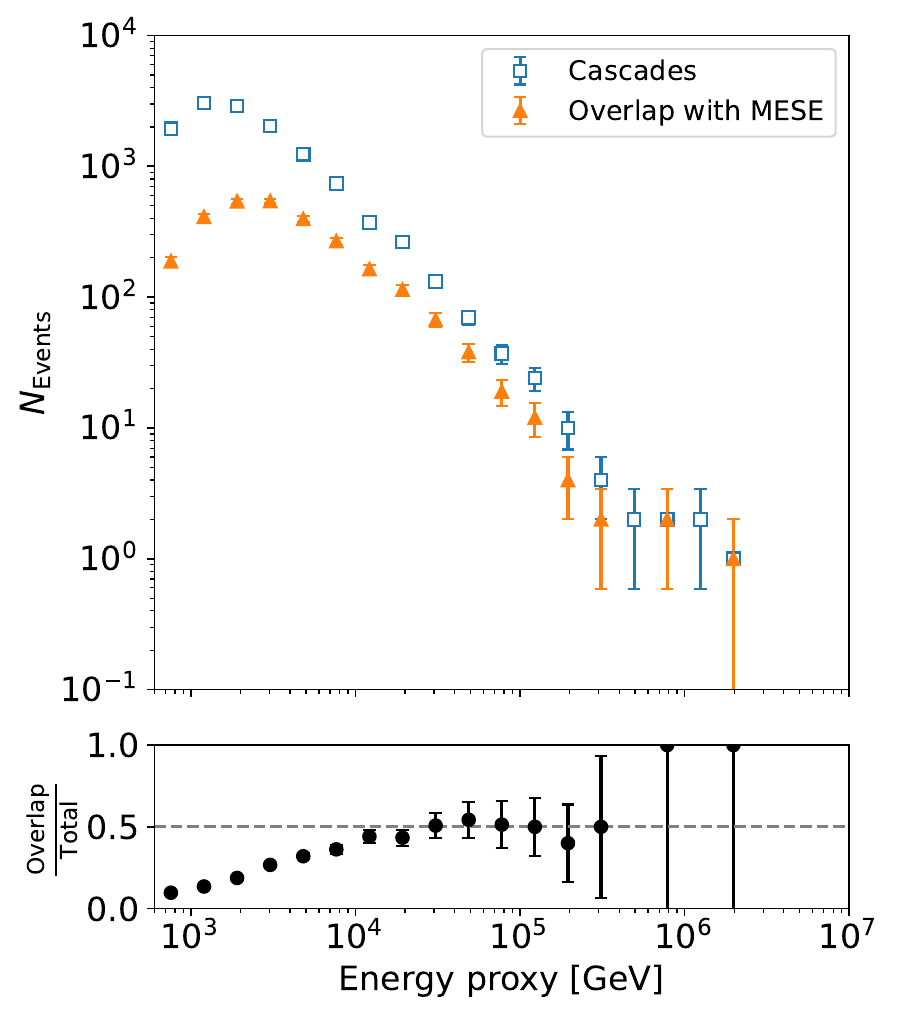}  
    \end{minipage}
    \begin{minipage}[b]{0.49\linewidth}
     \centering
        \includegraphics[width=\linewidth]{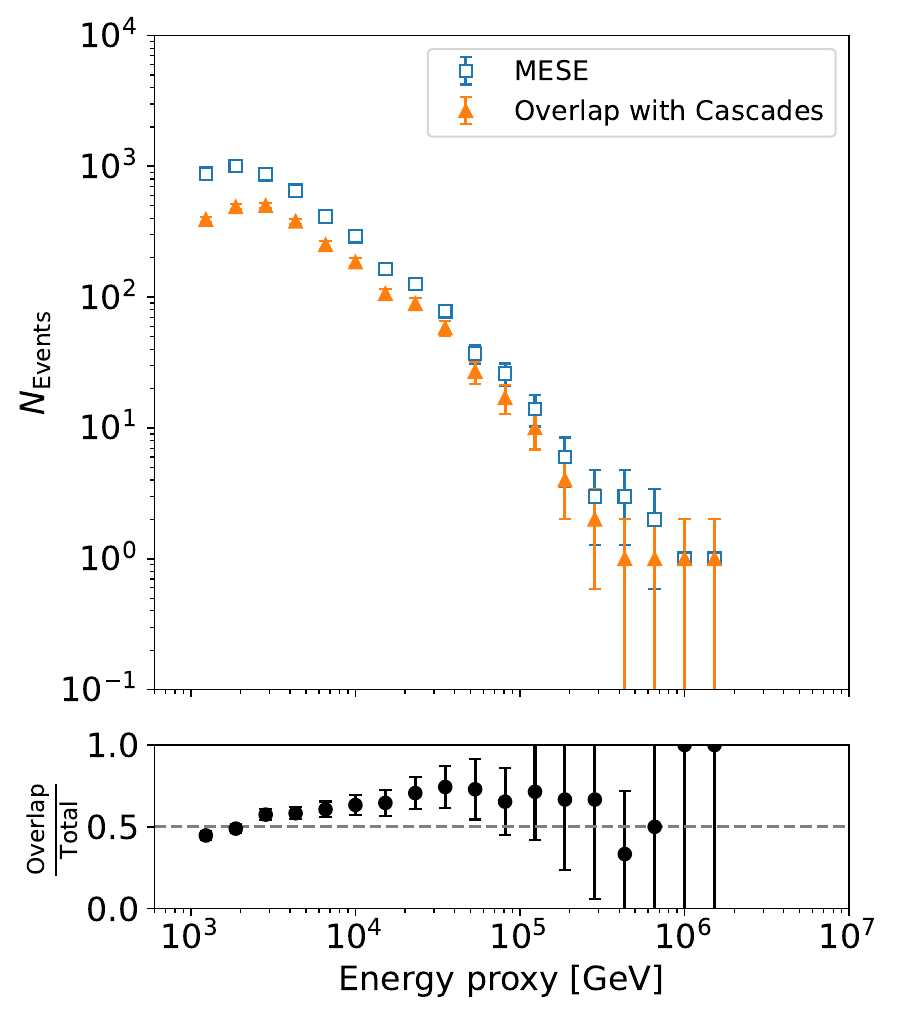}  
    \end{minipage}
\caption{\textbf{Overlap of MESE and Cascades datasets:} \SI{55}{\percent} of the cascade events in MESE overlap with the Cascades dataset and \SI{19}{\percent} of the events in Cascades dataset overlap with MESE. The energy dependence of the overlap is shown here.
}
\label{fig:Overlap}
\end{figure*}
The robustness of the analysis results was validated with numerous checks, both before and after the spectral parameters were fitted. The primary purpose of these checks was to ensure that the data was accurately modelled by the simulations and no significant biases are introduced by unaccounted or mismodelled systematic uncertainties. 
As the MESE selection and datasets used in CF utilize events with similar morphologies, especially in the cascades channel, a substantial fraction of events exist in both samples. Since both analyses consistently reject the SPL hypothesis, it is pertinent to evaluate the degree of overlap in the samples. 
The overlap of the cascade events as function of energy is illustrated by Fig.~\ref{fig:Overlap}. A total of \SI{55}{\percent}  of the MESE cascade events are also present in the cascades dataset used in the CF analysis, and  \SI{19}{\percent} of the cascade events of the CF analysis are present in MESE. 
At energies below \SI{10}{TeV},  there is only a small fraction of MESE cascades present in the cascades dataset of the CF analysis, and this fraction rises quickly to about \SI{50}{\percent} at energies beyond \SI{20}{TeV} (see Fig.~\ref{fig:Overlap} (left)). 
On the other hand, the overlap percentage consistently lies at $\sim$ \SI{50}{\percent} for events from the CF cascades dataset which are also in the MESE cascade sample, at all energy scales as seen in Fig.~\ref{fig:Overlap} (right). Only \SI{0.8}{\percent}  of MESE track events  exist in the through-going tracks dataset while these overlapping track events account for \SI{39}{\percent} of the events within the MESE tracks sample (see Tab.~\ref{tab:event_numbers}).

In accordance with IceCube's strict blind analysis procedures, numerous validation checks were performed on the fit before the physics parameters were derived.
Other validation tests have been performed post-unblinding. 
In this case the robustness of the fit results for spectral parameters with respect to changes in the simulation model and/or input data has been verified. 
Both MESE and CF went through a series of checks individually defined for the respective analysis. A selection of these robustness tests and their outcomes are reported below.

\subsection{Dataset splits}
Several dataset split tests were performed for the MESE analysis where the dataset was divided into disjoint sub-samples based on various metrics, and then the fits were rerun on these sub-samples. These fits were then compared to the results obtained with the full dataset. 
These metrics included separating events into (a) the Antarctic summer vs the winter, testing for seasonal variations in the data; (b) events from the southern sky vs the northern sky, since we have different background rates in the two hemispheres; (c) events with interaction vertices above and below the dust band, as the ice below the dust band shows on average less absorption and scattering than that above; and (d) events from 2011-2017 compared to 2017-2022, in order to check for variations in the data from the first and second half of the total observation period. 
The compatibility of these splits was verified by comparing the 1D profile likelihoods of the nuisance parameters where it was required that at least 7 out of the 15 nuisance parameters are within the $1\,\sigma$ regions of the best fit with the full dataset, to ensure that none of them were being pulled away from the best fit with the full dataset. 
All observed variations of the physics parameters are within $3\,\sigma$ of the best fit with the full dataset, as shown in Tab.~\ref{tab:splits}. 
A summer vs winter split was also performed with the CF analysis. Here, instead of performing separate fits for the summer and winter data, a joint fit was conducted but with separate modeling of the baseline atmospheric flux for the summer and winter seasons instead of an annual average. This was done only for the cascades sample of the CF analysis, since the excess at \SI{30}{TeV} is driven by the cascades. This check also finds good agreement of the spectral parameters between the annual and the seasonal fits. The seasonal fit shows a marginal improvement of the significance of BPL with respect to SPL by $0.1\,\sigma$.

\subsection{Data/MC agreement of through-going track sample} 
\label{sec:bkg_data_mc}
The CF through-going track dataset, exhibits small, percent-level deviations between data and simulations at energies below \SI{10}{TeV}. 
This region in energy is dominated by the atmospheric muons, and is defined as the background region for the CF analysis. These deviations have, so far, not been explained by known systematic uncertainties. A data-driven method was developed and applied to evaluate potential systematic effects of such deviations on the fit results, in particular, on the parameters describing the spectrum of the astrophysical neutrino flux.
The difference between data and simulation in the reconstructed energy vs. cos($\theta_{\rm{reco}}$) distribution of the through-going tracks for $E < \SI{10}{TeV} $ is decomposed into its Fourier modes using the analysis binning (by discrete Fourier transform). A correction to the baseline model is then calculated using only the $n$ lowest modes in the back-transformation. Pseudo-experiments are drawn from the corrected model for various choices of the cutoff mode $n$, which are then fit using the expectation values from the baseline simulation. The impact on the astrophysical parameters was observed to be minimal for all tested modes. The results of the tests confirmed that the physical parameters of interest in the analysis are not impacted by the small data / simulation discrepancies in the track dataset. More details on this method can be found in~\citep{EGansterThesis,RNaabThesis}. 

\subsection{Mean inelasticity of the neutrino interactions} Since the CF analysis does not explicitly include uncertainties on the inelasticity model used within the simulation in the baseline fit procedure, a test on this was performed on the sample after unblinding. 
For this test, a scaling factor for the mean inelasticity (from CSMS~\cite{CSMS}) of the neutrino interactions is included as an additional nuisance parameter in the fit using a procedure outlined in~\cite{binder2017measurements}.
A Gaussian prior of $\pm \SI{10}{\percent} $ is included in the fit.
The best-fit inelasticity scale was 1.01 times the nominal mean inelasticities of the DIS cross sections~\cite{CSMS} used in IceCube simulations, only slightly above the baseline.
No bias in the signal parameters was observed (see Tab.~\ref{tab:CF_checks}). The likelihood of the best fit improved only by 0.96 units for the fit with an additional inelasticity scale parameter, implying that the best fit inelasticity scale was compatible with the nominal value within 1\,$\sigma$.

The MESE analysis includes a scale parameter for uncertainties on the mean inelasticity as a fit parameter in the baseline fit. 
The best-fit value for this nuisance parameter pulls away from the baseline assumption from CSMS. The best-fit value is $\sim \SI{11}{\percent} $ lower for the BPL fit and $\sim \SI{18}{\percent} $ for the SPL fit when compared to the baseline value from CSMS, across all energies. 
This deviation is larger than the uncertainties typically quoted for the mean inelasticities in theoretical calculations~\citep{Candido:2023utz,Xie:2023suk}. 
However, it is possible that this parameter is absorbing some unmodeled systematic effects. A reduction in the scale parameter for the mean inelasticity (while holding all other parameters fixed) results in a lower number of observed events in both the cascades and tracks channels, as fewer events pass the charge threshold of the MESE event selection.  The inelasticity scaling parameter is only weakly correlated to the astrophysical flux parameters. Hence the astrophysical flux measurement is robust against the observed deviation.

\subsection{Neutrino cross section}
Another test performed on the CF sample tests for possible variations in the overall scales of the baseline neutrino DIS cross sections, to identify any effects on the spectral parameters. Here a cross section dependent scaling parameter is included in the fit. A Gaussian prior of $\pm \SI{10}{\percent} $ for the scaling factor is included in the fit, with the best-fit value of the cross section scaling factor at 1.1 ($\SI{10}{\percent}$ higher than baseline). Note that to first order, cross section uncertainties are absorbed by other systematic parameters, such as DOM efficiency. Any actual cross section uncertainty is only different from the systematic effect from DOM efficiency via the zenith distribution, as it affects both the absorption and detection of neutrinos for up-going events. With the inclusion of this parameter, the improvement in the likelihood of the best fit is only minimal at $\sim 1.4$ and the best fit values remain consistent with the original measurement (see Tab.~\ref{tab:CF_checks}), indicating that this does not have a major impact on the measurement and that the result is consistent with nominal cross sections.

\subsection{Neutrinos from the galactic plane} IceCube has recently detected a flux of neutrinos from the Milky Way\cite{GalacticPlaneScience}, which contributes to the diffuse neutrino spectrum measured by the two analyses. 
The non-isotropic flux of neutrinos from the galactic plane could affect the fitted parameters of the spectral models used to describe the diffuse neutrino flux, which is assumed to be isotropic.  
For the MESE analysis the impact on the spectral fits is tested by adding a galactic component to our spectral model, using the emission template and spectrum from~\cite{GalacticPlaneScience} that showed the best agreement with IceCube data (``Fermi-$\pi^0$''). The normalization of the galactic-emission template becomes an additional free parameter in the fit to data.
We find that in this case the spectral parameters for the diffuse astrophysical neutrino flux remain stable at their baseline best-fit values, and that our best fit of the galactic plane flux normalization is zero (see Tab.~\ref{tab:mese_checks}). 

A similar check was performed for the CF analysis using a  template for the galactic neutrino flux calculated in~\cite{CRINGE}. There, the galactic plane flux normalization was a free parameter in the fit, under the constraints of a Gaussian prior. The physics parameters did not significantly change from their baseline best fit values, i.e. when fitting without a galactic component in the fit. The best fit value of $\gamma_1$, reduced by 0.3, which is well within the $1\sigma$ limits. The likelihood for the best fit degraded by merely $0.8$ units (see Tab.~\ref{tab:CF_checks}).

\subsection{Energy scale for tracks}
The reconstruction used for the tracks sample in CF analysis may incorrectly estimate the muon energy if simulations do not correctly or completely describe the energy losses.
This would cause the energy scale of the track events to be incorrect. An additional parameter is introduced just for the track like events which accounts for their optical efficiency, parameterized identically to the DOM efficiency. A Gaussian prior of $\pm\SI{5}{\percent}$ is included in the fit, as this is the maximum uncertainty we would expect for the muon energy scale. This extension of the model did not affect the measured spectral features and the likelihood of the best fit changed only by $0.06$ (see Tab.~\ref{tab:CF_checks}). The best fit for the optical efficiency scale of tracks was $\SI{99}{\percent}$, showing minimal deviation from the nominal.

\subsection{Atmospheric neutrino flux modeling} Atmospheric neutrinos are the most dominant background in this analysis, where we search for the flux of astrophysical neutrinos. Any mismodeling in its flux could thus impact the astrophysical flux measurement. The robustness of the baseline atmospheric-neutrino template is checked in both MESE and CF. A cross check was done with the MESE dataset, where the atmospheric neutrino flux template was changed. A fit was performed with updated parameterizations of the atmospheric neutrino spectrum using the DAta-drivEn MuOn-calibrated Neutrino Flux model (DaemonFlux)~\cite{daemonflux}. DaemonFlux uses data-driven models of the cosmic-ray composition~\cite{GSF6} and the secondary particle yields~\cite{DDM} as inputs to the \textsc{MCEq} code~\cite{MCEq}, to produce the most constraining  atmospheric neutrino flux model to date. We perform the spectral model fits assuming DaemonFlux to be our model for the atmospheric neutrino flux and evaluate the effect on the physics parameters by comparing this fit to the baseline (H4a with Sibyll 2.3c). We observe that the physics parameters do not vary significantly compared to the baseline, while the nuisance parameters describing the uncertainties in the DaemonFlux predictions fit well within their expected ranges (see Tab.~\ref{tab:mese_checks}). We observe that the best-fit physics parameters using DaemonFlux are within $1\,\sigma$ of the physics parameters obtained from the baseline fit. This test confirms that our results do not depend significantly on the detailed CR composition and hadronic interaction models. 

The CF analysis performed checks on the atmospheric neutrino spectrum through the cosmic-ray interpolation parameter. 
This was updated as an interpolation between the global spline fit (GSF) model and the H4a model, which can account for variations in the atmospheric neutrino flux arising from variations in the primary cosmic-ray and hadronic-interaction models. The fit was repeated with this updated parameter and the likelihood improved by only $0.21$ units where we observe minimal deviation ($<1\,\sigma$) in the astrophysical flux parameters (see Tab.~\ref{tab:CF_checks}).

\begin{table}[htbp]
\caption{Results for systematics checks performed in the MESE analysis. The uncertainties are derived from 1D profile likelihood scans, assuming Wilks' theorem applies. The flux is measured in units of\SI{e-18}{GeV^{-1} cm^{-2} s^{-1} sr^{-1}}. All flux normalizations are at \SI{100}{TeV}.}
\label{tab:mese_checks}
\renewcommand{\arraystretch}{1.5}
\centering
\small
\begin{tabular}{c|c}

Fit type & 
BPL fit values \\ 
\hline

\makecell[tc]{Unblinded Fit} &  
\begin{tabular}[t]{@{} l l @{}}
    $\phi_0$ & $= 2.28^{+0.22}_{-0.20}$ \\
    $\gamma_1$ & $= 1.72^{+0.26}_{-0.35}$ \\
    $\gamma_2$ & $= 2.839^{+0.11}_{-0.091}$ \\
    $\log_{10}(\frac{\mathrm{E}_\mathrm{break}}{\mathrm{GeV}})$ & $= 4.524^{+0.097}_{-0.087}$ \\
\end{tabular} \\ 
\cline{1-2} 
\makecell[tc]{Fit with \\galactic plane flux} & 
\begin{tabular}[t]{@{} l l @{}}
    $\phi_0$ & $= 2.28^{+0.32}_{-0.48}$ \\
    $\gamma_1$ & $= 1.72^{+0.34}_{-0.72}$ \\
    $\gamma_2$ & $= 2.84^{+0.14}_{-0.13}$ \\
    $\log_{10}(\frac{\mathrm{E}_\mathrm{break}}{\mathrm{GeV}})$ & $= 4.52^{+0.17}_{-0.11}$ \\
\end{tabular} \\\cline{1-2} 
\makecell[tc]{Fit with \\DaemonFlux} & 
\begin{tabular}[t]{@{} l l @{}}
    $\phi_0$ & $= 2.22^{+0.22}_{-0.21}$ \\
    $\gamma_1$ & $= 1.88^{+0.23}_{-0.32}$ \\
    $\gamma_2$ & $= 2.843^{+0.12}_{-0.099}$ \\
    $\log_{10}(\frac{\mathrm{E}_\mathrm{break}}{\mathrm{GeV}})$ & $= 4.55^{+0.11}_{-0.10}$ \\
\end{tabular}  \\\cline{1-2} 
\makecell[tc]{Fit with \\Unbounded Prompt} & 
\begin{tabular}[t]{@{} l l @{}}
    $\phi_0$ & $= 2.38^{+0.35}_{-0.32}$ \\
    $\gamma_1$ & $= 1.88^{+0.30}_{-0.52}$ \\
    $\gamma_2$ & $= 2.86^{+0.19}_{-0.11}$ \\
    $\log_{10}(\frac{\mathrm{E}_\mathrm{break}}{\mathrm{GeV}})$ & $= 4.53^{+0.16}_{-0.14}$ \\
    $\phi_{\mathrm{prompt}} $ & $= -1.18^{+1.68}_{-1.90}$ \\
\end{tabular} 
\end{tabular}

\end{table}
\begin{table}[htbp]
\caption{Results for systematics checks performed in the CF analysis. The uncertainties are derived from 1D profile likelihood scans, assuming Wilks' theorem applies. The flux is measured in units of $\rm{10^{-18}/GeV/cm^2/s/sr}$. All flux normalizations are at 100 TeV.}
\label{tab:CF_checks}
\renewcommand{\arraystretch}{1.5}
\centering
\small
\begin{tabular}{c|c}

Fit type & 
BPL fit values \\ 
\hline

\makecell[tc]{Unblinded Fit} &  
\begin{tabular}[t]{@{} l l @{}}
    $\phi_0$ & $= 1.77^{+0.19}_{-0.18}$ \\
    $\gamma_1$ & $= 1.31^{+0.51}_{-1.30}$ \\
    $\gamma_2$ & $= 2.74^{+0.067}_{-0.075}$ \\
    $\log_{10}(\frac{\mathrm{E}_\mathrm{break}}{\mathrm{GeV}})$ & $= 4.39^{+0.1}_{-0.1}$ \\
\end{tabular} \\ 
\cline{1-2} 
\makecell[tc]{Inelasticity} & 
\begin{tabular}[t]{@{} l l @{}}
    $\phi_0$ & $= 1.80^{+0.18}_{-0.13}$ \\
    $\gamma_1$ & $= 1.11^{+0.59}_{-1.11}$ \\
    $\gamma_2$ & $= 2.72^{+0.055}_{-0.071}$ \\
    $\log_{10}(\frac{\mathrm{E}_\mathrm{break}}{\mathrm{GeV}})$ & $= 4.38^{+0.11}_{-0.10}$ \\
\end{tabular} \\\cline{1-2} 
\makecell[tc]{galactic plane} & 
\begin{tabular}[t]{@{} l l @{}}
    $\phi_0$ & $= 1.68^{+0.15}_{-0.15}$ \\
    $\gamma_1$ & $= 1.03^{+0.54}_{-1.03}$ \\
    $\gamma_2$ & $= 2.72^{+0.061}_{-0.087}$ \\
    $\log_{10}(\frac{\mathrm{E}_\mathrm{break}}{\mathrm{GeV}})$ & $= 4.39^{+0.091}_{-0.12}$ \\
\end{tabular}  \\\cline{1-2} 
\makecell[tc]{Cross section} & 
\begin{tabular}[t]{@{} l l @{}}
    $\phi_0$ & $= 1.62^{+0.16}_{-0.22}$ \\
    $\gamma_1$ & $= 1.29^{+0.43}_{-1.29}$ \\
    $\gamma_2$ & $= 2.73^{+0.065}_{-0.080}$ \\
    $\log_{10}(\frac{\mathrm{E}_\mathrm{break}}{\mathrm{GeV}})$ & $= 4.39^{+0.085}_{-0.099}$ \\
\end{tabular} \\
\cline{1-2}
\makecell[tc]{Energy scale} & 
\begin{tabular}[t]{@{} l l @{}}
    $\phi_0$ & $= 1.79^{+0.18}_{-0.15}$ \\
    $\gamma_1$ & $= 1.26^{+0.49}_{-1.26}$ \\
    $\gamma_2$ & $= 2.73^{+0.067}_{-0.074}$ \\
    $\log_{10}(\frac{\mathrm{E}_\mathrm{break}}{\mathrm{GeV}})$ & $= 4.39^{+0.092}_{-0.098}$ \\
\end{tabular} \\
\cline{1-2}
\makecell[tc]{Atmospheric \\neutrino flux} & 
\begin{tabular}[t]{@{} l l @{}}
    $\phi_0$ & $= 1.74^{+0.19}_{-0.19}$ \\
    $\gamma_1$ & $= 1.26^{+0.58}_{-1.26}$ \\
    $\gamma_2$ & $= 2.73^{+0.051}_{-0.076}$ \\
    $\log_{10}(\frac{\mathrm{E}_\mathrm{break}}{\mathrm{GeV}})$ & $= 4.39^{+0.12}_{-0.10}$ \\
\end{tabular} 
\\
\cline{1-2}
\makecell[tc]{\textsc{MuonGun} KDE} & 
\begin{tabular}[t]{@{} l l @{}}
    $\phi_0$ & $= 1.81^{+0.18}_{-0.17}$ \\
    $\gamma_1$ & $= 1.40^{+0.47}_{-1.40}$ \\
    $\gamma_2$ & $= 2.74^{+0.058}_{-0.076}$ \\
    $\log_{10}(\frac{\mathrm{E}_\mathrm{break}}{\mathrm{GeV}})$ & $= 4.38^{+0.10}_{-0.11}$ \\
\end{tabular} 
\\
\cline{1-2}
\makecell[tc]{Simultaneous \\modifications} & 
\begin{tabular}[t]{@{} l l @{}}
    $\phi_0$ & $= 1.54^{+0.22}_{-0.20}$ \\
    $\gamma_1$ & $= 1.04^{+0.69}_{1.04}$ \\
    $\gamma_2$ & $= 2.68^{+0.068}_{-0.066}$ \\
    $\log_{10}(\frac{\mathrm{E}_\mathrm{break}}{\mathrm{GeV}})$ & $= 4.37^{+0.11}_{-0.11}$ \\
\end{tabular}
\end{tabular}
\end{table}
\subsection{Statistics of \textsc{MuonGun} simulations} The baseline fit of the CF analysis uses a template for the CR muon component that is generated using simulations with the \textsc{MuonGun} package. However, the number of muons that pass the selection criteria is low, and the template is therefore strongly dominated by statistical fluctuations. In contrast, for the tracks sample used in CF analysis and the MESE dataset, a kernel density estimator (KDE) is used to create a smoothed template approximating the true muon distribution, already for the baseline fits.
The CF analysis was therefore rerun with the KDE smoothed template in place of the original \textsc{MuonGun} histogram, resulting in a TS~=~$\Delta\mathcal{L}$ of 3.2, 
and minimal changes in the nuisance parameters. As a further test of the robustness of the fit outcomes with respect to the predicted CR muon contamination of the cascade sample the fit was repeated excluding the vertically down-going bins, where  most muons are expected to come from. A refit of the BPL with this truncated sample, yields physics parameters that remain within the \SI{68}{\percent} uncertainty intervals of the baseline fit.

\subsection{Simultaneous modifications}
Further checks were performed in the CF analysis where several modifications were allowed in the fit simultaneously. These included the mean neutrino inelasticity, galactic plane flux, neutrino cross section and the track energy scale modifications discussed above. A fit that included all these modifications resulted in an improvement in the best fit likelihood by 3.92 units. The astrophysical flux normalization decreased by less than $2\,\sigma$ while the other three flux parameters varied by less than $1\,\sigma$ (see Tab.~\ref{tab:CF_checks}). This demonstrates that the measurement of the astrophysical neutrino flux
with the CF is robust.

\subsection{Prompt atmospheric neutrinos}
\label{sec:prompt}
Profile likelihood scans of the physics parameters were used to ensure that the best fit values were not in unphysical regions of the parameter space. This is particularly important if a specific parameter's best fit is close to, or at, a boundary in the baseline fit.
As the best-fit value for the normalization of the prompt atmospheric neutrino flux is zero in the MESE analysis, a check was performed to ensure that the prompt normalization hitting the boundary was not biasing the remaining physics parameters. The fit was therefore re-run by removing the physical bound on the prompt flux normalization and allowing negative values in the prompt flux normalization.
The best fit prompt normalization for the unbounded fit was at $-1.1\times\SI{e-18}{GeV^{-1} cm^{-2} s^{-1} sr^{-1}}$, which is less than $1\,\sigma$ away from zero, the result of the baseline fit.
We also note that  the result of a negative prompt flux did not induce any significant bias in any of the other physics parameters.\\
\begin{figure}
    \begin{center}
    \includegraphics[width=1\linewidth]{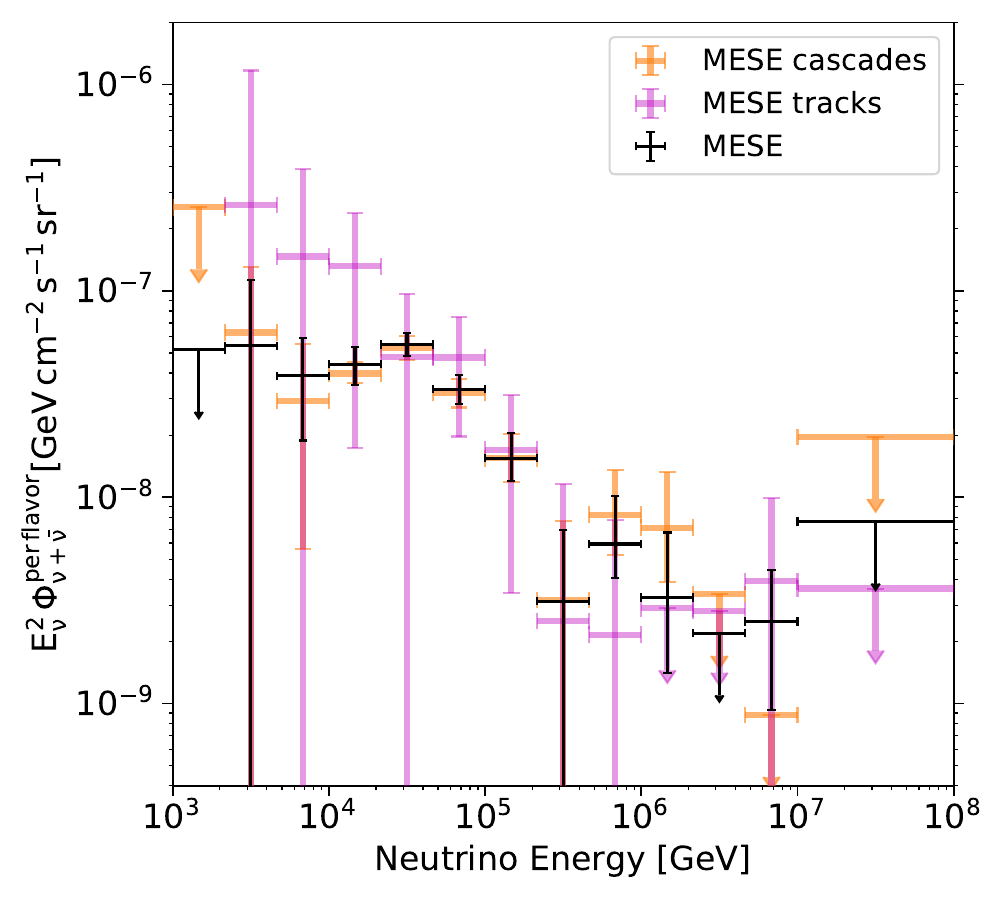}
    \end{center}
    \caption{MESE segmented fits with only MESE cascades and only MESE tracks compared to the
segmented fit including both detection channels.}
    \label{fig:mese_segmented_casc_tra}
\end{figure}
\subsection{Effect of nuisance parameters} We estimate the impact of each nuisance parameter on the fit by fixing the given parameter, calculating its 1D profile likelihood, and comparing it to the nominal 1D profile likelihood. This is repeated for each parameter of interest for the BPL fit. We observe minimal pull to these profile likelihoods from this procedure for both CF and MESE analyses. The largest pull is seen on the astrophysical normalization by the prompt atmospheric flux normalization for the CF analysis and on $\gamma_2$ by the ice anisotropy parameter for the MESE analysis.\\
\subsection{Cascades-only and tracks-only fits} 
\begin{figure*}[t!]
    \begin{minipage}[b]{0.49\linewidth}
     \centering
        \includegraphics[width=1\linewidth]{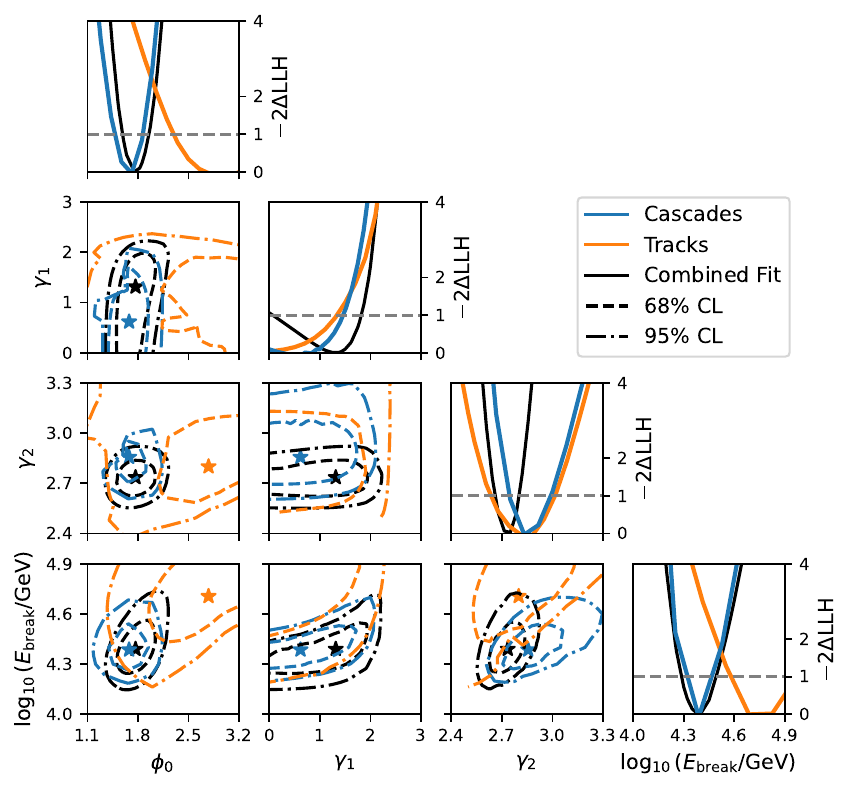}  
    \end{minipage}
    \begin{minipage}[b]{0.49\linewidth}
     \centering
        \includegraphics[width=1\linewidth]{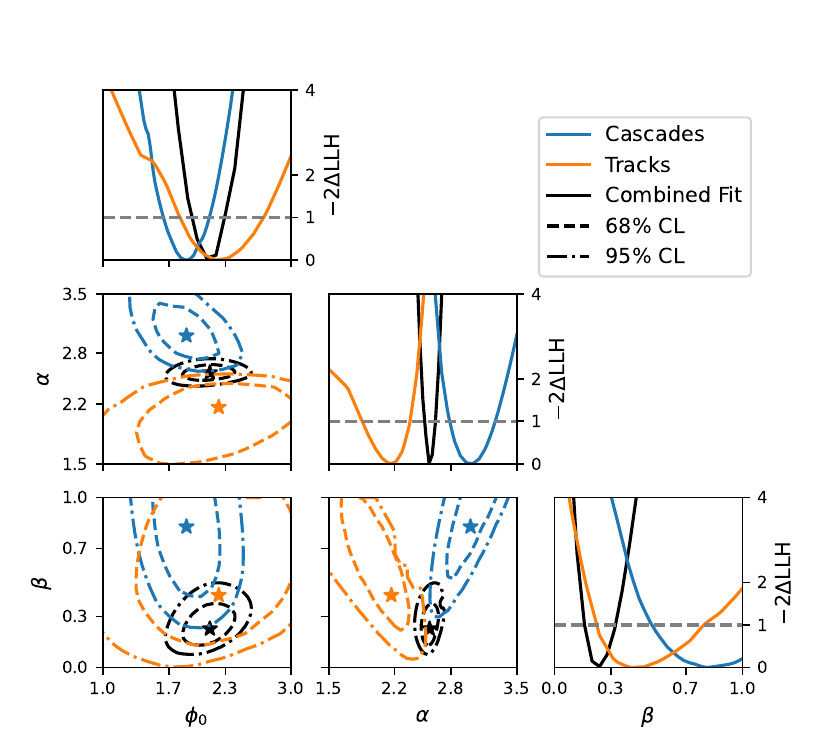}  
    \end{minipage}
\caption{\textbf{BPL \& LP cascade-only and track-only fits spectral parameter comparison for CF:} Two-dimensional profile likelihood scans of all physics parameters in the BPL and LP model fits, with the blue and orange contours representing cascades-only and tracks-only fits respectively. The star markers indicate the best fit parameter values for each fit.  The contours represent the \SI{68}{\percent} and \SI{95}{\percent}  confidence regions for the parameters based on Wilks’ theorem. 
}
\label{fig:BPLandLPContours_CF_CvT}
\end{figure*}

As MESE and CF both classify events into cascades and tracks subsamples, which are then fit simultaneously, it is interesting to study the subsamples individually as well. For the MESE sample, we performed individual fits using the segmented neutrino flux model on the cascades and tracks samples, and compared them to the fit with the whole dataset. As seen in Fig.~\ref{fig:mese_segmented_casc_tra}, the primary MESE fit is compatible with both the MESE cascades and tracks results, although the wide error bars on the tracks-only fit indicate the comparatively lower sensitivity from this channel. In particular, the sensitivity to the change in the spectral shape at \SI{30}{TeV} is driven by the MESE cascades. It is also noteworthy that the fit point at $\sim \SI{10}{PeV}$ is driven by the sensitivity of the tracks sample, consistent with the observation of the highest energy event being a starting track. 
The MESE tracks result is also compatible with the ESTES diffuse measurement~\cite{ESTES}, which solely used starting tracks for measuring the neutrino flux above \SI{1}{TeV}.
Both the tracks-only result from MESE and the results from ESTES indicate that the tracks  morphology alone does not exhibit a break in the astrophysical spectrum. A measurement that also includes the cascades channel is required to observe this structure in the spectrum. Further investigations into this difference are underway (see App.~\ref{sec:estes_crosschecks}).
For CF, similar cascades-only and tracks-only fits were performed for the BPL and LP flux models, and we compare the 2D-profile likelihoods for the model parameters in Fig. \ref{fig:BPLandLPContours_CF_CvT}. The figures demonstrate the strength of combining information from both tracks and cascades in observing the structure in the spectrum.
\begin{figure}[tbh!]
    \begin{center}
    \includegraphics[width=0.99\linewidth]{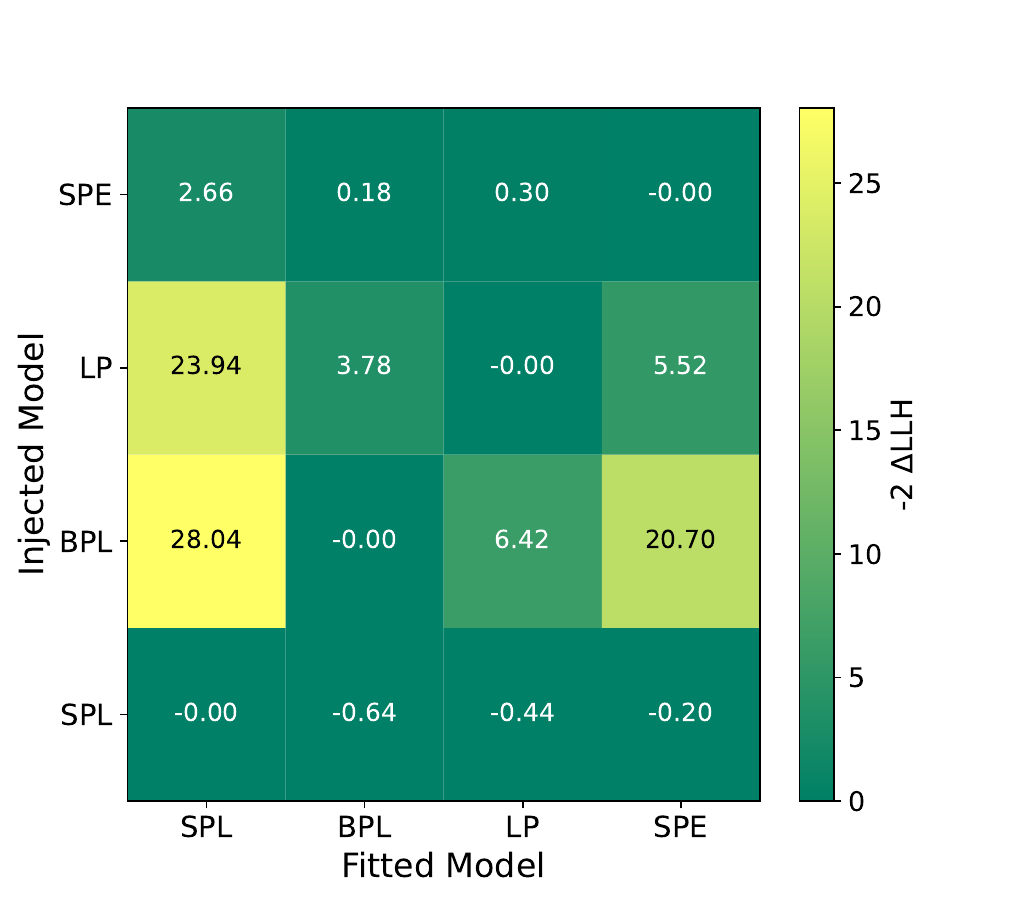}
    \end{center}
    \caption{\textbf{Cross-fit test statistic for the MESE analysis:} The best fit for a given spectral model is injected and subsequently fitted with another spectral model. The TS for these fits are shown in the color scale where the LLH of the fitted model is compared to the LLH of the fit from the injected model. The TS values indicate the power of the analysis to distinguish between different underlying models.}
    \label{fig:crossfit}
\end{figure}

\subsection{Test statistic} Cross-model fitting tests were performed on the MESE samples, evaluating how well different spectral models fit the Asimov data set  when the best fit was derived under the assumption of another model. We used the best-fit values obtained from data for each model for these cross-fit tests.
The test statistics obtained are illustrated in Fig. \ref{fig:crossfit}, where we see that the BPL and LP models do not exhibit a strong $\Delta\mathrm{ln}\mathcal{L}$ difference when injecting each other as the alternate best fit, but allow for a significant rejection of the SPL hypothesis, consistent with what we observe in data. We also note that the TS value for the rejection of the SPL model is on similar scales to what we observe in data.

\subsection{Check on $\gamma_1$ envelope} A comparison of the best fit BPL parameters for the two analyses was provided in Fig. \ref{fig:BPLContours}, where the two-dimensional profile likelihood scans show the complementarity of MESE and CF. The closing of the MESE $\gamma_1$-$\gamma_2$ contour along the $\gamma_1$ axis illustrates that MESE has a better sensitivity to the low-energy astrophysical neutrino flux, while the narrower contour of CF along the $\gamma_2$ axis indicates superior high-energy sensitivity. A validation check was performed on the error envelope of $\gamma_1$ using the MESE dataset. The test aimed at identifying how often a fit of $\gamma_1$ similar to that observed with data occurred.
A sample of pseudo data were generated and a sub-sample of events whose best-fit parameters are close to the data fit were selected, so that their contours are broadly compatible with that of the data.
For these pseudo data, we further fixed the $\gamma_1$ parameter to the $\pm \SI{95}{\percent} $ values in the data contour and evaluate the likelihood value while allowing the remaining parameters to fit freely. The $\Delta\mathcal{L}$ distributions, where $\Delta\mathcal{L}$ = $\mathrm{LLH_{fixed\,\gamma_1}}-\mathrm{LLH_{free\,fit}}$, were determined for these realizations of pseudo data and were compared to the $\Delta\mathcal{L}$ from observed data.
A substantial fraction of the pseudo data had $\Delta\mathcal{L}$ greater than that of the observed data, indicating that if the observed data follows those realizations, the likelihood contour would close for them as well. This gives us confidence that the closed contour is not the effect of unmodeled systematics that artificially force the data fit to exclude zero.

\section{Discussion and Conclusion}
\label{sec:discussion}
We have presented new results from the measurement of the diffuse astrophysical neutrino spectrum using IceCube data, combining information from the two main observation channels, cascades and tracks. Two complementary analyses  have been developed  to conduct this measurement, one focusing on the combination of event samples that have been used in previous IceCube analyses (CF analysis), the other focused on a novel sample of starting events (MESE analysis). \textbf{Both analyses reject the spectral model of a single power law between \SI{5}{TeV} and \SI{10}{PeV} by greater than $\mathbf{4\,\sigma}$}. This differs from the most recent IceCube spectral measurement, which favored a single power law flux~\cite{ESTES}.
While some prior IceCube measurements have indicated possible tensions of the single power law model with the IceCube observations~(e.g., \cite{MESE_2yr,SBUCascades}), the work presented here marks the first time that this model can be rejected at high statistical confidence. 
Both analyses consistently favor a model with a soft spectrum with a power-law index of $\sim$~2.8 at energies above $\sim \SI{30}{TeV} $, and a hard spectrum with a power-law index $\lesssim 2$ at lower energies. 
While the CF analysis marks a first step towards a joint fit with existing datasets, incorporating the MESE dataset and other IceCube samples in a future joint fit could yield stronger constraints on the cosmic neutrino spectrum than those presented here.
The best fit spectral model is a broken power law. A log parabola spectral model and a model where the neutrino emission has a local peak at $\sim$~20~TeV on top of a single power law spectrum can also be used to reject the SPL hypothesis with at least 4\,$\sigma$ significance. Likelihood ratio tests comparing the alternative models to the BPL find a preference for the latter, but only at a 2\,$\sigma$ level, so we are unable to firmly distinguish between the various spectral shapes with analyses using currently available data. We estimate that with 10 years of additional data and with no improvements in performance, these analyses will be able to distinguish between the BPL and LP models at the best-fit values obtained in this paper, with a median expected significance of $\sim 5\,\sigma$. An analysis that performs a joint fit with all existing IceCube event samples, built upon the concept of the CF analysis, can be the next step towards improving the statistical strength of this test with currently available data. 

Various cross-checks have been performed to validate the robustness of the result. In particular, we observe that incorporating a non-isotropic contribution from the galactic plane as an additional component of the diffuse neutrino flux in the fit does not affect the fit results significantly, disfavoring the hypothesis that the observed spectral features  around $\sim \SI{30}{TeV} $  are caused by neutrinos from the galactic plane. Fitting the spectrum with an atmospheric neutrino component that is derived from a data-driven CR composition and hadronic interaction model (DaemonFlux) also does not significantly change the fit results. 

The harder spectral index for energies below $\sim$~30~TeV implies a lower flux of astrophysical neutrinos below 10 TeV than the one expected for the SPL model. This has significant implications in searches for neutrino emitters from a multi-messenger perspective. The extrapolation of the SPL spectrum from prior measurements of IceCube toward lower energies leads to tensions with the observed intensity of the extragalactic gamma-ray background at GeV energies. The tension can be resolved if a substantial fraction of the astrophysical neutrinos is produced in environments that are opaque to GeV gamma rays~\cite{murase_hidden_2016,capanema_new_2020}. The change in the spectral index measured here alleviates some of these tensions, putting less constraints on the environments that accelerate cosmic rays and produce neutrinos. The observed feature in the spectral shape will also be a powerful input for future modeling of the origin of astrophysical neutrinos, potentially constraining source population properties. The observed diffuse flux can provide insight on possible source classes and inform the search for individual neutrino sources, e.g. models on the population of AGN predict a natural break in the neutrino spectrum~\citep{SeyfertPopulations,Murase:2019vdl}. The relevant energy range for each possible source class can also be informed by the diffuse flux, e.g. starburst galaxies are predicted as possible sources at hundreds of TeV, while a lower energy neutrino flux can be composed of sources with a galactic origin~\citep{MultiComponentAstroFlux, Palladino:2018bqf,Peretti:2019vsj,2024MNRAS.529.4137C}. They also serve as a test of 
maximum CR acceleration energies in such sources~\cite{SeyfertPopulations}, and test neutrino production target properties~\cite{fang_tev_2022}.

\begin{acknowledgements}
The authors gratefully acknowledge the support from the following agencies and institutions:
USA {\textendash} U.S. National Science Foundation-Office of Polar Programs,
U.S. National Science Foundation-Physics Division,
U.S. National Science Foundation-EPSCoR,
U.S. National Science Foundation-Office of Advanced Cyberinfrastructure,
Wisconsin Alumni Research Foundation,
Center for High Throughput Computing (CHTC) at the University of Wisconsin{\textendash}Madison,
Open Science Grid (OSG),
Partnership to Advance Throughput Computing (PATh),
Advanced Cyberinfrastructure Coordination Ecosystem: Services {\&} Support (ACCESS),
Frontera and Ranch computing project at the Texas Advanced Computing Center,
U.S. Department of Energy-National Energy Research Scientific Computing Center,
Particle astrophysics research computing center at the University of Maryland,
Institute for Cyber-Enabled Research at Michigan State University,
Astroparticle physics computational facility at Marquette University,
NVIDIA Corporation,
and Google Cloud Platform;
Belgium {\textendash} Funds for Scientific Research (FRS-FNRS and FWO),
FWO Odysseus and Big Science programmes,
and Belgian Federal Science Policy Office (Belspo);
Germany {\textendash} Bundesministerium f{\"u}r Bildung und Forschung (BMBF),
Deutsche Forschungsgemeinschaft (DFG),
Helmholtz Alliance for Astroparticle Physics (HAP),
Initiative and Networking Fund of the Helmholtz Association,
Deutsches Elektronen Synchrotron (DESY),
and High Performance Computing cluster of the RWTH Aachen;
Sweden {\textendash} Swedish Research Council,
Swedish Polar Research Secretariat,
Swedish National Infrastructure for Computing (SNIC),
and Knut and Alice Wallenberg Foundation;
European Union {\textendash} EGI Advanced Computing for research;
Australia {\textendash} Australian Research Council;
Canada {\textendash} Natural Sciences and Engineering Research Council of Canada,
Calcul Qu{\'e}bec, Compute Ontario, Canada Foundation for Innovation, WestGrid, and Digital Research Alliance of Canada;
Denmark {\textendash} Villum Fonden, Carlsberg Foundation, and European Commission;
New Zealand {\textendash} Marsden Fund;
Japan {\textendash} Japan Society for Promotion of Science (JSPS)
and Institute for Global Prominent Research (IGPR) of Chiba University;
Korea {\textendash} National Research Foundation of Korea (NRF);
Switzerland {\textendash} Swiss National Science Foundation (SNSF).
\end{acknowledgements}

\newpage
\appendix
\section{Nuisance parameters in both analyses}
A summary of the nuisance parameters used in both analyses are shown in Tab.~\ref{tab:systs_MESE} and Tab.~\ref{tab:systs_CombinedFit}.
\begin{table*}[t!]
\setlength{\tabcolsep}{6pt} 
\renewcommand{\arraystretch}{1.5}
\caption{Summary of all nuisance parameters used in the measurement of the astrophysical diffuse flux using MESE. All parameters are assumed to be independent. The flux is measured in units of \SI{e-18}{GeV^{-1} cm^{-2} s^{-1} sr^{-1}}. All flux normalizations are at \SI{100}{TeV}.}
\label{tab:systs_MESE}
\begin{tabular}{c c c c c c} 
 \hline
 \hline
 Parameter & Range & Prior  & Nominal Value& BPL Best Fit & Description \\ [0.5ex] 
 \hline
 \multicolumn{6}{>{\bfseries}p{\textwidth}}{Flux Parameters}\\
 $\phi_\mathrm{muon}$ & $[0,\infty)$ & Gaussian & $ 1.43 \pm 0.75$ & $ 1.08^{+0.13}_{-0.15}$ &  Atmospheric muon flux normalization \\ 
 $\phi_\mathrm{conv}$ & $[0,\infty)$ & Gaussian & $ 1.00 \pm 0.25$& $ 1.20^{+0.17}_{-0.13}$ & Atmospheric conventional neutrino flux normalization\\ 
 $\phi_\mathrm{prompt}$ & $[0,4)$ & - & 1.00 & $0.00^{+0.50}$ & Atmospheric prompt neutrino flux normalization\\
 $\Delta\gamma$ & $[-1,1]$ & Gaussian & $ 0.00 \pm 0.055$& $ 0.040^{+0.045}_{-0.034}$ & Variations in the primary cosmic-ray spectral index \\
 $\eta_{\mathrm{CR}}$ & [-1,+2] & Gaussian & $ 0.00 \pm 1.00$ &$ 0.50^{+0.59}_{-0.55}$ & H4a-GST4 cosmic-ray flux model interpolation \\
  $\mathrm{H}$ & [-0.8,0.8] & Gaussian & $ 0.00 \pm 0.15$& $ 0.00^{+0.15}_{-0.14}$ & pion uncertainty parameterization from~\cite{barr_uncertainties_2006} \\
$\mathrm{W}$ & [-0.6,0.6] & Gaussian & $ 0.00 \pm 0.40$ &$ 0.24^{+0.36}_{-0.37}$ & kaon uncertainty parameterization from~\cite{barr_uncertainties_2006}\\
$\mathrm{Y}$ & [-0.6,0.6] & Gaussian & $ 0.00 \pm 0.30$ & $ 0.19^{+0.26}_{-0.21}$ & kaon uncertainty parameterization from~\cite{barr_uncertainties_2006}\\
$\mathrm{Z}$ & [-0.244,0.6] &  Gaussian & $ 0.00 \pm 0.12$ & $ 0.009^{+0.097}_{-0.12}$ & kaon uncertainty parameterization from~\cite{barr_uncertainties_2006} \\
  $\eta_{\mathrm{inelasticity}}$ & [-2.0, 2.0] & Gaussian & $ 0.00 \pm 1.00$ &$ -1.20^{+0.58}_{-0.38}$ &  Variations from the inelasticity of neutrino interactions \\
 $\eta_{\mathrm{Self-Veto}}$ & [-5, 15] & Gaussian & $ 0.00 \pm 3.00$ &$ 0.79^{+2.82}_{-2.21}$ & Self-veto interpolation term  \\
 \hline
 \multicolumn{6}{>{\bfseries}p{\textwidth}}{Detector Systematic Parameters}\\
 $\epsilon_\mathrm{Scattering}$ & [0.9,1.1] & Gaussian & $ 1.00 \pm 0.050$ &$ 1.00^{+0.015}_{-0.018}$ & Bulk ice model scattering coefficient scaling \\ 
  $\epsilon_\mathrm{Anisotropy}$ & [0.0,2.0] & Uniform & $ 1.00 $ & $ 0.52^{+0.21}_{-0.26}$& Bulk ice model anisotropy variation \\ 
 $\epsilon_\mathrm{Absorption}$ & [0.9,1.1] & Gaussian & $ 1.00 \pm 0.050$ &$ 1.020^{+0.012}_{-0.002}$ & Bulk ice model absorption coefficient scaling  \\ 
 $\epsilon_\mathrm{HoleIce(p_{0})}$ & [-0.84,0.3] & Uniform  & $ -0.27 $& $ -0.34^{+0.11}_{-0.13}$ & Hole ice angular acceptance parameter p0\\ 
 $\epsilon_\mathrm{HoleIce(p_{1})}$ & [-0.134,0.05] & Uniform  & $ -0.042$ & $ -0.042^{+0.013}_{-0.019}$ &  Hole ice angular acceptance parameter p1\\ 
 $\epsilon_\mathrm{DOM}$& [0.9,1.1] & Uniform & $ 1.00 $ &$ 0.98^{+0.022}_{-0.021} $ & DOM efficiency\\ 
 \hline
 \hline
\end{tabular}

\end{table*}
\begin{table*}[tbh!]
\setlength{\tabcolsep}{6pt} 
\renewcommand{\arraystretch}{1.5} 
\caption{Summary of all nuisance parameters used in the measurement of the astrophysical diffuse flux using CF. All parameters are assumed to be independent. The flux is measured in units of \SI{e-18}{GeV^{-1} cm^{-2} s^{-1} sr^{-1}}. All flux normalizations are at \SI{100}{TeV}.}
\label{tab:systs_CombinedFit}
\begin{tabular}{c c c c c p{7cm}} 
 \hline
 \hline
 Parameter & Range & Prior  & Nominal Value & BPL Best Fit & Description \\ [0.5ex] 
 \hline
 \multicolumn{6}{>{\bfseries}p{\textwidth}}{Atmospheric Flux Parameters}\\
 $\phi_\mathrm{MuonGun}$ & $[0,\infty)$ & -& 1.00& $1.12^{+0.021}_{-0.033}$ & Atmospheric muon flux normalization (cascades) \\ 
 $\phi_\mathrm{muon\,template}$ & $[0,\infty)$ & Gaussian &  $ 1.00 \pm 0.50$&  $ 1.50^{+0.45}_{-0.47}$  & Atmospheric muon template flux normalization (tracks) \\ 
 $\phi_\mathrm{conv}$ & $[0,\infty)$ & - & 1.00& $1.25^{+0.11}_{-0.18}$ & Atmospheric conventional neutrino flux normalization\\ 
 $\phi_\mathrm{prompt}$ & $[0,\infty)$ & - & 1.00& $1.10^{+1.2}_{-1.1}$ & Atmospheric prompt neutrino flux normalization\\
 $\Delta\gamma$ & $[-1,1]$ & Uniform & 0.00& $0.049^{+0.030}_{-0.031}$ & Variations in the 
spectral index of the primary cosmic-ray spectrum \\
 $\eta_{\mathrm{CR}}$ & [-1,+2] & Gaussian & $ 0.00 \pm 1.00$ & $ 1.22^{+0.23}_{-0.32}$& H4a-GST4 cosmic-ray flux model interpolation \\
  $\mathrm{H}$ & [-0.8,0.8] & Gaussian & $ 0.00 \pm 0.15$ & $ -0.056^{+0.16}_{-0.12}$ & pion uncertainty parameterization from~\cite{barr_uncertainties_2006} \\
$\mathrm{W}$ & [-0.6,0.6] & Gaussian & $ 0.00 \pm 0.40$ &$ -0.21^{+0.42}_{-0.30}$ & kaon uncertainty parameterization from~\cite{barr_uncertainties_2006}\\
$\mathrm{Y}$ & [-0.6,0.6] & Gaussian & $ 0.00 \pm 0.30$ &$ 0.029^{+0.23}_{-0.17}$ &kaon uncertainty parameterization from~\cite{barr_uncertainties_2006}\\
$\mathrm{Z}$ & [-0.244,0.6] &  Gaussian & $ 0.00 \pm 0.12$ &$ 0.043^{+0.12}_{-0.78}$ &kaon uncertainty parameterization from~\cite{barr_uncertainties_2006} \\
  
 $\eta_{\mathrm{Self-Veto}}$ & [5, 2000] & Uniform & 0.00 & $467.74^{+138.04}_{-167.82}$ &Self-veto effective threshold (GeV) (cascades)  \\
 \hline
 \multicolumn{6}{>{\bfseries}p{\textwidth}}{Detector Systematic Parameters}\\
 $\epsilon_\mathrm{Scattering}$ & [0.9,1.1] & Uniform & $ 1.00 $& $ 1.016^{+0.006}_{-0.011}$ & Bulk ice model scattering coefficient scaling \\ 
  $\epsilon_\mathrm{Absorption}$ & [0.9,1.1] & Uniform & $ 1.00 $& $ 0.99^{+0.0049}_{-0.0090}$ & Bulk ice model absorption coefficient scaling  \\ 
 $\epsilon_\mathrm{HoleIce(p_{0})}$ & [-0.84,0.3] & Uniform  & $ -0.27$& $ -0.26^{+0.042}_{-0.037}$ & Hole ice angular acceptance parameter p0\\ 
 $\epsilon_\mathrm{HoleIce(p_{1})}$ & [-0.134,0.05] & Uniform  & $ -0.042 $& $ -0.065^{+0.0081}_{-0.0042}$ &  Hole ice angular acceptance parameter p1\\ 
 $\epsilon_{\mathrm{DOM}}$& [0.9,1.1] & Uniform & $ 1.00  $ & $ 1.020^{+0.0052}_{-0.0047} $ & DOM efficiency\\ 
 \hline
 \hline
\end{tabular}
\end{table*}
%
%

%
%

\section{Segmented Flux Normalizations}\label{sec:SegmentedFluxParameters}
The best fit parameters for the segmented flux model are provided in Tab.~\ref{tab:segmented_fit_results} 
\begin{table*}[h!]
\caption{Results for the segmented power law tested in the MESE and CF analyses. The uncertainties are derived from 1D profile likelihood scans, assuming Wilks' theorem applies.
The flux is measured in units of \SI{e-18}{GeV^{-1} cm^{-2} s^{-1} sr^{-1}}  and all normalization components are fit simultaneously, assuming a power-law spectrum with an index of 2 in each neutrino energy band.}
    \label{tab:segmented_fit_results}
    \renewcommand{\arraystretch}{2}
\centering
\begin{adjustwidth}{}{}

\begin{tabularx}{2\columnwidth}{
        >{\centering\arraybackslash}X|
        >{\centering\arraybackslash}X|
        >{\centering\arraybackslash}X}
        \hline\hline
        \multicolumn{1}{c|}{\textbf{Energy Bins (TeV)}} & 
        \multicolumn{1}{c|}{\textbf{MESE}} & 
        \multicolumn{1}{c}{\textbf{CF}} \\
        \multicolumn{1}{c|}{\small (range, center)} & 
        \multicolumn{1}{c|}{} & 
        \multicolumn{1}{c}{} \\
        \hline

  (1.0, 2.15), 1.47       & $0^{+5.2}$& $0^{+13}$     \\ \hline
  (2.15, 4.64), 3.16       & $5.4^{+5.8}_{-5.4}$& $0^{+1.9}$     \\ \hline
  (4.64, 10.0), 6.81       & $3.9^{+2.1}_{-2.0}$& $3.1^{+2.1}_{-1.5}$     \\ \hline
  (10.0, 21.5), 14.7       & $4.41^{+0.93}_{-0.90}$& $3.36^{+1.10}_{-0.63}$     \\ \hline
  (21.5, 46.4), 31.6       & $5.51^{+0.72}_{-0.66}$& $4.42^{+0.74}_{-0.48}$    \\ \hline
  (46.4, 100), 68.1       & $3.34^{+0.63}_{-0.52}$& $2.03^{+0.40}_{-0.39}$    \\ \hline
  (100, 215.4), 146.8       & $1.55^{+0.50}_{-0.35}$& $1.81^{+0.38}_{-0.37}$     \\ \hline
  (215.4, 464.2), 316.2       & $0.31^{+0.33}_{-0.31}$& $0.089^{+0.350}_{-0.089}$     \\ \hline
  (464.2, 1000), 681.3       & $0.59^{+0.42}_{-0.23}$& $0.85^{+0.43}_{-0.42}$    \\ \hline
  (1000, 2154.4), 1467.8       & $0.327^{+0.32}_{-0.059}$& $0.41^{+0.36}_{-0.25}$    \\ \hline
  (2154.4, 4641.6), 3162.3       & $0.0^{+0.19}$& $0^{+0.20}$     \\ \hline
  (4641.6, 10000), 6812.9       & $0.25^{+0.24}_{-0.16}$& $0.017^{+0.073}_{-0.017}$     \\ \hline
(10000, 100000), 31622.8       & $0^{+0.76}$ & $0.069^{+0.38}_{-0.069}$    \\ \hline
\end{tabularx}
\end{adjustwidth}
\end{table*}
\section{Results from dataset split cross check studies with MESE}
The fit values from all cross-check studies performed with MESE is shown in Tab.~\ref{tab:splits}.
\begin{turnpage}
\begin{table*}[h!]
\caption{Results for the fits with dataset splits in the MESE analysis. The uncertainties are derived from 1D profile likelihood scans, assuming Wilks' theorem applies. The flux normalization is measured in units of \SI{e-18}{GeV^{-1} cm^{-2} s^{-1} sr^{-1}}. All flux normalizations are at \SI{100}{TeV}.}
\label{tab:splits}
\renewcommand{\arraystretch}{1.5}
\centering
\small
\begin{adjustwidth}{0cm}{}
\begin{tabularx}{\textwidth}{p{0.8cm} c|c|c|c|c|c|c|c|c|c}
 && Full Data & \makecell[c]{Deep \\(below dust band)} & \makecell[c]{Shallow \\(above dust band)} & Upgoing & Downgoing & \makecell[c]{Early \\(2011-June 2017)} &\makecell[c]{Late \\(June 2017-2022)} & Summer & Winter \\ 
 \cline{1-11} 
\makecell[tl]{\textbf{SPL}} &
\begin{tabular}[t]{@{} c @{}}
     $ \phi_0$ \\
     $ \gamma$
\end{tabular} 
& 
\begin{tabular}[t]{@{} l @{}}
     $ 2.13^{+0.18}_{-0.17}$ \\
     $ 2.548^{+0.039}_{-0.041}$
\end{tabular} & 
\begin{tabular}[t]{@{} l @{}}
     $ 2.18^{+0.28}_{-0.26}$ \\
    $ 2.503^{+0.066}_{-0.067}$
\end{tabular} &
\begin{tabular}[t]{@{} l @{}}
     $ 1.78^{+0.22}_{-0.20}$ \\
    $ 2.585^{+0.046}_{-0.053}$
\end{tabular} & 
\begin{tabular}[t]{@{} l @{}}
     $ 2.06^{+0.49}_{-0.84}$ \\
    $2.464^{+0.082}_{-0.17}$
   
\end{tabular} & 
\begin{tabular}[t]{@{} l @{}}
    $ 2.12^{+0.23}_{-0.22}$ \\
    $ 2.652^{+0.067}_{-0.057}$
\end{tabular} & 
\begin{tabular}[t]{@{} l @{}}
    $ 2.22^{+0.24}_{-0.24}$ \\
     $ 2.545^{+0.059}_{-0.056}$
     
\end{tabular} &
\begin{tabular}[t]{@{} l @{}}
     $ 1.89^{+0.23}_{-0.24}$ \\
    $ 2.537^{+0.064}_{-0.061}$
\end{tabular} &
\begin{tabular}[t]{@{} l @{}}
     $ 2.09^{+0.24}_{-0.24}$ \\
    $ 2.523^{+0.059}_{-0.055}$
\end{tabular} &
\begin{tabular}[t]{@{} l @{}}
    $ 2.08^{+0.23}_{-0.22}$ \\
    $ 2.536^{+0.056}_{-0.057}$
\end{tabular} \\ \cline{1-11}
\makecell[tl]{\textbf{SPE}} &
\begin{tabular}[t]{@{} c @{}}
     $ \phi_0$ \\
     $\gamma$\\$\log_{10}(\frac{\mathrm{E}_\mathrm{cutoff}}{\mathrm{GeV}})$
\end{tabular} 
& 
\begin{tabular}[t]{@{} l l @{}}
    & $ 3.9^{+1.2}_{-1.2}$ \\
    & $ 2.16^{+0.11}_{-0.16}$ \\
     & $ 5.52^{+0.39}_{-0.35}$

\end{tabular} & 
\begin{tabular}[t]{@{} l l @{}}
    & $ 9.10^{+1.10}_{-3.60}$ \\
    & $ 1.73^{+0.22}_{-0.18}$ \\
    & $ 5.00^{+0.07}_{-0.33}$
    
\end{tabular} &
\begin{tabular}[t]{@{} l l @{}}
    & $ 5.88^{+0.88}_{-2.20}$ \\
    & $ 1.95^{+0.27}_{-0.13}$ \\
    & $ 5.0^{+0.24}_{-0.39}$

\end{tabular} & 
\begin{tabular}[t]{@{} l l @{}}
    & $ 1.95^{+0.43}_{-0.73}$ \\
    & $ 2.45^{+0.10}_{-0.15}$ \\
    & $ 10^{}_{-4.82}$
    
\end{tabular} & 
\begin{tabular}[t]{@{} l l @{}}
    & $ 2.57^{+0.38}_{-0.34}$ \\
    & $ 2.51^{+0.10}_{-0.11}$ \\
    & $ 6.08^{+0.47}_{-0.27}$
\end{tabular} & 
\begin{tabular}[t]{@{} l l @{}}
    & $2.86^{+0.45}_{-0.45}$ \\
    & $2.393^{+0.095}_{-0.096}$ \\
    & $ 5.99^{+0.41}_{-0.13}$
\end{tabular} &
\begin{tabular}[t]{@{} l l @{}}
    & $ 7.12^{+5.91}_{-1.48}$ \\
    & $ 1.84^{+0.21}_{-0.05}$ \\
    & $ 5^{+0.30}_{-0.09}$
\end{tabular} &
\begin{tabular}[t]{@{} l l @{}}
    & $ 7.6^{+1.2}_{-2.4}$ \\
    & $ 1.764^{+0.23}_{-0.08}$ \\
    & $ 5.00^{+0.30}_{-0.17}$
\end{tabular} &
\begin{tabular}[t]{@{} l l @{}}
    & $3.3^{+1.1}_{-1.4}$ \\
    & $2.28^{+0.09}_{-0.13}$ \\
    & $ 5.58^{+0.26}_{-0.14}$
    
\end{tabular} \\ \cline{1-11}
\makecell[tl]{\textbf{BPL}} &
\begin{tabular}[t]{@{} c @{}}
     $ \phi_0$ \\
     $\gamma_1$\\$\gamma_2$\\$\log_{10}(\frac{\mathrm{E}_\mathrm{break}}{\mathrm{GeV}})$
\end{tabular} 
& 
\begin{tabular}[t]{@{} l l @{}}
    & $ 2.28^{+0.22}_{-0.20}$ \\
    & $ 1.72^{+0.26}_{-0.35}$ \\
    & $ 2.839^{+0.11}_{-0.091}$ \\
    & $ 4.524^{+0.097}_{-0.087}$

\end{tabular} & 
\begin{tabular}[t]{@{} l l @{}}
    & $ 4.12^{+0.93}_{-0.45}$ \\
    & $ 2.07^{+0.15}_{-0.16}$ \\
    & $ 4.26^{+5.8}_{-0.56}$ \\
    & $ 5.07^{+0.13}_{-0.10}$

\end{tabular} &
\begin{tabular}[t]{@{} l l @{}}
    & $ 1.71^{+0.31}_{-0.26}$ \\
    & $ 1.59^{+0.43}_{-0.77}$ \\
    & $ 3.09^{+0.13}_{-0.21}$ \\
    & $ 4.53^{+0.15}_{-0.13}$
  
\end{tabular} & 
\begin{tabular}[t]{@{} l l @{}}
    & $ 2.26^{+0.30}_{-0.67}$ \\
    & $ 1.81^{+0.35}_{-1.81}$ \\
    & $ 2.66^{+0.12}_{-0.19}$ \\
    & $ 4.388^{+0.072}_{-0.19}$ 
      
\end{tabular} &
\begin{tabular}[t]{@{} l l @{}}
    & $ 2.33^{+0.38}_{-0.29}$ \\
    & $ 1.99^{+0.30}_{-0.43}$ \\
    & $ 2.98^{+0.12}_{-0.12}$ \\
    & $ 4.65^{+0.17}_{-0.11}$ 
    
\end{tabular} & 
\begin{tabular}[t]{@{} l l @{}}
    & $ 2.34^{+0.30}_{-0.29}$ \\
    & $ 1.73^{+0.36}_{-0.65}$ \\
    & $ 2.80^{+0.14}_{-0.10}$ \\
    & $ 4.46^{+0.17}_{-0.14}$
   
\end{tabular} & 
\begin{tabular}[t]{@{} l l @{}}
    & $ 1.97^{+0.36}_{-0.32}$ \\
    & $ 1.69^{+0.30}_{-0.47}$ \\
    & $ 3.14^{+0.71}_{-0.26}$ \\
    & $ 4.61^{+0.17}_{-0.12}$ 
     
\end{tabular} &
\begin{tabular}[t]{@{} l l @{}}
    & $ 2.25^{+0.29}_{-0.28}$ \\
    & $ 1.22^{+0.53}_{-0.99}$ \\
    & $ 2.88^{+0.16}_{-0.11}$ \\
    & $ 4.47^{+0.11}_{-0.08}$
\end{tabular} & 
\begin{tabular}[t]{@{} l l @{}}
    & $ 2.21^{+0.29}_{-0.27}$ \\
    & $ 1.98^{+0.26}_{-0.69}$ \\
    & $ 2.79^{+0.17}_{-0.21}$ \\
    & $ 4.54^{+0.17}_{-0.54}$
\end{tabular}\\ \cline{1-11}
\makecell[tl]{\textbf{LP}} &
\begin{tabular}[t]{@{} c @{}}
     $ \phi_0$ \\
     $\alpha_\mathrm{LP}$\\$\beta_\mathrm{LP}$
\end{tabular} 
&
\begin{tabular}[t]{@{} l l @{}}
    & $ 2.58^{+0.26}_{-0.26}$ \\
     & $ 2.668^{+0.12}_{-0.061}$ \\
     & $ 0.359^{+0.11}_{-0.082}$ 
\end{tabular} &
\begin{tabular}[t]{@{} l l @{}}
    & $ 2.90^{+0.44}_{-0.39}$ \\
    & $ 2.91^{+0.17}_{-0.16}$ \\
    & $ 0.82^{+0.24}_{-0.22}$ 
\end{tabular} &
\begin{tabular}[t]{@{} l l @{}}
    & $ 2.02^{+0.27}_{-0.24}$ \\
    & $ 3.0^{+0.35}_{-0.085}$ \\
    & $ 0.69^{+0.22}_{-0.24}$
\end{tabular}&
\begin{tabular}[t]{@{} l l @{}}
    & $ 2.44^{+0.38}_{-0.38}$ \\
    & $ 2.80^{+0.21}_{-0.47}$ \\
    & $ 0.50^{+0.27}_{-0.49}$
\end{tabular}& 
\begin{tabular}[t]{@{} l l @{}}
    & $ 2.48^{+0.31}_{-0.28}$ \\
    & $ 2.705^{+0.084}_{-0.061}$ \\
    & $ 0.31^{+0.16}_{-0.12}$
\end{tabular} &
\begin{tabular}[t]{@{} l l @{}}
    & $ 2.68^{+0.33}_{-0.33}$ \\
    & $ 2.658^{+0.12}_{-0.092}$ \\
    & $ 0.307^{+0.10}_{-0.081}$ 
\end{tabular}&
\begin{tabular}[t]{@{} l l @{}}
    & $ 2.34^{+0.34}_{-0.30}$ \\
    & $ 3.00^{+0.34}_{-0.17}$ \\
    & $ 0.78^{+0.18}_{-0.20}$
\end{tabular} &
\begin{tabular}[t]{@{} l l @{}}
    & $ 2.66^{+0.36}_{-0.32}$ \\
    & $ 2.632^{+0.26}_{-0.074}$ \\
    & $ 0.38^{+0.43}_{-0.10}$
\end{tabular} & 
\begin{tabular}[t]{@{} l l @{}}
    & $ 2.43^{+0.31}_{-0.29}$ \\
    & $ 2.70^{+0.15}_{-0.12}$ \\
    & $ 0.35^{+0.17}_{-0.17 }$ 
\end{tabular} \\  \cline{1-11}

\makecell[tl]{\textbf{SPB}} &
\begin{tabular}[t]{@{} c @{}}
     $ \phi_0$ \\
     $\gamma$\\$\log_{10}(\frac{\mathrm{E}_\mathrm{bump}}{\mathrm{GeV}})$\\$\log_{10}(\frac{\sigma_\mathrm{bump}}{\mathrm{GeV}})$\\$\phi^{\mathrm{bump}} / C$
\end{tabular} 
&
\begin{tabular}[t]{@{} l l @{}}
    & $ 1.42^{+0.21}_{-0.20} $ \\
     & $ 2.512^{+0.059}_{-0.067}$ \\
     & $4.30^{+0.13}_{}$ \\
     & $ 4.421^{+0.097}_{-0.15}$ \\
    & $ 24.4^{+13}_{-8.4}$
\end{tabular} &
\begin{tabular}[t]{@{} l l @{}}
    & $ 1.76^{+0.37}_{-0.38}$ \\
    & $ 2.480^{+0.10}_{-0.0067}$ \\
    & $ 4.80^{+0.037}_{-0.037}$ \\
    & $ 4.745^{+0.023}_{-0.057}$ \\
    & $ 28.0^{+18}_{-15}$
\end{tabular}&
\begin{tabular}[t]{@{} l l @{}}
    & $ 0.88^{+0.34}_{-0.18}$ \\
    & $ 2.520^{+0.12}_{-0.069}$ \\
    & $ 4.452^{+0.11}_{-0.077}$ \\
    & $ 4.393^{+0.092}_{-0.27}$ \\
    & $ 34^{+12}_{-14}$

\end{tabular}&
\begin{tabular}[t]{@{} l l @{}}
    & $ 1.68^{+0.47}_{-0.87}$ \\
    & $ 2.440^{+0.078}_{-0.24}$ \\
    & $ 4.30^{+0.20}_{}$ \\
    & $ 4.13^{+0.24}_{-0.78}$ \\
    & $ 42^{+29}_{-22}$
\end{tabular}& 
\begin{tabular}[t]{@{} l l @{}}
    & $ 1.65^{+0.24}_{-0.26}$ \\
    & $ 2.640^{+0.078}_{-0.085}$ \\
    & $ 4.0^{+0.049}_{1.3}$ \\
    & $ 4.464^{+0.071}_{-0.23}$ \\
    & $ 16.0^{+13}_{-5.03}$
\end{tabular} &
\begin{tabular}[t]{@{} l l @{}}
    & $ 1.65^{+0.31}_{-0.23}$ \\
    & $ 2.52^{+0.07}_{-0.09}$ \\
    & $ 4.30^{+0.1}_{}$ \\
    & $ 4.39^{+0.13}_{-0.32}$ \\
    & $ 40.0^{+9.9}_{-20}$
\end{tabular}&
\begin{tabular}[t]{@{} l l @{}}
    & $ 1.04^{+0.20}_{-0.23}$ \\
    & $ 2.520^{+0.081}_{-0.110}$ \\
    & $ 4.48^{+0.10}_{-0.12}$ \\
    & $ 4.464^{+0.043}_{-0.160}$ \\
    & $ 30^{+17}_{-11}$
\end{tabular}  &
\begin{tabular}[t]{@{} l l @{}}
    & $ 1.28^{+0.30}_{-0.29}$ \\
    & $ 2.48^{+0.02}_{-0.06}$ \\
    & $ 4.30^{+0.05}_{}$ \\
    & $ 4.37^{+0.27}_{-0.19}$ \\
    & $ 34.55^{+20.34}_{-13.61}$
\end{tabular} & 
\begin{tabular}[t]{@{} l l @{}}
    & $ 1.48^{+0.31}_{-0.25}$ \\
    & $ 2.52^{+0.09}_{-0.08}$ \\
    & $ 4.30^{+0.18}_{}$ \\
    & $ 4.46^{+0.08}_{-0.26}$ \\
    & $ 18.0^{+14.}_{-9.5}$
\end{tabular}
\end{tabularx}
\end{adjustwidth}
\end{table*}
\end{turnpage}
\section{Crosschecks with previous results}\label{sec:estes_crosschecks}
\begin{figure}[h!]
    \begin{center}
    \includegraphics[width=1\linewidth]{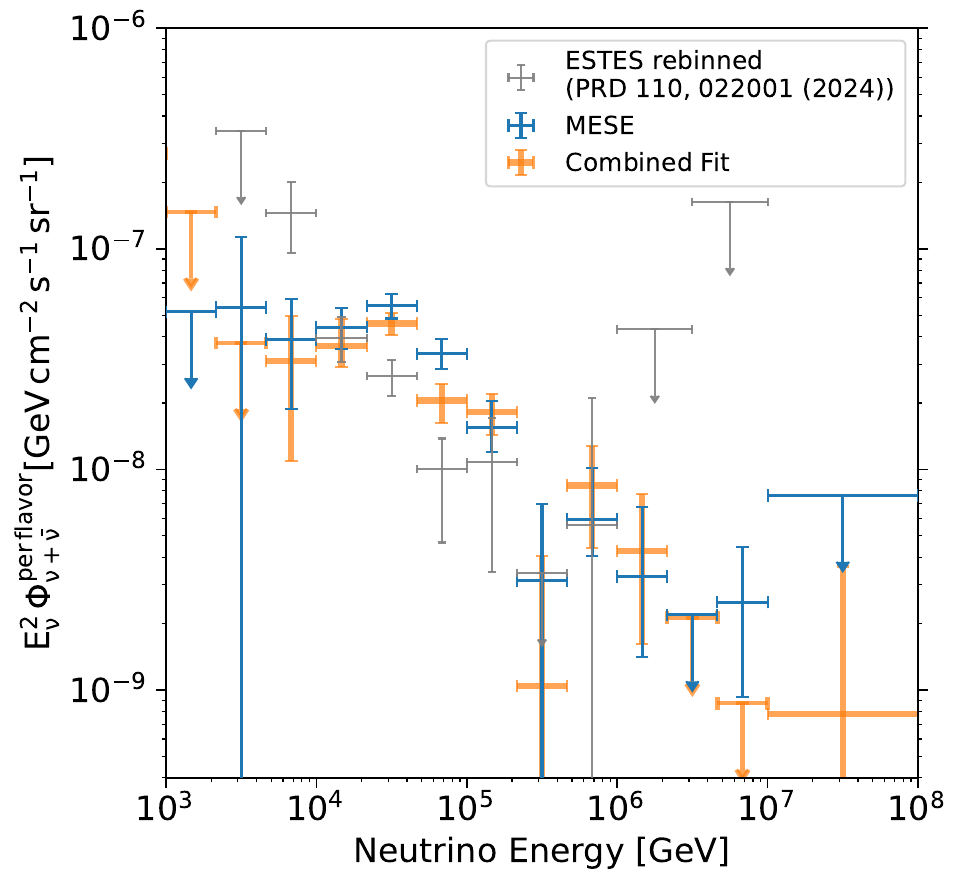}
    \end{center}
    \caption{Segmented fits for the MESE, CF, and ESTES \cite{ESTES} with similar energy bins. The ESTES fits shown here represent the rebinned post-unblinding fit reported in \cite{ESTES}, which has a binning scheme more consistent with that of MESE and CF.}
    \label{fig:segmentedfit_cf_estes}
\end{figure}

The segmented spectra presented here are consistent with previously reported results with HESE~\cite{HESE7.5}, the cascades sample~\cite{SBUCascades}, the tracks sample~\cite{NorthernTracks} and the 2-year MESE sample~\cite{MESE_2yr}. However, there is a large overlap in the those data selections with the data samples shown in this paper.
An exception is the Enhanced Starting Track Event Selection (ESTES)~\cite{ESTES}, where only $\sim \SI{30}{\percent}$ of MESE tracks are included in the ESTES sample and $\sim \SI{0.7}{\percent}$ of events in the tracks sample exist in ESTES. There is no overlap with the cascade sample of the CF and the MESE cascades, both of which dominate the segmented spectra of the respective analysis. The best fit astrophysical neutrino flux with the ESTES sample was consistent with a single power law~\cite{ESTES}, which was also the case for an earlier measurement with a starting event sample developed to measure inelasticity~\cite{IceCube:2018pgc}.
A comparison of the segmented fits of MESE and CF with that of ESTES is shown in Fig \ref{fig:segmentedfit_cf_estes}. The segmented fit from ESTES was rebinned to match the previous cascades analysis~\cite{SBUCascades} and reported in~\cite{ESTES}, and we use this for the comparison shown in the figure.
An apparent difference is the non-observation of a break in ESTES, which, however, relies on a single data point at the bin of 4-10~TeV. This point is in tension with the results shown in this paper at the 2$\,\sigma$ level. 
More significant at the 3$\,\sigma$ level is the deviation at \SIrange{21}{46}{TeV}, where the ESTES flux is smaller than the results reported here.
There exist several notable differences between the ESTES analysis and the analyses presented here.
The ESTES analysis was based on a single flavor fit of muon neutrinos with different angular acceptances and backgrounds when compared to the MESE and the CF analyses. ESTES used a different energy reconstruction when compared to the tracks sample of the CF, which can result in a shift of energy scale,
a different treatment of systematic uncertainties, and a different parameterization of the atmospheric and self-veto uncertainties. 
With these differences, the origin of the tension is not fully understood. 
A comparison of the results obtained with the MESE and CF analyses reported in this paper with previous measurements from IceCube is shown in Tab.~\ref{tab:mese_results_icecube_comparison}.
Future analyses will revisit these differences and make use of a more consistent systematics treatment across analyses, or even integrate the ESTES data into a fully combined fit with the data samples used here.
\begin{table}[h!]
\caption{A comparison of past results from IceCube and the results from the Combined Fit and MESE analysis. The uncertainties are derived from 1D profile likelihood scans, assuming Wilks' theorem. 
The flux is measured in units of $\SI{e-18}{GeV^{-1} cm^{-2} s^{-1} sr^{-1}}$ and the normalization is determined at an energy of \SI{100}{TeV}.}
    \label{tab:mese_results_icecube_comparison}
    \renewcommand{\arraystretch}{1.5}
\centering
\begin{tabular}{c|c}
Analysis & Best Fit Model Parameters \\ \hline 
Cascades SPL~\cite{SBUCascades}          &   \multicolumn{1}{c}{$\begin{array}{ccl}
          \phi_0 \, &=&1.66^{+0.25}_{-0.27} \\
          \gamma&=&2.53^{+0.07}_{-0.07} \\
         \end{array}$}      \\ \hline
HESE SPL~\cite{HESE7.5}        &    \multicolumn{1}{c}{$\begin{array}{ccl}
          \phi_0 \, &=&6.37^{+1.46}_{-1.62} \\
          \gamma&=&2.87^{+0.21}_{-0.19} \\
         \end{array}$}     \\ \hline
Tracks SPL~\cite{NorthernTracks}           &   \multicolumn{1}{c} {$\begin{array}{ccl}
          \phi_0 \, &=&1.44^{+0.25}_{-0.26} \\
          \gamma&=&2.37^{+0.09}_{-0.09} \\
         \end{array}$} 
         \\ \hline
ESTES SPL~\cite{ESTES}           &   \multicolumn{1}{c} {$\begin{array}{ccl}
          \phi_0 \, &=&1.68^{+0.19}_{-0.22} \\
          \gamma&=&2.58^{+0.10}_{-0.09} \\
         \end{array}$} 
         \\ \hline
CF BPL (this work)           &   \multicolumn{1}{c} {$\begin{array}{ccl}
          \phi_0 \, &=&1.77^{+0.19}_{-0.18} \\    \gamma_1&=&1.31^{+0.51}_{-1.30} \\
\gamma_2&=&2.735^{+0.067}_{-0.075} \\
\\log_{10}(\frac{\mathrm{E}_\mathrm{break}}{\mathrm{GeV}})&=&4.39^{+0.1}_{-0.1} \\
         \end{array}$} 
         \\ \hline
MESE BPL (this work)           &   \multicolumn{1}{c} {$\begin{array}{ccl}
          \phi_0 \, &=&2.28^{+0.22}_{-0.20} \\    \gamma_1&=&1.72^{+0.26}_{-0.35} \\
\gamma_2&=&2.839^{+0.11}_{-0.091} \\
\\log_{10}(\frac{\mathrm{E}_\mathrm{break}}{\mathrm{GeV}})&=&4.524^{+0.097}_{-0.087} \\
         \end{array}$} 
         \\ \hline
\end{tabular}
\end{table}

%
%
\nocite{*}
\bibliography{references}

\end{document}